\documentclass[american]{llncs}
\usepackage[T1]{fontenc}
\usepackage{babel}

\makeatletter

\providecommand{\LyX}{L\kern-.1667em\lower.25em\hbox{Y}\kern-.125emX\@}

\usepackage{latexsym}
\usepackage{amssymb}

\makeatother
\begin{document}

\vfill{}
\title{Transformation Rules\\
for Locally Stratified Constraint Logic Programs}
\vfill{}

\author{Fabio Fioravanti\( ^{1} \), Alberto Pettorossi{\small \( ^{2} \)},
Maurizio Proietti{\small \( ^{3} \) }}

\institute{(1) Dipartimento di Informatica, Università dell'Aquila, L'Aquila,
Italy\\
\texttt{fioravan@univaq.it}\\
(2) DISP, University of Tor Vergata, Roma, Italy\texttt{}~\\
\texttt{adp@iasi.rm.cnr.it}\\
(3) IASI-CNR, Roma, Italy\texttt{}~\\
\texttt{proietti@iasi.rm.cnr.it}}

\maketitle
\begin{abstract}
We propose a set of transformation rules for constraint logic programs
with negation. We assume that every program is locally stratified
and, thus, it has a unique perfect model. We give sufficient conditions
which ensure that the proposed set of transformation rules preserves
the perfect model of the programs. Our rules extend in some respects
the rules for logic programs and constraint logic programs already
considered in the literature and, in particular, they include a rule
for unfolding a clause with respect to a negative literal.
\end{abstract}

\section{Introduction\label{sec:intro}}

Program transformation is a very powerful methodology for developing
correct and efficient programs from formal specifications. This methodology
is particularly convenient in the case of declarative programming
languages, where programs are formulas and program transformations
can be viewed as replacements of formulas by new, equivalent formulas. 

The main advantage of using the program transformation methodology
for program development is that it allows us to address the correctness
and the efficiency issues at separate stages. Often little effort
is required for encoding formal specifications (written by using equational
or logical formalisms) as declarative programs (written as functional
or logic programs). These programs are correct by construction, but
they are often computationally inefficient. Here is where program
transformation comes into play: from a correct (and possibly inefficient)
initial program version we can derive a correct and efficient program
version by means of a sequence of program transformations that preserve
correctness. We say that a program transformation preserves correctness,
or it is \emph{correct}, if the semantics of the initial program is
equal to the semantics of the derived program.

A very popular approach followed when applying the program transformation
methodology, is the one based on \emph{transformation rules}
and \emph{strategies}~\cite{BuD77}:
the rules are elementary transformations that preserve the program
semantics and the strategies are (possibly nondeterministic) procedures
that guide the application of transformation rules with the objective
of deriving efficient programs. Thus, a program transformation is
realized by a sequence \( P_{0},\ldots ,P_{n} \) of programs, called
a \emph{transformation sequence}, where, for \( i=0,\ldots ,n\! -\! 1 \),
\( P_{k+1} \) is derived from \( P_{k} \) by applying a transformation
rule according to a given transformation strategy. A transformation
sequence is said to be \emph{correct} if the programs \( P_{0},\ldots ,P_{n} \)
have the same semantics.

Various sets of program transformation rules have been proposed in
the literature for several declarative programming languages, such
as, functional~\cite{BuD77,San96}, logic~\cite{TaS84}, constraint~\cite{BeG98,EtG96,Mah93},
and functional-logic languages~\cite{Al&99}. In this paper we consider
a constraint logic programming language with negation~\cite{JaM94,MaS98}
and we study the correctness of a set of transformation rules that
extends the sets which were already considered for constraint logic
programming languages. We will not deal here with transformation strategies,
but we will show through some examples (see Section~\ref{sec:examples})
that the transformation rules can be applied in a rather systematic
(yet not fully automatic) way.

We assume that constraint logic programs are \emph{locally stratified}~\cite{ApB94,Prz87}.
This assumption simplifies our treatment because the semantics of
a locally stratified program is determined by its unique \emph{perfect
model} which is equal to its unique \emph{stable model}, which is
also its unique, total \emph{well-founded model}~\cite{ApB94,Prz87}.
(The definitions of locally stratified programs, perfect models, and
other notions used in this paper are recalled in Section~\ref{sec:preliminaries}.)

The set of transformation rules we consider in this paper includes
the \emph{unfolding} and \emph{folding} rules (see, for instance,
\cite{BeG98,EtG96,GaS91,GeK94,KaH87,Mah93,PeP94,PeP00a,Ro&02,Ro&99a,Sat92,Sek91,Sek93,TaS84}).
In order to understand how these rules work, let us first consider
propositional programs. The \emph{definition} of an atom \( a \)
in a program is the set of clauses that have \( a \) as head. The
atom \( a \) is also called the \emph{definiendum}. The disjunction
of the bodies of the clauses that constitute the definition of \( a \),
is called the \emph{definiens}. Basically, the application of the
unfolding rule consists in replacing an atom occurring in the body
of a clause by its definiens and then applying, if necessary, some
suitable boolean laws to obtain clauses. For instance, given the following
programs \( P_{1} \) and \( P_{2} \): 
\smallskip{}

{\noindent \centering \begin{tabular}{llll}
\( P_{1} \):~~&
\( p\leftarrow q\wedge r \) ~~~~~~~~~~~~~~~~~~~~~~&
\( P_{2} \):~~&
\( p\leftarrow \neg a\wedge r \)\\
&
\( q\leftarrow \neg a \)&
&
\( p\leftarrow b\wedge r \)\\
&
\( q\leftarrow b \)&
&
\( q\leftarrow \neg a \)\\
&
&
&
\( q\leftarrow b \)\\
\end{tabular}\par}
\smallskip{}

\noindent we have that by unfolding the first clause of program \( P_{1} \)
we get program \( P_{2} \).

Folding is the inverse of unfolding and consists in replacing an occurrence
of a definiens by the corresponding occurrence of the definiendum
(before this replacement we may apply suitable boolean laws). For
instance, by folding the first two clauses of \( P_{2} \) using the
definition of \( q \), we get program \( P_{1} \). An important
feature of the folding rule is that the definition used for folding
may occur in a previous program in the transformation sequence. The
formal definitions of the unfolding and folding transformation rules
for constraint logic programs will be given in Section~\ref{sec:rules}.
The usefulness of the program transformation approach based on the
unfolding and folding rules, is now very well recognized in the scientific
community as indicated by a large number of papers (see \cite{PeP94}
for a survey).

A relevant property we will prove in this paper is that the unfolding
of a clause w.r.t.~an atom occurring in a negative literal, also
called \emph{negative unfolding}, preserves the perfect model of a
locally stratified program. This property is interesting, because
negative unfolding is useful for program transformation, but it may
\emph{not} preserve the perfect models (nor the stable models, nor
the well-founded model) if the programs are not locally stratified.
For instance, let us consider the following programs \( P_{1} \)
and \( P_{2} \):
\smallskip{}

{\noindent \centering \begin{tabular}{llll}
\( P_{1} \):~~&
\( p\leftarrow \neg q \) ~~~~~~~~~~~~~~~~~~~~~~&
\( P_{2} \):~~&
\( p\leftarrow p \)\\
&
\( q\leftarrow \neg p \)&
&
\( q\leftarrow \neg p \)\\
\end{tabular}\par}
\smallskip{}

\noindent Program \( P_{2} \) can be obtained by unfolding the first
clause of \( P_{1} \) (i.e., by first replacing \( q \) by the body
\( \neg p \) of the clause defining \( q \), and then replacing
\( \neg \neg p \) by \( p \)). Program \( P_{1} \) has two perfect
models: \( \{p\} \) and \( \{q\} \), while program \( P_{2} \)
has the unique perfect model \( \{q\} \). 

In this paper we consider the following transformation rules (see
Section~\ref{sec:rules}): definition introduction and definition
elimination (for introducing and eliminating definitions of predicates),
positive and negative unfolding, positive and negative folding (that
is, unfolding and folding w.r.t.~a positive and a negative occurrence
of an atom, respectively), and also rules for applying boolean laws
and rules for manipulating constraints.
\smallskip{}

Similarly to other sets of transformation rules presented in the literature
(see, for instance, \cite{Al&99,BeG98,BuD77,EtG96,Mah93,San96,TaS84}),
a transformation sequence constructed by arbitrary applications of
the transformation rules presented in this paper, may be incorrect.
As customary, we will ensure the correctness of transformation sequences
only if they satisfy suitable properties: we will call them \emph{admissible}
sequences (see Section~\ref{sec:correctness}). Although our transformation
rules are extensions or adaptations of transformation rules already
considered for stratified logic programs or logic programs, in general,
for our correctness proof we cannot rely on already known results.
Indeed, the definition of an admissible transformation sequence depends
on the interaction among the rules and, in particular, correctness
may not be preserved if we modify even one rule only.

To see that known results do not extend in a straightforward way when
adding negative unfolding to a set of transformation rules, let us
consider the transformation sequences constructed by first (1)~unfolding
all clauses of a definition \( \delta  \) and then (2)~folding some
of the resulting clauses by using the definition \( \delta  \) itself.
If at Step~(1) we use positive unfolding only, then the perfect model
semantics is preserved~\cite{Ro&02,Sek91}, while this semantics
may not be preserved if we use negative unfolding, as indicated by
the following example.

\begin{example}
\label{ex:last_intro}Let us consider the transformation sequence
\( P_{0},P_{1},P_{2} \), where:
\smallskip{}

\noindent \begin{tabular}{llllll}
\( P_{0} \):~ &
\( p(X)\leftarrow \neg q(X) \)~~~~&
\( P_{1} \):~&
\( p(X)\leftarrow X\! <\! 0\wedge \neg q(X) \)~~~~&
\( P_{2} \):~ &
\( p(X)\leftarrow X\! <\! 0\wedge p(X) \)\\
&
\( q(X)\leftarrow X\! \geq 0 \)&
&
\( q(X)\leftarrow X\! \geq 0 \)&
&
\( q(X)\leftarrow X\! \geq 0 \)\\
&
\( q(X)\leftarrow q(X) \)&
&
\( q(X)\leftarrow q(X) \)&
&
\( q(X)\leftarrow q(X) \)\\
\end{tabular} 
\smallskip{}

\noindent Program \( P_{1} \) is derived by unfolding the first clause
of \( P_{0} \) w.r.t.~the negative literal \( \neg q(X) \) (that
is, by replacing the definiendum \( q(X) \) by its definiens \( X\! \geq 0\vee q(X) \),
and then applying De Morgan's law). Program \( P_{2} \) is derived
by folding the first clause of \( P_{1} \) using the definition \( p(X)\leftarrow \neg q(X) \)
in \( P_{0} \). We have that, for any \( a\! <\! 0 \), the atom
\( p(a) \) belongs to the perfect model of \( P_{0} \), while \( p(a) \)
does not belong to the perfect model of \( P_{2} \).
\end{example}
The main result of this paper (see Theorem~\ref{th:corr_of_rules}
in Section~\ref{sec:correctness}) shows the correctness of a transformation
sequence constructed by first (1)~unfolding all clauses of a (non-recursive)
definition \( \delta  \) w.r.t.~a \emph{positive} literal, then
(2)~unfolding zero or more clauses w.r.t.~a \emph{negative} literal,
and finally (3)~folding some of the resulting clauses by using the
definition \( \delta  \). The correctness of such transformation
sequences cannot be established by the correctness results presented
in~\cite{Ro&02,Sek91}.

The paper is structured as follows. In Section~\ref{sec:preliminaries}
we present the basic definitions of locally stratified constraint
logic programs and perfect models. In Section~\ref{sec:rules} we
present our set of transformation rules and in Section~\ref{sec:correctness}
we give sufficient conditions on transformation sequences that ensure
the preservation of perfect models. In Section~\ref{sec:examples}
we present some examples of program derivation using our transformation
rules. In all these examples the negative unfolding rule plays a crucial
role. Finally, in Section~\ref{sec:related_work} we discuss related
work and future research.

\section{Preliminaries \label{sec:preliminaries}}

In this section we recall the syntax and semantics of constraint logic
programs with negation. In particular, we will give the definitions
of locally stratified programs and perfect models. For notions not
defined here the reader may refer to \cite{Apt90,ApB94,JaM94,Ja&98,Llo87}.

\subsection{Syntax of Constraint Logic Programs \label{subsec:clp_syntax}}

We consider a first order language \( \mathcal{L} \) generated by
an infinite set \emph{Vars} of \emph{variables}, a set \emph{Funct}
of \emph{function symbols} with arity, and a set \emph{Pred} of \emph{predicate
symbols} (or predicates, for short) with arity. \emph{}We assume that
\emph{Pred} is the union of two disjoint sets: (i) the set \emph{Pred\( _{c} \)}
of \emph{constraint} predicate symbols, including the \emph{equality}
symbol \( = \), and (ii) the set \emph{Pred}\( _{u} \) of \emph{user
defined} predicate symbols\emph{.} 

A \emph{term} of \( \mathcal{L} \) is either a variable or an expression
of the form \( f(t_{1},\ldots ,t_{n}) \), where \( f \) is an \( n \)-ary
function symbol and \( t_{1},\ldots ,t_{n} \) are terms. An \emph{atomic
formula} is an expression of the form \( p(t_{1},\ldots ,t_{n}) \)
where \( p \) is an \( n \)-ary predicate symbol and \( t_{1},\ldots ,t_{n} \)
are terms. A \emph{formula} of \( \mathcal{L} \) is either an atomic
formula or a formula constructed from atomic formulas by means of
connectives (\( \neg  \), \( \wedge  \), \( \vee  \), \( \rightarrow  \),
\( \leftarrow  \), \( \leftrightarrow  \)) and quantifiers (\( \exists  \),
\( \forall  \)).

Let \( e \) be a term, or a formula, or a set of terms or formulas.
The set of variables occurring in \( e \) is denoted by \( \mathit{vars}(e) \).
Given a formula \( \varphi  \), the set of the \emph{free variables}
occurring in \( \varphi  \) is denoted by \( FV(\varphi ) \). A
term or a formula is \emph{ground} iff it \emph{}does not contain
variables. Given a set \( X=\{X_{1},\ldots ,X_{n}\} \) of \( n \)
variables, by \( \forall X\, \varphi  \) we denote the formula \( \forall X_{1}\ldots \forall X_{n}\, \varphi  \).
By \( \forall (\varphi ) \) we denote the \emph{universal closure}
of \( \varphi  \), that is, the formula \( \forall X\, \varphi  \),
where \( FV(\varphi )=X \). Analogous notations will be adopted for
the existential quantifier \( \exists  \).

A \emph{primitive constraint} is an atomic formula \( p(t_{1},\ldots ,t_{n}) \)
where \( p \) is a predicate symbol in \emph{Pred\( _{c} \)}. The
set \( \mathcal{C} \) of \emph{constraints} is the smallest set of
formulas of \( \mathcal{L} \) that \emph{}contains all primitive
constraints and is closed w.r.t.~negation, conjunction, and existential
quantification. This closure assumption simplifies our treatment,
but as we will indicate at the end of this section, we can do without
it.

An \emph{atom} is an atomic formula \( p(t_{1},\ldots ,t_{n}) \)
where \( p \) is an element of \emph{Pred\( _{u} \)} and \( t_{1},\ldots ,t_{n} \)
are terms. A \emph{literal} is either an atom \( A \), also called
\emph{positive literal}, or a negated atom \( \neg \, A \), also
called \emph{negative literal}. Given any literal \( L \), by \( \overline{{L}} \)
we denote: (i)~\( \neg A \), if \( L \) is the atom \( A \), and
(ii)~\( A \), if \( L \) is the negated atom \( \neg A \). A \emph{goal}
is a (possibly empty) conjunction of literals (here we depart from
the terminology used in \cite{Apt90,Llo87}, where a goal is defined
as the negation of a conjunction of literals). A \emph{constrained
literal} is the conjunction of a constraint and a literal. A \emph{constrained
goal} is the conjunction of a constraint and a goal. 

A \emph{clause \( \gamma  \)} is a formula of the form \( H\leftarrow c\wedge G \),
where: (i) \( H \) is an atom, called the \emph{head} of \( \gamma  \)
and denoted \( hd(\gamma ) \), and (ii) \( c\wedge G \) is a constrained
goal, called the \emph{body} of \( \gamma  \) and denoted \( bd(\gamma ) \).
A conjunction of constraints and/or literals may be empty (in which
case it is equivalent to \emph{true}). A clause of the form \( H\leftarrow c \),
where \( c \) is a constraint and the goal part of the body is the
empty conjunction of literals, is called a \emph{constrained fact}.
A clause of the form \( H\leftarrow  \), whose body is the empty
conjunction, is called a \emph{fact.}

A \emph{constraint logic program} (or \emph{program}, for short) is
a finite set of clauses. A \emph{definite clause} is a clause whose
body has no occurrences of negative literals. A \emph{definite program}
is a finite set of definite clauses. 

Given two atoms \( p(t_{1},\ldots ,t_{n}) \) and \( p(u_{1},\ldots ,u_{n}) \),
we denote by \( p(t_{1},\ldots ,t_{n}) \) = \( p(u_{1},\ldots ,u_{n}) \)
the constraint: \( t_{1}\! =\! u_{1}\wedge \ldots \wedge t_{n}\! =\! u_{n} \).
For the notion of \emph{substitution} and for the application of a
substitution to a term we refer to \cite{Apt90,Llo87}. Given a formula
\( \varphi  \) and a substitution \( \{X_{1}/t_{1},\ldots ,X_{n}/t_{n}\} \)
we denote by \( \varphi \{X_{1}/t_{1},\ldots ,X_{n}/t_{n}\} \) the
result of simultaneously replacing in \( \varphi  \) all free occurrences
of \( X_{1},\ldots ,X_{n} \) by \( t_{1},\ldots ,t_{n} \). 

We say that a predicate \( p \) \emph{immediately depends on} a predicate
\( q \) in a program \( P \) iff there exists in \( P \) a clause
of the form \( p(\ldots )\leftarrow B \) and \( q \) occurs in \( B \).
We say that \( p \) \emph{depends on} \( q \) in \( P \) iff there
exists a sequence \( p_{1},\ldots ,p_{n} \), with \( n\! >\! 1 \),
of predicates such that: (i)~\( p_{1}=p \), (ii)~\( p_{n}=q \),
and (iii)~for \( i=1,\ldots ,n\! -\! 1 \), \( p_{i} \) immediately
depends on \( p_{i+1} \). Given a user defined predicate \( p \)
and a program \( P \), the \emph{definition of \( p \) in} \( P \),
denoted \( \mathit{Def}(p,P) \), is the set of clauses \( \gamma  \)
in \( P \) such that \( p \) is the predicate symbol of \( hd(\gamma ) \).

A \emph{variable renaming} is a bijective mapping from \emph{Vars}
to \emph{Vars}. The application of a variable renaming \( \rho  \)
to a formula \( \varphi  \) returns the formula \( \rho (\varphi ) \),
which is said to be a \emph{variant} of \( \varphi  \), obtained
by replacing each (bound or free) variable occurrence \( X \) in
\( \varphi  \) by the variable \( \rho (X) \). A variant of a set
\( \{\varphi _{1},\ldots ,\varphi _{n}\} \) of formulas is the set
\( \{\rho (\varphi _{1}),\ldots ,\rho (\varphi _{n})\} \), also denoted
\( \rho (\{\varphi _{1},\ldots ,\varphi _{n}\}) \). During program
transformation we will feel free to silently apply variable renamings
to clauses and to sets of clauses because, as the reader may verify,
they preserve program semantics (see Section~\ref{subsec:clp_semantics}).
Moreover, we will feel free to change the names of the bound variables
occurring in constraints, as usually done in predicate calculus.

\subsection{Semantics of Constraint Logic Programs \label{subsec:clp_semantics}}

\smallskip{}
In this section we present the definition of the semantics of constraint
logic programs with negation. This definition extends similar definitions
given in the literature for definite constraint logic programs~\cite{JaM94}
and logic programs with negation~\cite{ApB94,Prz87}.

We proceed as follows: (i)~we define an \emph{interpretation for
the constraints}, following the approach used in first order logic
(see, for instance, \cite{Apt90}), (ii)~we introduce the notion
of \( \mathcal{D} \)\emph{-model}, that is, a model for constraint
logic programs which is parametric w.r.t.~the interpretation \emph{\( \mathcal{D} \)}
for the \emph{}constraints, (iii)~we introduce the notion of \emph{locally
stratified} \emph{program}, and finally, (iv)~we define the \emph{perfect}
\( \mathcal{D} \)\emph{-model} (also called \emph{perfect model},
for short) of locally stratified programs.

An \emph{interpretation} \( \mathcal{D} \) \emph{for the} \emph{constraints}
consists of: (1)~a non-empty set \( D \), called \emph{carrier},
(2)~an assignment of a function \( f \){\small $\!_\mathcal D$}\( :D^{n}\rightarrow D \)
to each \emph{n}-ary function symbol \( f \) in \emph{Funct}, and
(3)~an assignment of a relation \( p \){\small $_\mathcal D$} over
\( D^{n} \) to each \emph{n}-ary predicate symbol in \( \mathit{Pred}_{c} \).
In particular, \( \mathcal{D} \) assigns the set \( \{\langle d,d\rangle \, |\, d\in D\} \)
to the equality symbol \( = \).

We assume that \( D \) is a set of ground terms. This is not restrictive
because we may add suitable 0-ary function symbols to \( \mathcal{L} \).

Given a formula \( \varphi  \) whose predicate symbols belong to
\( \mathit{Pred}_{c} \), we consider the satisfaction relation \( \mathcal{D}\models \varphi  \),
which is defined as usual in first order predicate calculus (see,
for instance, \cite{Apt90}). A constraint \( c \) is said to be
\emph{satisfiable} iff its existential closure is satisfiable, that
is, \( \mathcal{D}\models \exists (c) \). If \( \mathcal{D}\not \models \exists (c) \),
then \( c \) is said to be \emph{unsatisfiable} in \( \mathcal{D} \).

Given an interpretation \( \mathcal{D} \) for the constraints, a
\( \mathcal{D} \)-\emph{interpretation} \emph{I} assigns a relation
over \( D^{n} \) to each \emph{n}-ary user defined predicate symbol
in \( \mathit{Pred}_{u} \), that is, \( I \) can be identified with
a subset of the set \( \mathcal{B}_{\mathcal{D}} \) of ground atoms
defined as follows:

\smallskip{}
{\centering \( \mathcal{B}_{\mathcal{D}}=\{p(d_{1},\ldots ,d_{n})\, |\, p \)
is a predicate symbol in \( \mathit{Pred}_{u} \) and \( (d_{1},\ldots ,d_{n})\in D^{n}\}. \)\par}
\smallskip{}

\noindent A \emph{valuation} is a function \( v \): \( \mathit{Vars}\rightarrow D \).
We extend the domain of a valuation \( v \) to terms, constraints,
literals, and clauses as we now indicate. Given a term \( t \), we
inductively define the term \( v(t) \) as follows: (i) if \( t \)
is a variable \( X \) then \( v(t)=v(X) \), and (ii) if \( t \)
is \( f(t_{1},\ldots ,t_{n}) \) then \( v(t)=f \){\small $\!_\mathcal D$}\( (v(t_{1}),\ldots ,v(t_{n})) \).
Given a constraint \( c \), \( v(c) \) is the constraint obtained
by replacing every free variable \( X\in FV(c) \) by the ground term
\( v(X) \). Notice that \( v(c) \) is a closed formula which may
be not ground. Given a literal \( L \), (i) if \( L \) is the atom
\( p(t_{1},\ldots ,t_{n}) \), then \( v(L) \) is the ground atom
\( p(v(t_{1}),\ldots ,v(t_{n})) \), and (ii) if \( L \) is the negated
atom \( \neg A \), then \( v(L) \) is the ground, negated atom \( \neg v(A) \).
Given a clause \( \gamma  \): \( H\leftarrow c\wedge L_{1}\wedge \ldots \wedge L_{m} \),
\( v(\gamma ) \) is the clause \( v(H)\leftarrow v(c)\wedge v(L_{1})\wedge \ldots \wedge v(L_{m}) \).

Let \( I \) be a \( \mathcal{D} \)-interpretation and \( v \) a
valuation. Given a literal \( L \), we say that \( v(L) \) \emph{is
true in} \( I \) iff either (i)~\( L \) is an atom and \( v(L)\in I \),
or (ii)~\( L \) is a negated atom \( \neg A \) and \( v(A)\not \in I \).
We say that the literal \( v(L) \) \emph{is false in} \( I \) iff
it is not true in \( I \). Given a clause \( \gamma  \):~\( H\leftarrow c\wedge L_{1}\wedge \ldots \wedge L_{m} \),
\( v(\gamma ) \) \emph{is true in} \( I \) iff either (i)~\( v(H) \)
is true in \( I \), or (ii)~\( \mathcal{D}\not \models v(c) \),
or (iii)~there exists \( i\in \{1,\ldots ,m\} \) such that \( v(L_{i}) \)
is false in \( I \). 

A \( \mathcal{D} \)-interpretation \( I \) is a \( \mathcal{D} \)-\emph{model}
of a program \( P \) iff for every clause \( \gamma  \) in \( P \)
and for every valuation \( v \), we have that \( v(\gamma ) \) is
true in \( I \). It can be shown that every definite constraint logic
program \( P \) has a \emph{least} \( \mathcal{D} \)-model w.r.t.
set inclusion (see, for instance \cite{Ja&98}). 

Unfortunately, constraint logic programs which are not definite may
fail to have a least \( \mathcal{D} \)-model. For example, the program
consisting of the clause \( p\leftarrow \neg q \) has the two minimal
(not least) models \( \{p\} \) and \( \{q\} \). This fact has motivated
the introduction of the set of \emph{locally stratified} programs~\cite{ApB94,Prz87}.
For every locally stratified program one can associate a unique (minimal,
but not least, w.r.t.~set inclusion) model, called \emph{perfect}
\emph{model}, as follows.

A \emph{local stratification} is a function \( \sigma  \): \( \mathcal{B}_{\mathcal{D}}\rightarrow W \),
where \( W \) is the set of countable ordinals. If \( A\in \mathcal{B}_{\mathcal{D}} \)
and \emph{\( \sigma (A) \)} is the ordinal \emph{\( \alpha  \),}
we say that the \emph{stratum} of \( A \) is \( \alpha  \). Given
a clause \( \gamma  \) in a program \( P \), a valuation \( v \),
and a local stratification \( \sigma  \), we say that a clause \( v(\gamma ) \)
of the form: \( H\leftarrow c\wedge L_{1}\wedge \ldots \wedge L_{m} \)
is \emph{locally stratified} w.r.t.~\( \sigma  \) iff \emph{either}
\( \mathcal{D}\models \neg c \) \emph{or}, for \( i=1,\ldots ,m \),
if \emph{}\( L_{i} \) is an atom \( A \) then \emph{\( \sigma (H)\geq \sigma (A) \)}
else if \( L_{i} \) is a negated atom \( \neg A \) then \emph{\( \sigma (H)>\sigma (A) \)}.
Given a local stratification \( \sigma  \), we say that program \( P \)
is \emph{locally stratified w.r.t.~}\( \sigma  \), or \( \sigma  \)
is a \emph{local stratification for} \( P \), iff for every clause
\( \gamma  \) in \( P \) and for every valuation \( v \), the clause
\( v(\gamma ) \) is locally stratified w.r.t.~\( \sigma  \). A
program \( P \) is \emph{locally stratified} iff there exists a local
stratification \( \sigma  \) such that \( P \) is \emph{locally
stratified w.r.t.~}\( \sigma  \). For instance, let us consider
the following program \emph{Even}:
\smallskip{}

\( \mathit{even}(0)\leftarrow  \)

\( \mathit{even}(X)\leftarrow X\! =\! Y\! +\! 1\, \, \wedge \, \, \neg \mathit{even}(Y) \)
\smallskip{}

\noindent where the interpretation for the constraints is as follows:
(1)~the carrier is the set of the natural numbers, and (2)~the addition
function is assigned to the function symbol \( + \). The program
\emph{Even} is locally stratified w.r.t.~the stratification function
\( \sigma  \) such that for every natural number \( n \), \emph{\( \sigma (\mathit{even}(n))=n \).}

The perfect model of a program \( P \) which is locally stratified
w.r.t.~a stratification function \( \sigma  \) is the least \emph{\( \mathcal{D} \)-}model
of \( P \) w.r.t.~a suitable ordering based on \( \sigma  \), as
specified by the following definition. This ordering is, in general,
different from set inclusion.

\begin{definition}
\textup{(}Perfect Model\textup{)}~\textup{\cite{Prz87}}. Let \( P \)
be a locally stratified program, let \( \sigma  \) be any local stratification
for \( P \), and let \( I \), \( J \) be \( \mathcal{D} \)-interpretations.
We say that \( I \) is \emph{preferable} to \( J \), and we write
\( I\! \prec \! J \) iff for every \( A_{1}\in I\! -\! J \) there
exists \( A_{2}\in J\! -\! I \) such that \( \sigma (A_{1})>\sigma (A_{2}) \).
A \( \mathcal{D} \)-model \( M \) of \( P \) is called a \emph{perfect}
\( \mathcal{D} \)-\emph{model (}or a \emph{perfect model,} for short\emph{)}
iff for every \( \mathcal{D} \)-model \( N \) of \( P \) different
from \( M \), we have that \( M\! \prec \! N \).
\end{definition}
It can be shown that the perfect model of a locally stratified program
always exists and does not depend on the choice of the local stratification
function \( \sigma  \), as stated by the following theorem. 

\begin{theorem}
\textup{\cite{Prz87}} \label{th:unique_pm}Every locally stratified
program \( P \) has a unique perfect model \( M(P) \).
\end{theorem}
\noindent By Theorem~\ref{th:unique_pm}, \( M(P) \) is the least
\emph{\( \mathcal{D} \)-}model of \( P \) w.r.t.~the \( \prec  \)
ordering. For instance, the perfect model of the program consisting
of the clause \( p\leftarrow \neg q \) is \( \{p\} \) because \( \sigma (p)>\sigma (q) \)
and, thus, the \mbox{\emph{\( \mathcal{D} \)-}model} \( \{p\} \)
is preferable to the \mbox{\emph{\( \mathcal{D} \)-}model} \( \{q\} \)
(i.e., \( \{p\}\! \prec \! \{q\} \) ). Similarly, it can be verified
that the perfect model of the program \emph{Even} is \( M(\mathit{Even})=\{\mathit{even}(n)\, |\, n \)
is an even non-negative integer\( \} \). In Section~\ref{sec:correctness}
we will provide a method for constructing the perfect model of a locally
stratified program based on the notion of \emph{proof tree}.

Let us conclude this section by showing that the assumption that the
set \( \mathcal{C} \) of constraints is closed w.r.t.~negation,
conjunction, and existential quantification is not really needed.
Indeed, given a locally stratified clause \( H\leftarrow c\wedge G \),
where the constraint \( c \) is written by using negation, or conjunction,
or existential quantification, we can replace \( H\leftarrow c\wedge G \)
by an equivalent set of locally stratified clauses. For instance,
if \( c \) is \( \exists X\, d \) then we can replace \( H\leftarrow c\wedge G \)
by the two clauses:
\smallskip{}

\( H\leftarrow \mathit{newp}(Y_{1},\ldots ,Y_{n})\wedge G \)

\( \mathit{newp}(Y_{1},\ldots ,Y_{n})\leftarrow d \)
\smallskip{}

\noindent where \( \mathit{newp} \) is a new, user defined predicate
and \( \{Y_{1},\ldots ,Y_{n}\}=FV(\exists X\, d) \). Analogous replacements
can be applied in the case where a constraint is written by using
negation or conjunction.

\section{The Transformation Rules \label{sec:rules}}

In this section we present a set of rules for transforming locally
stratified constraint logic programs. We postpone to Section~\ref{sec:related_work}
the detailed comparison of our set of transformation rules with other
sets of rules which were proposed in the literature for transforming
logic programs and constraint logic programs. The application of our
transformation rules is illustrated by simple examples. More complex
examples will be given in Section~\ref{sec:examples}.

The transformation rules are used to construct a \emph{transformation
sequence}, that is, a sequence \( P_{0},\ldots ,P_{n} \) of programs.
We assume that \( P_{0} \) is locally stratified w.r.t. a fixed local
stratification function \( \sigma  \): \( \mathcal{B}_{\mathcal{D}}\rightarrow W \),
and we will say that \( P_{0},\ldots ,P_{n} \) is constructed \emph{using}
\( \sigma  \). We also assume that we are given a set \( \mathit{Pred}_{int}\subseteq \mathit{Pred}_{u} \)
of \emph{predicates of interest}. 
\medskip{}

A transformation sequence \( P_{0},\ldots ,P_{n} \) is constructed
as follows. Suppose that we have constructed a transformation sequence
\( P_{0},\ldots ,P_{k} \), for \( 0\! \leq \! k\! \leq \! n\! -\! 1 \),
the next program \( P_{k+1} \) in the transformation sequence is
derived from program \( P_{k} \) by the application of a transformation
rule among \mbox{R1--R10} defined below.

Our first rule is the \emph{definition introduction} rule, which is
applied for introducing a new predicate definition. Notice that by
this rule we can introduce a new predicate defined by \( m \) (\( \geq 1 \))
non-recursive clauses.
\smallskip{}

\noindent \textbf{R1. Definition Introduction.} Let us consider \( m \)
(\( \geq \! 1 \)) clauses of the form: 
\smallskip{}

\begin{tabular}{cl}
\( \delta _{1}: \)&
\( \mathit{newp}(X_{1},\ldots ,X_{h})\leftarrow c_{1}\wedge G_{1} \)\\
&
\( \, \, \, \, \, \, \, \, \, \, \, \ldots  \)\\
\( \delta _{m}: \)&
\( \mathit{newp}(X_{1},\ldots ,X_{h})\leftarrow c_{m}\wedge G_{m} \)\\
\end{tabular}
\smallskip{}

\noindent where: 

\noindent (i) \( \mathit{newp} \) is a predicate symbol not occurring
in \( \{P_{0},\ldots ,P_{k}\} \),

\noindent (ii) \( X_{1},\ldots ,X_{h} \) are distinct variables occurring
in \( \mathit{FV}(\{c_{1}\wedge G_{1},\ldots ,c_{m}\wedge G_{m}\}) \),

\noindent (iii) every predicate symbol occurring in \( \{G_{1},\ldots ,G_{m}\} \)
also occurs in \( P_{0} \), and

\noindent (iv) for every ground substitution \( \vartheta  \) with
domain \( \{X_{1},\ldots ,X_{h}\} \), 

\noindent \( \sigma (\mathit{newp}(X_{1},\ldots ,X_{h})\vartheta ) \)
is the least ordinal \( \alpha  \) such that, for every valuation
\( v \) and for every \( i=1,\ldots ,m \),\\
\emph{either} (iv.1) \( \mathcal{D}\models \neg v(c_{i}\vartheta ) \)
\emph{or} (iv.2) for every literal \emph{}\( L \) occurring in \( v(G_{i}\vartheta ) \),
if \( L \) is an atom \( A \) then \emph{\( \alpha \! \geq \! \sigma (A) \)}
else if \( L \) is a negated atom \( \neg A \) then \emph{\( \alpha \! >\! \sigma (A) \)}. 

\noindent By \emph{definition introduction} (or \emph{definition},
for short) from program \( P_{k} \) we derive the program \( P_{k+1}=P_{k}\cup \{\delta _{1},\ldots ,\delta _{m}\} \).
For \( k\geq 0 \), \( \mathit{Defs}_{k} \) denotes the set of clauses
introduced by the definition rule during the transformation sequence
\( P_{0},\ldots ,P_{k} \). In particular, \( \mathit{Defs}_{0}=\emptyset  \).
\medskip{}

Condition~(iv), which is needed to ensure that \( \sigma  \) is
a local stratification for each program in the transformation sequence
\( P_{0},\ldots ,P_{k+1} \) (see Proposition~\ref{prop:local_strat}),
is not actually restrictive, because \( \mathit{newp} \) is a predicate
symbol \emph{not} occurring in \( P_{0} \) and, thus, we can always
choose the local stratification \( \sigma  \) for \( P_{0} \) so
that Condition~(iv)~holds. As a consequence of Condition~(iv),
\( \sigma (\mathit{newp}(X_{1},\ldots ,X_{h})\vartheta ) \) is the
least upper bound of \( S_{p}\cup S_{n} \) w.r.t.~\( < \) where:
\smallskip{}

\begin{tabular}{ll}
\( S_{p}=\{\sigma (A) \)&
\( |\, 1\! \leq \! i\! \leq \! m,\, \,  \)\( v \) is a valuation,
\( A \) occurs in \( v(G_{i}\vartheta ),\, \,  \) \\
&
~~\( \mathcal{D}\models v(c_{i}\vartheta ) \)\}, and\\
\( S_{n}=\{\sigma (A)\! +\! 1 \)&
\( |\, 1\! \leq \! i\! \leq \! m,\, \,  \)\( v \) is a valuation,
\( \neg A \) occurs in \( v(G_{i}\vartheta ),\, \,  \) \\
&
~~\( \mathcal{D}\models v(c_{i}\vartheta ) \)\}.\\
\end{tabular}
\smallskip{}

\noindent In particular, if for \( i=1,\ldots ,m \), \( \mathcal{D}\models \neg \exists (c_{i}\vartheta ) \),
then \( S_{p}\cup S_{n}=\emptyset  \) and we have that \( \sigma (\mathit{newp}(X_{1},\ldots ,X_{h})\vartheta )=0 \).
\medskip{}

The \emph{definition elimination} rule is the inverse of the definition
introduction rule. It can be used to discard from a given program
the definitions of predicates which are not of interest. 
\smallskip{}

\noindent \textbf{R2. Definition Elimination.} Let \( p \) be a predicate
such that no predicate of the set \( \mathit{Pred}_{int} \) of the
predicates of interest depends on \( p \) in \( P_{k} \). By \emph{eliminating}
the definition of \( p \), from program \( P_{k} \) we derive the
new program \( P_{k+1}=P_{k}-\mathit{Def}(p,P_{k}) \).
\medskip{}

The \emph{unfolding} rule consists in: (i)~replacing an atom \( p(t_{1},\ldots ,t_{m}) \)
occurring in the body of a clause, by a suitable instance of the disjunction
of the bodies of the clauses which are the definition of \( p \),
and (ii)~applying suitable boolean laws for deriving clauses. The
suitable instance of Step~(i) is computed by adding a constraint
of the form \( p(t_{1},\ldots ,t_{m})\! =\! K \) for each head \( K \)
of a clause in \( \mathit{Def}(p,P_{k}) \). There are two unfolding
rules: (1)\emph{~}the positive unfolding rule, and (2)~the negative
unfolding rule, corresponding to the case where \( p(t_{1},\ldots ,t_{m}) \)
occurs positively and negatively, respectively, in the body of the
clause to be unfolded. In order to perform Step~(ii), in the case
of positive unfolding we apply the distributivity law, and in the
case of negative unfolding we apply De Morgan's, distributivity, and
double negation elimination laws. 
\smallskip{}

\noindent \textbf{R3. Positive Unfolding.} Let \( \gamma :\, \, H\leftarrow c\wedge G_{L}\wedge A\wedge G_{R} \)
be a clause in program \( P_{k} \) and let \( P'_{k} \) be a variant
of \( P_{k} \) without common variables with \( \gamma  \). Let
\smallskip{}

\( \begin{array}{l}
\gamma _{1}:\, \, K_{1}\leftarrow c_{1}\wedge B_{1}\\
\, \, \, \, \, \, \, \, \, \, \, \ldots \\
\gamma _{m}:\, \, K_{m}\leftarrow c_{m}\wedge B_{m}
\end{array} \)
\smallskip{}

\noindent where \( m\geq 0 \) and \( B_{1},\ldots ,B_{m} \) are
conjunction of literals, be all clauses of program \( P'_{k} \) such
that, for \( i=1,\ldots ,m \), \( \mathcal{D}\models \exists (c\wedge A\! =\! K_{i}\wedge c_{i}) \). 

\noindent By \emph{unfolding clause} \( \gamma  \) \emph{w.r.t.}~\emph{the
atom} \( A \) we derive the clauses
\smallskip{}

\( \begin{array}{l}
\eta _{1}:\, \, H\leftarrow c\wedge A\! =\! K_{1}\wedge c_{1}\wedge G_{L}\wedge B_{1}\wedge G_{R}\\
\, \, \, \, \, \, \, \, \, \, \, \ldots \\
\eta _{m}:\, \, H\leftarrow c\wedge A\! =\! K_{m}\wedge c_{m}\wedge G_{L}\wedge B_{m}\wedge G_{R}
\end{array} \)

\smallskip{}
\noindent and from program \( P_{k} \) we derive the program \( P_{k+1}=(P_{k}-\{\gamma \})\cup \{\eta _{1},\ldots ,\eta _{m}\} \). 
\medskip{}

Notice that if \( m\! =\! 0 \) then, by positive unfolding, clause
\( \gamma  \) is deleted from \( P_{k} \).

\begin{example}
Let \( P_{k} \) be the following program:
\smallskip{}

\begin{tabular}{ll}
1.&
\( p(X)\leftarrow X\! \geq \! 1\wedge q(X) \)\\
2.&
\( q(Y)\leftarrow Y\! =\! 0 \)\\
3.~~&
\( q(Y)\leftarrow Y\! =\! Z\! +\! 1\wedge q(Z) \)\\
\end{tabular}
\smallskip{}

\noindent where we assume that the interpretation for the constraints
is given by the structure \( \mathcal{R} \) of the real numbers.
Let us unfold clause~1 w.r.t.~the atom \( q(X) \). The constraint
\( X\! \geq \! 1\wedge X\! =\! Y\wedge Y\! =\! 0 \) constructed from
the constraints of clauses~1 and~2 is unsatisfiable, that is, \( \mathcal{R}\models \neg \exists X\exists Y(X\! \geq \! 1\wedge X\! =\! Y\wedge Y\! =\! 0) \),
while the constraint \( X\! \geq \! 1\wedge X\! =\! Y\wedge Y\! =\! Z\! +\! 1 \)
constructed from the constraints of clauses~1 and 3, is satisfiable.
Thus, we derive the following program \( P_{k+1} \):
\smallskip{}

\begin{tabular}{rl}
1u.~~&
\( p(X)\leftarrow X\! \geq \! 1\wedge X\! =\! Y\wedge Y\! =\! Z\! +\! 1\wedge q(Z) \)\\
2.~~&
\( q(Y)\leftarrow Y\! =\! 0 \)\\
3.~~&
\( q(Y)\leftarrow Y\! =\! Z\! +\! 1\wedge q(Z) \)\\
\end{tabular}\medskip{}

\end{example}
\noindent \textbf{R4. Negative Unfolding.} Let \( \gamma :\, \, H\leftarrow c\wedge G_{L}\wedge \neg A\wedge G_{R} \)
be a clause in program \( P_{k} \) and let \( P'_{k} \) be a variant
of \( P_{k} \) without common variables with \( \gamma  \). Let~

\( \begin{array}{l}
\gamma _{1}:\, \, K_{1}\leftarrow c_{1}\wedge B_{1}\\
\, \, \, \, \, \, \, \, \, \, \, \ldots \\
\gamma _{m}:\, \, K_{m}\leftarrow c_{m}\wedge B_{m}
\end{array} \)
\smallskip{}

\noindent where \( m\geq 0 \) and \( B_{1},\ldots ,B_{m} \) are
conjunction of literals, be all clauses of program \( P'_{k} \) such
that, for \( i=1,\ldots ,m \), \( \mathcal{D}\models \exists (c\wedge A\! =\! K_{i}\wedge c_{i}) \).
Suppose that, for \( i=1,\ldots ,m \), there exist an idempotent
substitution \( \vartheta _{i}=\{X_{i1}/t_{i1},\ldots ,X_{in}/t_{in}\} \)
and a constraint \( d_{i} \) such that the following conditions hold:
\smallskip{}

(i) \( \mathcal{D}\models \forall (c\rightarrow ((A\! =\! K_{i}\wedge c_{i})\leftrightarrow (X_{i1}\! =\! t_{i1}\wedge \ldots \wedge X_{in}\! =\! t_{in}\wedge d_{i}))) \),
\smallskip{}

(ii) \( \{X_{i1},\ldots ,X_{in}\}\subseteq V_{i} \), where \( V_{i}=FV(\gamma _{i}) \),
and
\smallskip{}

(iii) \( FV(d_{i}\wedge B_{i}\vartheta _{i})\subseteq FV(c\wedge A) \).
\smallskip{}

\noindent Then, from the formula
\smallskip{}

\( \psi _{0}: \)~~\( c\wedge G_{L}\wedge \neg (\exists V_{1}\, (A\! =\! K_{1}\wedge c_{1}\wedge B_{1})\vee \ldots \vee \exists V_{m}\, (A\! =\! K_{m}\wedge c_{m}\wedge B_{m}))\wedge G_{R} \)
\smallskip{}

\noindent we get an equivalent disjunction of constrained goals by
performing the following steps. In these steps we silently apply the
associativity of \( \wedge  \) and \( \vee  \).
\smallskip{}

\noindent Step 1. (\emph{Eliminate} \( \exists  \)) Since Conditions
(i), (ii), and (iii) hold, we derive from \( \psi _{0} \) the following
equivalent formula:
\smallskip{}

\( \psi _{1}: \)~~\( c\wedge G_{L}\wedge \neg ((d_{1}\wedge B_{1}\vartheta _{1})\vee \ldots \vee (d_{m}\wedge B_{m}\vartheta _{m}))\wedge G_{R} \)
\smallskip{}

\noindent Step 2. (\emph{Push \( \neg  \) inside}) We apply to \( \psi _{1} \)
as long as possible the following rewritings of formulas, where \( d \)
is a constraint, \( At \) is an atom, \( G \), \( G_{1} \), \( G_{2} \)
are goals, and \( D \) is a disjunction of constrained literals:
\smallskip{}

\smallskip{}
\begin{tabular}{lll}
\( \neg ((d\wedge G)\vee D) \)&
\( \, \longrightarrow \,  \)&
\( \neg (d\wedge G)\wedge \neg D \)\\
\( \neg (d\wedge G) \)&
\( \, \longrightarrow \,  \)&
\( \neg d\vee (d\wedge \neg G) \)\\
\( \neg (G_{1}\wedge G_{2}) \)&
\( \, \longrightarrow \,  \)&
\( \neg G_{1}\vee \neg G_{2} \)\\
\( \neg \neg At \)&
\( \, \longrightarrow \,  \)&
\( At \)\\
\end{tabular}
\smallskip{}

\noindent Thus, from \( \psi _{1} \) we derive the following equivalent
formula:
\smallskip{}

\begin{tabular}{llll}
\( \psi _{2}: \)~~\( c\wedge G_{L}\!  \)&
\( \wedge \, (\neg d_{1} \)&
\( \vee \, \, (d_{1}\wedge (\overline{{L_{11}}\vartheta _{1}} \)&
\( \! \vee \ldots \vee \, \overline{{L_{1p}}\vartheta _{1}})))\,  \)\\
&
\( \wedge  \)~~~\ldots{}&
&
\\
&
\( \wedge \, (\neg d_{m} \)&
\( \vee \, \, (d_{m}\wedge (\overline{{L_{m1}}\vartheta _{m}} \) &
\( \! \vee \ldots \vee \overline{{L_{mq}}\vartheta _{m}}))) \)\\
&
\( \wedge \, G_{R} \)&
&
\\
\end{tabular}
\smallskip{}

\noindent where \( L_{11}\wedge \ldots \wedge L_{1p} \) is \( B_{1} \),
\( \ldots  \), and \( L_{m1}\wedge \ldots \wedge L_{mq} \) is \( B_{m} \).
\smallskip{}

\noindent Step 3. (\emph{Push \( \vee  \) outside}) We apply to \( \psi _{2} \)
as long as possible the following rewriting of formulas, where \( \varphi _{1} \),
\( \varphi _{2} \), and \( \varphi _{3} \) are formulas:

\smallskip{}
\( \varphi _{1}\wedge (\varphi _{2}\vee \varphi _{3}) \) \( \, \longrightarrow \,  \)
\( (\varphi _{1}\wedge \varphi _{2})\vee (\varphi _{1}\wedge \varphi _{3}) \) 

\smallskip{}
\noindent and then we move constraints to the left of literals by
applying the commutativity of \( \wedge  \). Thus, from \( \psi _{2} \)
we get an equivalent formula of the form:
\smallskip{}

\( \psi _{3}: \)~~\( (c\wedge e_{1}\wedge G_{L}\wedge Q_{1}\wedge G_{R})\vee \ldots \vee (c\wedge e_{r}\wedge G_{L}\wedge Q_{r}\wedge G_{R}) \)
\smallskip{}

\noindent where \( e_{1},\ldots ,e_{r} \) are constraints and \( Q_{1},\ldots ,Q_{r} \)
are goals.
\smallskip{}

\noindent Step 4. (\emph{Remove unsatisfiable disjuncts}) We remove
from \( \psi _{3} \) every disjunct \( (c\wedge e_{j}\wedge G_{L}\wedge Q_{j}\wedge G_{R}) \),
with \( 1\! \leq \! j\! \leq \! r \), such that \( \mathcal{D}\models \neg \exists (c\wedge e_{j}) \),
thereby deriving an equivalent disjunction of constrained goals of
the form: 
\smallskip{}

\( \psi _{4}: \)~~\( (c\wedge e_{1}\wedge G_{L}\wedge Q_{1}\wedge G_{R})\vee \ldots \vee (c\wedge e_{s}\wedge G_{L}\wedge Q_{s}\wedge G_{R}) \)
\smallskip{}

\noindent By \textit{unfolding clause \( \gamma  \) w.r.t.~the negative
literal \( \neg A \)} we derive the clauses 
\smallskip{}

\( \begin{array}{l}
\eta _{1}:\, \, H\leftarrow c\wedge e_{1}\wedge G_{L}\wedge Q_{1}\wedge G_{R}\\
\, \, \, \, \, \, \, \, \, \, \, \ldots \\
\eta _{s}:\, \, H\leftarrow c\wedge e_{s}\wedge G_{L}\wedge Q_{s}\wedge G_{R}
\end{array} \)
\smallskip{}

\noindent and from program \( P_{k} \) we derive the program \( P_{k+1}=(P_{k}-\{\gamma \})\cup \{\eta _{1},\ldots ,\eta _{s}\} \). 
\medskip{}

\noindent Notice that: (i) if \( m=0 \), that is, if we unfold clause
\( \gamma  \) w.r.t.~a negative literal \( \neg A \) such that
the constraint \( c\wedge A\! =\! K_{i}\wedge c_{i} \) is satisfiable
for no clause \( K_{i}\leftarrow c_{i}\wedge B_{i} \) in \( P'_{k} \),
then we get the new program \( P_{k+1} \) by deleting \( \neg A \)
from the body of clause \( \gamma  \), and (ii) if we unfold clause
\( \gamma  \) w.r.t.~a negative literal \( \neg A \) such that
for some clause \( K_{i}\leftarrow c_{i}\wedge B_{i} \) in \( P'_{k} \),
\( \mathcal{D}\models \forall (c\rightarrow \exists V_{i}\, (A\! =\! K_{i}\wedge c_{i})) \)
and \( B_{i} \) is the empty conjunction, then we derive the new
program \( P_{k+1} \) by deleting clause \( \gamma  \) from \( P_{k} \). 

An application of the negative unfolding rule is illustrated by the
following example.

\begin{example}
Suppose that the following clause belongs to program \( P_{k} \):
\smallskip{}

\( \gamma :\, \, h(X)\leftarrow X\! \geq \! 0\wedge \neg p(X) \)
\smallskip{}

\noindent and let 
\smallskip{}

\( p(Y)\leftarrow Y\! =\! Z\! +\! 1\wedge Z\! \geq \! 0\wedge q(Z) \)

\( p(Y)\leftarrow Y\! =\! Z\! -\! 1\wedge Z\! \geq \! 1\wedge q(Z)\wedge \neg r(Z) \)
\smallskip{}

\noindent be the definition of \( p \) in \( P_{k} \). Suppose also
that the constraints are interpreted in the structure \( \mathcal{R} \)
of the real numbers. Now let us unfold clause \( \gamma  \) w.r.t.~\( \neg p(X) \).
We start off from the formula:
\smallskip{}

\begin{tabular}{cl}
\( \psi _{0}: \)~~\( X\! \geq \! 0\wedge \neg ( \)&
\( \exists Y\, \exists Z\, (X\! =\! Y\wedge Y\! =\! Z\! +\! 1\wedge Z\! \geq \! 0\wedge q(Z))\vee  \)\\
&
\( \exists Y\, \exists Z\, (X\! =\! Y\wedge Y\! =\! Z\! -\! 1\wedge Z\! \geq \! 1\wedge q(Z)\wedge \neg r(Z)) \))\\
\end{tabular}
\smallskip{}

\noindent Then we perform the four steps indicated in rule R4 as follows. 

\noindent Step 1. Since we have that:
\smallskip{}

\begin{tabular}{ll}
\( \mathcal{R}\models \forall X\, \forall Y\, \forall Z\, (X\! \geq \! 0\rightarrow ( \)&
\( (X\! =\! Y\wedge Y\! =\! Z\! +\! 1\wedge Z\! \geq \! 0)\leftrightarrow  \)\\
&
\( (Y\! =\! X\wedge Z\! =\! X\! -\! 1\wedge X\! \geq \! 1))) \)\\
\end{tabular}

\noindent and
\smallskip{}

\begin{tabular}{ll}
\( \mathcal{R}\models \forall X\, \forall Y\, \forall Z\, (X\! \geq \! 0\rightarrow ( \)&
\( (X\! =\! Y\wedge Y\! =\! Z\! -\! 1\wedge Z\! \geq \! 1)\leftrightarrow  \)\\
&
\( (Y\! =\! X\wedge Z\! =\! X\! +\! 1))) \)\\
\end{tabular}
\smallskip{}

\noindent we derive the formula:
\smallskip{}

\( \psi _{1}: \)~~\( X\! \geq \! 0\wedge \neg ((X\! \geq \! 1\wedge q(X\! -\! 1))\vee (q(X\! +\! 1)\wedge \neg r(X\! +\! 1))) \)
\smallskip{}

\noindent Steps 2 and 3. By applying the rewritings indicated in rule
R4 we derive the following formula:
\smallskip{}

\begin{tabular}{cl}
\( \psi _{3}: \)~~&
\( (X\! \geq \! 0\wedge \neg X\! \geq \! 1\wedge \neg q(X\! +\! 1))\vee  \)\\
&
\( (X\! \geq \! 0\wedge \neg X\! \geq \! 1\wedge r(X\! +\! 1))\vee  \)\\
&
\( (X\! \geq \! 0\wedge X\! \geq \! 1\wedge \neg q(X\! -\! 1)\wedge \neg q(X\! +\! 1))\vee  \)\\
&
\( (X\! \geq \! 0\wedge X\! \geq \! 1\wedge \neg q(X\! -\! 1)\wedge r(X\! +\! 1)) \)\\
\end{tabular}
\smallskip{}

\noindent Step 4. Since all constraints in the formula derived at
the end of Steps 2 and 3 are satisfiable, no disjunct is removed.
\smallskip{}

\noindent Thus, by unfolding \( h(X)\leftarrow X\! \geq \! 0\wedge \neg p(X) \)
w.r.t.~\( \neg p(X) \) we derive the following clauses:
\smallskip{}

\( h(X)\leftarrow X\! \geq \! 0\wedge \neg X\! \geq \! 1\wedge \neg q(X\! +\! 1) \)

\( h(X)\leftarrow X\! \geq \! 0\wedge \neg X\! \geq \! 1\wedge r(X\! +\! 1) \)

\( h(X)\leftarrow X\! \geq \! 0\wedge X\! \geq \! 1\wedge \neg q(X\! -\! 1)\wedge \neg q(X\! +\! 1) \)

\( h(X)\leftarrow X\! \geq \! 0\wedge X\! \geq \! 1\wedge \neg q(X\! -\! 1)\wedge r(X\! +\! 1) \)
\end{example}
\noindent The validity of Conditions (i), (ii), and (iii) in the negative
folding rule allows us to eliminate the existential quantifiers as
indicated at Step 1. If these conditions do not hold and nonetheless
we eliminate the existential quantifiers, then negative unfolding
may be incorrect, as illustrated by the following example.

\begin{example}
\label{ex:incorrect_negative_unfolding}Let us consider the following
programs \( P_{0} \) and \( P_{1} \), where \( P_{1} \) is obtained
by negative unfolding from \( P_{0} \), but Conditions (i)--(iii)
do not hold:
\smallskip{}

\begin{tabular}{llll}
\( P_{0} \):~~&
\( p\leftarrow \neg q \) ~~~~~~~~~~~~~~~~~~~~~~&
\( P_{1} \):~~&
\( p\leftarrow \neg r(X) \)\\
&
\( q\leftarrow r(X) \)&
&
\( q\leftarrow r(X) \)\\
&
\( r(X)\leftarrow X\! =\! 0 \)&
&
\( r(X)\leftarrow X\! =\! 0 \)\\
\end{tabular}
\smallskip{}

\noindent We have that: \( p\not \in M(P_{0}) \) while \( p\in M(P_{1}) \).
(We assume that the carrier of the interpretation for the constraints
contains at least one element different from \( 0 \).) 
\end{example}
The reason why the negative unfolding step of Example~\ref{ex:incorrect_negative_unfolding}
is incorrect is that the clause \( q\leftarrow r(X) \) is, as usual,
implicitly universally quantified at the front, and \( \forall X\, (q\leftarrow r(X)) \)
is logically equivalent to \( q\leftarrow \exists X\, r(X) \). Now,
a correct negative unfolding rule should replace the clause \( p\leftarrow \neg q \)
in program \( P_{0} \) by \( p\leftarrow \neg \exists X\, r(X) \),
while in program \( P_{1} \) we have derived the clause \( p\leftarrow \neg r(X) \)
which, by making the quantification explicit at the front of the body,
can be written as \( p\leftarrow \exists X\, \neg r(X) \).
\medskip{}

The \emph{folding} rule consists in replacing instances of the bodies
of the clauses that are the definition of a predicate by the corresponding
head. As for unfolding, we have a positive folding and a negative
folding rule, depending on whether folding is applied to positive
or negative occurrences of (conjunctions of) literals. Notice that
by the positive folding rule we may replace \( m\, (\geq \! 1) \)
clauses by one clause only.
\smallskip{}

\noindent \textbf{R5. Positive Folding.} Let \( \gamma _{1},\ldots ,\gamma _{m} \),
with \( m\! \geq \! 1 \), be clauses in \( P_{k} \) and let \( \mathit{Defs}'_{k} \)
be a variant of \( \mathit{Defs}_{k} \) without common variables
with \( \gamma _{1},\ldots ,\gamma _{m} \). Let the definition of
a predicate in \( \mathit{Defs}'_{k} \) consist of the clauses

\smallskip{}
\( \begin{array}{l}
\delta _{1}:\, \, K\leftarrow d_{1}\wedge B_{1}\\
\, \, \, \, \, \, \, \, \, \, \, \ldots \\
\delta _{m}:\, \, K\leftarrow d_{m}\wedge B_{m}
\end{array} \)
\smallskip{}

\noindent where, for \( i=1,\ldots ,m \), \( B_{i} \) is a non-empty
conjunction of literals. Suppose that there exists a substitution
\( \vartheta  \) such that, for \( i=1,\ldots ,m \), clause \( \gamma _{i} \)
is of the form \( H\leftarrow c\wedge d_{i}\vartheta \wedge G_{L}\wedge B_{i}\vartheta \wedge G_{R} \)
and, for every variable \( X \) in the set \( FV(d_{i}\wedge B_{i})-FV(K) \),
the following conditions hold: (i) \( X\vartheta  \) is a variable
not occurring in \( \{H,c,G_{L},G_{R}\} \), and (ii)~\( X\vartheta  \)
does not occur in the term \( Y\vartheta  \), for any variable \( Y \)
occurring in \( d_{i}\wedge B_{i} \) and different from \( X \).

\noindent By \textit{folding} \emph{clauses \( \gamma _{1},\ldots ,\gamma _{m} \)
using clauses} \( \delta _{1},\ldots ,\delta _{m} \) we derive the
clause \( \eta  \): \( H\leftarrow c\wedge G_{L}\wedge K\vartheta \wedge G_{R} \)
and from program \( P_{k} \) we derive the program \( P_{k+1}=(P_{k}-\{\gamma _{1},\ldots ,\gamma _{m}\})\cup \{\eta \} \). 
\medskip{}

The following example illustrates an application of rule R5.

\begin{example}
Suppose that the following clauses belong to \( P_{k} \):
\smallskip{}

\( \gamma _{1} \):~~\( h(X)\leftarrow X\! \geq \! 1\wedge Y\! =\! X\! -\! 1\wedge p(Y,1) \) 

\( \gamma _{2} \):~~\( h(X)\leftarrow X\! \geq \! 1\wedge Y\! =\! X\! +\! 1\wedge \neg q(Y) \) 
\smallskip{}

\noindent and suppose that the following clauses constitute the definition
of a predicate \( \mathit{new} \) in \( \mathit{Defs}_{k} \):
\smallskip{}

\( \delta _{1} \):~~\( \mathit{new}(Z,C)\leftarrow V\! =\! Z\! -\! C\wedge p(V,C) \) 

\( \delta _{2} \):~~\( \mathit{new}(Z,C)\leftarrow V\! =\! Z\! +\! C\wedge \neg q(V) \) 
\smallskip{}

\noindent For \( \vartheta =\{V/Y,Z/X,C/1\} \), we have that \( \gamma _{1}=h(X)\leftarrow X\! \geq \! 1\wedge (V\! =\! Z\! -\! C\wedge p(V,C))\vartheta  \)
and \( \gamma _{2}=h(X)\leftarrow X\! \geq \! 1\wedge (V\! =\! Z\! +\! C\wedge \neg q(V))\vartheta  \),
and the substitution \( \vartheta  \) satisfies Conditions (i) and
(ii) of the positive folding rule. By folding clauses \( \gamma _{1} \)
and \( \gamma _{2} \) using clauses \( \delta _{1} \) and \( \delta _{2} \)
we derive:
\smallskip{}

\( \eta  \):~~\( h(X)\leftarrow X\! \geq \! 1\wedge \mathit{new}(Z,1) \) 
\end{example}
\noindent \textbf{R6. Negative Folding.} Let \( \gamma  \) be a clause
in \( P_{k} \) and let \( \mathit{Defs}'_{k} \) be a variant of
\( \mathit{Defs}_{k} \) without common variables with \( \gamma  \).
Suppose that there exists a predicate in \( \mathit{Defs}'_{k} \)
whose definition consists of a single clause \( \delta : \) \( K\leftarrow d\wedge A \),
where \( A \) is an atom. Suppose also that there exists a substitution
\( \vartheta  \) such that clause \( \gamma  \) is of the form:
\( H\leftarrow c\wedge d\vartheta \wedge G_{L}\wedge \neg A\vartheta \wedge G_{R} \)
and \( FV(K)=FV(d\wedge A) \).
\smallskip{}

\noindent By \textit{folding} \emph{clause \( \gamma  \) using clause}
\( \delta  \) we derive the clause \( \eta  \): \( H\leftarrow c\wedge d\vartheta \wedge G_{L}\wedge \neg K\vartheta \wedge G_{R} \)
and from program \( P_{k} \) we derive the program \( P_{k+1}=(P_{k}\! -\! \{\gamma \})\cup \{\eta \} \).~
\medskip{}

The following is an example of application of the negative folding
rule.

\begin{example}
\label{ex:negative_folding}Let the following clause belong to \( P_{k} \):
\smallskip{}

\( \gamma  \):~~\( h(X)\leftarrow X\! \geq \! 0\wedge q(X)\wedge \neg r(X,0) \)
\smallskip{}

\noindent and let \( \mathit{new} \) be a predicate whose definition
in \( \mathit{Defs}_{k} \) consists of the clause:
\smallskip{}

\( \delta  \):~~\( \mathit{new}(X,C)\leftarrow X\! \geq \! C\wedge r(X,C) \)
\smallskip{}

\noindent By folding \( \gamma  \) using \( \delta  \) we derive:
\smallskip{}

\( \eta  \):~~\( h(X)\leftarrow X\! \geq \! 0\wedge q(X)\wedge \neg \mathit{new}(X,0) \)
\end{example}
\noindent The positive and negative folding rule are not fully symmetric
for the following three reasons. 

\noindent (1) By positive folding we can fold several clauses at a
time by using \emph{several} clauses whose body may contain several
literals, while by negative folding we can fold a \emph{single} clause
at a time by using a single clause whose body contains precisely one
atom. This is motivated by the fact that a conjunction of more than
one literal cannot occur inside negation in the body of a clause.

\noindent (2) By positive folding, for \( i=1,\ldots ,m \), the constraint
\( d\vartheta _{i} \) occurring in the body of clause \( \gamma _{i} \)
is removed, while by negative folding the constraint \( d\vartheta  \)
occurring in the body of clause \( \gamma  \) is not removed. Indeed,
the removal of the constraint \( d\vartheta  \) would be incorrect.
For instance, let us consider the program \( P_{k} \) of Example~\ref{ex:negative_folding}
above and let us assume that \( \gamma  \) is the only clause defining
the predicate \( h \). Let us also assume that the predicates \( q \)
and \( r \) are defined by the following two clauses: \( q(X)\leftarrow X\! <\! 0 \)
and \( r(X,0)\leftarrow X\! <\! 0 \). We have that \( h(-1)\not \in M(P_{k}) \).
Suppose that we apply the negative folding rule to clause \( \gamma  \)
and we remove the constraint \( X\! \geq \! 0 \), thereby deriving
the clause \( h(X)\leftarrow q(X)\wedge \neg \mathit{new}(X,0) \),
instead of clause~\( \eta  \). Then we obtain a program whose perfect
model has the atom \( h(-1) \).

\noindent (3) The conditions on the variables occurring in the clauses
used for folding are less restrictive in the case of positive folding
(see Conditions~(i) and~(ii) of R5) than in the case of negative
folding (see the condition \( FV(K)=FV(d\wedge A) \)). Notice that
a negative folding rule where the condition \( FV(K)=FV(d\wedge A) \)
is replaced by Conditions~(i) and~(ii) of R5 would be incorrect,
in general. To see this, let us consider the following example which
may be viewed as the inverse derivation of Example~\ref{ex:incorrect_negative_unfolding}.

\begin{example}
Let us consider the following programs \( P_{0} \), \( P_{1} \),
and \( P_{2} \), where \( P_{1} \) is obtained from \( P_{0} \)
by definition introduction, and \( P_{2} \) is obtained from \( P_{1} \)
by incorrectly folding \( p\leftarrow \neg r(X) \) using \( q\leftarrow r(Y) \).
Notice that \( FV(q)\! \neq \! FV(r(X \))) but Conditions~(i) and~(ii)
are satisfied by the substitution \( \{Y/X\} \).
\smallskip{}

\begin{tabular}{llllll}
\( P_{0} \):~~&
\( p\leftarrow \neg r(X) \)&
\( P_{1} \):~~&
\( p\leftarrow \neg r(X) \)&
\( P_{2} \):~~&
\( p\leftarrow \neg q \) \\
&
\( r(X)\leftarrow X\! =\! 0 \) ~~~~~&
&
\( r(X)\leftarrow X\! =\! 0 \) ~~~~~&
&
\( r(X)\leftarrow X\! =\! 0 \) ~~~~~\\
&
&
&
\( q\leftarrow r(Y) \)&
&
\( q\leftarrow r(Y) \)\\
\end{tabular}
\smallskip{}

\noindent We have that: \( p\in M(P_{0}) \) while \( p\not \in M(P_{2}) \).
(We assume that the carrier of the interpretation for the constraints
contains at least one element different from \( 0 \).) 
\end{example}
If we consider the folding and unfolding rules outside the context
of a transformation sequence, either rule can be viewed as the inverse
of the other. However, given a transformation sequence \( P_{0},\ldots ,P_{n} \),
it may be the case that from a program \( P_{k} \) in that sequence
we derive program \( P_{k+1} \) by folding, and from program \( P_{k+1} \)
we \emph{cannot} derive by unfolding a program \( P_{k+2} \) which
is equal to \( P_{k} \). This is due to the fact that in the transformation
sequence \( P_{0},\ldots ,P_{k},P_{k+1} \), in order to fold some
clauses in program \( P_{k} \), we may use clauses in \( \mathit{Defs}_{k} \)
which are neither in \( P_{k} \) nor in \( P_{k+1} \), while for
unfolding program \( P_{k+1} \) we can only use clauses which belong
to \( P_{k+1} \). Thus, according to the terminology introduced in
\cite{PeP94}, we say that folding is, in general, not \emph{reversible}.
This fact is shown by the following example.

\begin{example}
Let us consider the transformation sequence:
\smallskip{}

\begin{tabular}{llllllll}
\( P_{0} \):~~ &
\( p\leftarrow q \)~~~~&
~~~~~\( P_{1} \):~~&
\( p\leftarrow q \)~~~~&
~~~~~\( P_{2} \):~~ &
\( p\leftarrow q \)~~~~&
~~~~~\( P_{3} \):~~ &
\( p\leftarrow r \)\\
&
\( q\leftarrow  \)&
&
\( q\leftarrow  \)&
&
\( q\leftarrow  \)&
&
\( q\leftarrow  \)\\
&
&
&
\( r\leftarrow q \)&
&
\( r\leftarrow  \)&
&
\( r\leftarrow  \)\\
\end{tabular} 
\smallskip{}

\noindent where \( P_{1} \) is derived from \( P_{0} \) by introducing
the definition \( r\leftarrow q \), \( P_{2} \) is derived from
\( P_{1} \) by unfolding the clause \( r\leftarrow q \), and \( P_{3} \)
is derived from \( P_{2} \) by folding the clause \( p\leftarrow q \)
using the definition \( r\leftarrow q \). We have that from program
\( P_{3} \) we cannot derive a program equal to \( P_{2} \) by applying
the positive unfolding rule. 
\end{example}
Similarly, the unfolding rules are not reversible in general. In fact,
if we derive a program \( P_{k+1} \) by unfolding a clause in a program
\( P_{k} \) and we have that \( \mathit{Defs}_{k}=\emptyset  \),
then we cannot apply the folding rule and derive a program \( P_{k+2} \)
which is equal to \( P_{k} \), simply because no clause in \( \mathit{Defs}_{k} \)
is available for folding.~
\medskip{}

The following \emph{replacement} rule can be applied to replace a
set of clauses with a new set of clauses by using laws based on equivalences
between formulas. In particular, we consider: (i)~boolean laws, (ii)~equivalences
that can be proved in the chosen interpretation \( \mathcal{D} \)
for the constraints, and (iii)~properties of the equality predicate. 
\smallskip{}

\noindent \textbf{R7. Replacement Based on Laws.} Let us consider
the following rewritings \( \Gamma _{1}\Rightarrow \Gamma _{2} \)
between sets of clauses (we use \( \Gamma _{1}\Leftrightarrow \Gamma _{2} \)
as a shorthand for the two rewritings \( \Gamma _{1}\Rightarrow \Gamma _{2} \)
and \( \Gamma _{2}\Rightarrow \Gamma _{1} \)). Each rewriting is
called a \emph{law}.
\smallskip{}

\noindent \emph{Boolean Laws\nopagebreak}
\smallskip{}

\noindent \begin{tabular}{llcl}
(1)~~~&
\( \{H\leftarrow c\wedge A\wedge \neg A\wedge G\} \)&
\( \, \, \Leftrightarrow \, \,  \)&
\( \emptyset  \)\smallskip{}
\\
(2)~~~&
\( \{H\leftarrow c\wedge H\wedge G\} \)&
\( \, \, \Leftrightarrow \, \,  \)&
\( \emptyset  \)\smallskip{}
\\
(3)~~~&
\( \{H\leftarrow c\wedge G_{1}\wedge A_{1}\wedge A_{2}\wedge G_{2}\} \)&
\( \, \, \Leftrightarrow \, \,  \)&
\( \{H\leftarrow c\wedge G_{1}\wedge A_{2}\wedge A_{1}\wedge G_{2}\} \)\smallskip{}
\\
(4)~~~&
\( \{H\leftarrow c\wedge A\wedge A\wedge G\} \)&
\( \, \, \Rightarrow \, \,  \)&
\( \{H\leftarrow c\wedge A\wedge G\} \)\smallskip{}
\\
(5)~~~&
\begin{tabular}{l}
\( \{H\leftarrow c\wedge G_{1}, \)\\
\( \, \, \, H\leftarrow c\wedge d\wedge G_{1}\wedge G_{2}\} \)\\
\end{tabular}&
\( \, \, \Leftrightarrow \, \,  \)&
\( \{H\leftarrow c\wedge G_{1}\} \)\smallskip{}
\\
(6)~~~&
\begin{tabular}{l}
\( \{H\leftarrow c\wedge A\wedge G, \)\\
\( \, \, \, H\leftarrow c\wedge \neg A\wedge G\} \)\\
\end{tabular}\smallskip{}
&
\( \, \, \Rightarrow \, \,  \)&
\( \{H\leftarrow c\wedge G\} \)\smallskip{}
\\
\end{tabular}
\medskip{}

\noindent \emph{Laws of Constraints}

\noindent \begin{tabular}{ccl}
(7)~~~&
\( \{H\leftarrow c\wedge G\} \)&
\( \, \, \Leftrightarrow \, \, \,  \)\( \emptyset  \)\\
&
&
~~~~~~if the constraint \( c \) is unsatisfiable, that is,
\( \mathcal{D}\models \neg \exists (c) \)\smallskip{}
\\
(8)~~~&
\( \{H\leftarrow c_{1}\wedge G\} \)&
\( \, \, \Leftrightarrow \, \, \,  \)\( \{H\leftarrow c_{2}\wedge G\} \)\\
&
&
~~~~~~if \( \mathcal{D}\models \forall \, (\exists Y\, c_{1}\leftrightarrow \exists Z\, c_{2}) \),
where:\\
&
&
~~~~~~~~~(i) \( Y=\mathit{FV}(c_{1})\! -\! FV(\{H,G\}) \),
and \\
&
&
~~~~~~~~~(ii) \( Z=\mathit{FV}(c_{2})\! -\! FV(\{H,G\}) \)\smallskip{}
\\
(9)~~~&
\( \{H\leftarrow c\wedge G\} \)&
\( \, \, \Leftrightarrow \, \, \,  \)\( \{H\leftarrow c_{1}\wedge G,\, \, H\leftarrow c_{2}\wedge G\} \)\\
&
&
~~~~~~if \( \mathcal{D}\models \forall \, (c\leftrightarrow (c_{1}\vee c_{2})) \)\smallskip{}
\\
\end{tabular}
\medskip{}

\noindent \emph{Laws of Equality}

\noindent \begin{tabular}{ccl}
(10)~~~&
\( \{(H\leftarrow \mathit{c}\wedge G)\{X/t\}\} \)&
\( \, \, \Leftrightarrow \, \, \,  \)\( \{H\leftarrow X\! =\! t\wedge \mathit{c}\wedge G\} \)\\
&
&
~~if the variable \( X \) does not occur in the term \( t \) \\
&
&
~~and \( t \) is free for \( X \) in \( c \).\\
\end{tabular}
\medskip{}

\noindent Let \( \Gamma _{1} \) and \( \Gamma _{2} \) be sets of
clauses such that: (i)~\( \Gamma _{1}\Rightarrow \Gamma _{2} \),
and (ii)~\( \Gamma _{2} \) is locally stratified w.r.t.~the fixed
local stratification function \( \sigma  \). By \emph{replacement}
from \( \Gamma _{1} \) we derive \( \Gamma _{2} \) and from program
\( P_{k} \) we derive the program \( P_{k+1}=(P_{k}-\Gamma _{1})\cup \Gamma _{2} \).~
\medskip{}

Condition (ii) on~\( \Gamma _{2} \) is needed because a replacement
based on laws (1), (2), (5), and (7), used from right to left, may
not preserve local stratification. For instance, the first law may
be used to introduce a clause of the form \( p\leftarrow p\wedge \neg p \),
which is not locally stratified. We will see at the end of Section~\ref{sec:correctness}
that if we add the reverse versions of the boolean laws (4) or (6),
then the correctness result stated in Theorem~\ref{th:corr_of_rules}
does not hold.
\medskip{}

The following definition is needed for stating rule R8 below. The
set of \emph{useless predicates} in a program \( P \) is the maximal
set \( U \) of predicate symbols occurring in \( P \) such that
a predicate \( p \) is in \( U \) iff every clause \( \gamma  \)
in \( \mathit{Def}(p,P) \) is of the form \( H\leftarrow c\wedge G_{1}\wedge q(\ldots )\wedge G_{2} \)
for some \( q \) in \( U \). For example, in the following program:~
\smallskip{}

\( p(X)\leftarrow q(X)\wedge \neg r(X) \)

\( q(X)\leftarrow p(X) \)

\( r(X)\leftarrow X\! >\! 0 \) 
\smallskip{}

\noindent \( p \) and \( q \) are useless predicates, while \( r \)
is not useless.
\smallskip{}

\noindent \textbf{R8. Deletion of Useless Predicates.} If \( p \)
is a useless predicate in \( P_{k} \), then from program \( P_{k} \)
we derive the program \( P_{k+1}=P_{k}-\mathit{Def}(p,P_{k}) \). 
\medskip{}

Neither of the rules R2 and R8 subsumes the other. Indeed, on one
hand the definition of a predicate \( p \) on which no predicate
of interest depends, can be deleted by rule R2 even if \( p \) is
not useless. On the other hand, the definition of a useless predicate
\( p \) can be deleted by rule R8 even if a predicate of interest
depends on \( p \). 
\medskip{}

The \emph{constraint addition} rule R9 which we present below, can
be applied to add to the body of a clause a constraint which is implied
by that body. Conversely, the \emph{constraint deletion} rule R10,
also presented below, can be applied to delete from the body of a
clause a constraint which is implied by the rest of the body. Notice
that these implications should hold in the perfect model of program
\( P_{k} \), while the applicability conditions of rule R7 (see,
in particular, the replacements based on laws 7--9) are independent
of \( P_{k} \). Thus, for checking the applicability conditions of
rules R9 and R10 we may need a program analysis based, for instance,
on abstract interpretation~\cite{Ga&96}.
\medskip{}

\noindent \textbf{R9. Constraint Addition.} Let \( \gamma _{1}:\, \, H\leftarrow c\wedge G \)
be a clause in \( P_{k} \) and let \( d \) be a constraint such
that \( M(P_{k})\models \forall ((c\wedge G)\rightarrow \exists X\, d) \),
where \( X=FV(d)-FV(\gamma _{1}) \). By \emph{constraint addition}
from clause \( \gamma _{1} \) we derive the clause \( \gamma _{2}:\, \, H\leftarrow c\wedge d\wedge G \)
and from program \( P_{k} \) we derive the program \( P_{k+1}=(P_{k}-\{\gamma _{1}\})\cup \{\gamma _{2}\} \). 
\medskip{}

The following example shows an application of the constraint addition
rule that cannot be realized by applying laws of constraints according
to rule R7.

\begin{example}
Let us consider the following program \( P_{k} \):
\smallskip{}

\begin{tabular}{ll}
1. &
\( \mathit{nat}(0)\leftarrow  \)\\
2. &
\( \mathit{nat}(N)\leftarrow N\! =\! M\! +\! 1\wedge \mathit{nat}(M) \) \\
\end{tabular}
\smallskip{}

\noindent Since \( M(P_{k})\models \forall M\, (nat(M)\rightarrow M\! \geq \! 0) \),
we can add the constraint \( M\! \geq \! 0 \) to the body of clause
2. This constraint addition improves the termination of the program
when using a top-down strategy.
\end{example}
\noindent \textbf{R10. Constraint Deletion.} Let \( \gamma _{1}:\, \, H\leftarrow c\wedge d\wedge G \)
be a clause in \( P_{k} \) and let \( d \) be a constraint such
that \( M(P_{k})\models \forall ((c\wedge G)\rightarrow \exists X\, d) \),
where \( X=FV(d)-FV(H\leftarrow c\wedge G) \). Suppose that the clause
\( \gamma _{2}:\, \, H\leftarrow c\wedge G \) is locally stratified
w.r.t.~the fixed \( \sigma  \). By \emph{constraint deletion} from
clause \( \gamma _{1} \) we derive clause \( \gamma _{2} \) and
from program \( P_{k} \) we derive the program \( P_{k+1}=(P_{k}-\{\gamma _{1}\})\cup \{\gamma _{2}\} \). 
\medskip{}

We assume that \( \gamma _{2} \) is locally stratified w.r.t.~\( \sigma  \)
because otherwise, the constraint deletion rule may not preserve local
stratification. For instance, let us consider the following program
\( P \):
\smallskip{}

\begin{tabular}{l}
\( p(X)\leftarrow  \)\\
\( p(X)\leftarrow X\! \neq \! X\wedge \neg p(X) \)\\
\end{tabular}
\smallskip{}

\noindent \( P \) is locally stratified because for all elements
\( d \) in the carrier of the interpretation \( \mathcal{D} \) for
the constraints, we have that \( \mathcal{D}\models d\! =\! d \).
We also have that \( M(P)\models \forall X\, (\neg p(X)\rightarrow X\! \neq \! X) \).
However, if we delete the constraint \( X\! \neq \! X \) from the
second clause of \( P \) we derive the clause \( p(X)\leftarrow \neg p(X) \)
which is not locally stratified w.r.t.~any local stratification function.

\section{Preservation of Perfect Models \label{sec:correctness}}

In this section we present some sufficient conditions which ensure
that a transformation sequence constructed by applying the transformation
rules listed in Section \ref{sec:rules}, preserves the perfect model
semantics.

\smallskip{}
We will prove our correctness theorem for \emph{admissible} transformation
sequences, that is, transformation sequences constructed by applying
the rules according to suitable restrictions. The reader who is familiar
with the program transformation methodology, will realize that most
transformation strategies can, indeed, be realized by means of admissible
transformation sequences. In particular, all examples of Section~\ref{sec:examples}
are worked out by using this kind of transformation sequences. 

We proceed as follows. (i) First we show that the transformation rules
preserve local stratification. (ii)~Then we introduce the notion
of an \emph{admissible} transformation sequence. (iii)~Next we introduce
the notion of a \emph{proof tree} for a ground atom \( A \) and a
program \( P \) and we show that \( A\in M(P) \) iff there exists
a proof tree for \( A \) and \( P \). Thus, the notion of proof
tree provides the operational counterpart of the perfect model semantics.
(iv)~Then, we prove that given any admissible transformation sequence
\( P_{0},\ldots ,P_{n} \), any set \( \mathit{Pred}_{int} \) of
predicates of interest, and any ground atom \( A \) whose predicate
is in \( \mathit{Pred}_{int} \), we have that for \( k=0,\ldots ,n \),
there exists a proof tree for \( A \) and \( P_{k} \) iff there
exists a proof tree for \( A \) and \( P_{0}\cup \mathit{Defs}_{n} \).
(v)~Finally, by using the property of proof trees considered at Point~(iii),
we conclude that an admissible transformation sequence preserves the
perfect model semantics (see Theorem~\ref{th:corr_of_rules}).
\medskip{}

Let us start off by showing that the transformation rules preserve
the local stratification function \( \sigma  \) which was fixed for
the initial program \( P_{0} \) at the beginning of the construction
of the transformation sequence.

\begin{proposition}
\emph{\label{prop:local_strat} {[}Preservation of Local} \emph{Stratification{]}}.
Let \( P_{0} \) be a locally stratified program, let \( \sigma : \)
\( \mathcal{B}_{\mathcal{D}}\rightarrow W \) be a local stratification
function for \( P_{0} \), and let \( P_{0},\ldots ,P_{n} \) be a
transformation sequence using \( \sigma  \). Then the programs \( P_{0},\ldots ,P_{n} \),
and \( P_{0}\cup \mathit{Defs}_{n} \) are locally stratified w.r.t.~\( \sigma  \).
\end{proposition}
The proof of Proposition~\ref{prop:local_strat} is given in Appendix~A. 

An admissible transformation sequence is a transformation sequence
that satisfies two conditions: (1)~every clause used for positive
folding is unfolded w.r.t.~a positive literal, and (2)~the definition
elimination rule cannot be applied before any other transformation
rule. An admissible transformation sequence is formally defined as
follows.

\begin{definition}
\emph{{[}Admissible Transformation Sequence{]}} A transformation sequence
\( P_{0},\ldots ,P_{n} \) is said to be \emph{admissible} iff the
following two conditions hold:

\noindent \textup{(1)} for \( k=0,\ldots ,n\! -\! 1 \), if \( P_{k+1} \)
is derived from \( P_{k} \) by applying the positive folding rule
to clauses \( \gamma _{1},\ldots ,\gamma _{m} \) using clauses \( \delta _{1},\ldots ,\delta _{m} \),
then for \( i=1,\ldots ,m \) there exists \( j \), with \( 0\! <\! j\! <\! n \),
such that \( \delta _{i}\in P_{j} \) and program \( P_{j+1} \) is
derived from \( P_{j} \) by positive unfolding of clause \( \delta _{i} \),
and 

\noindent \textup{(2)} if for some \( m\! <\! n \), \( P_{m+1} \)
is derived from \( P_{m} \) by the definition elimination rule then
for all \( k=m,\ldots ,n\! -\! 1 \), \( P_{k+1} \) is derived from
\( P_{k} \) by applying the definition elimination rule.
\end{definition}
When proving our correctness theorem (see Theorem~\ref{th:corr_of_rules}
below), we will find it convenient to consider transformation sequences
which are admissible and satisfy some extra suitable properties. This
motivates the following notion of \emph{ordered} transformation sequences.

\begin{definition}
\emph{\label{def:ordered}{[}Ordered Transformation Sequence{]}} A
transformation sequence \( P_{0},\ldots ,P_{n} \) is said to be \emph{ordered}
iff it is of the form:
\smallskip{}

\( P_{0},\ldots ,P_{i},\ldots ,P_{j},\ldots ,P_{m},\ldots ,P_{n} \) 
\smallskip{}

\noindent where:

\noindent \textup{(1)} the sequence \( P_{0},\ldots ,P_{i} \), with
\( i\! \geq \! 0 \), is constructed by applying \( i \) times the
definition introduction rule, that is, \( P_{i}=P_{0}\cup \mathit{Defs}_{i} \);

\noindent \textup{(2)} the sequence \( P_{i},\ldots ,P_{j} \) is
constructed by unfolding w.r.t.~a positive literal each clause in
\( \mathit{Defs}_{i} \) which is used for applications of the folding
rule in \( P_{j},\ldots ,P_{m} \);

\noindent \textup{(3)} the sequence \( P_{j},\ldots ,P_{m} \), with
\( j\! \leq \! m \), is constructed by applying any rule, except
the definition introduction and definition elimination rules; and

\noindent \textup{(4)} the sequence \( P_{m},\ldots ,P_{n} \), with
\( m\! \leq \! n \), is constructed by applying the definition elimination
rule.
\end{definition}
Notice that in an ordered transformation sequence we have that \( \mathit{Defs}_{i}=\mathit{Defs}_{n} \).
Every ordered transformation sequence is admissible, because of Points~(2)
and (4) of Definition \ref{def:ordered}. Conversely, by the following
Proposition~\ref{prop:ordered}, in our correctness proofs we will
assume, without loss of generality, that any admissible transformation
sequence is ordered. 

\begin{proposition}
\label{prop:ordered} For every admissible transformation sequence
\( P_{0},\ldots ,P_{n} \), there exists an ordered transformation
sequence \( Q_{0},\ldots ,Q_{r} \) (with \( r \) possibly different
from \( n \)), such that: (i)~\( P_{0}=Q_{0} \), (ii)~\( P_{n}=Q_{r} \),
and (iii)~the set of definitions introduced during \( P_{0},\ldots ,P_{n} \)
is equal to the set of definitions introduced during \( Q_{0},\ldots ,Q_{r} \).
\end{proposition}
\noindent The easy proof of Proposition~\ref{prop:ordered} is omitted
for reasons of space. It is based on the fact that the applications
of some transformation rules can be suitably rearranged without changing
the initial and final programs in a transformation sequence.

Now we present the operational counterpart of the perfect model semantics,
that is, the notion of a proof tree. A proof tree for a ground atom
\( A \) and a locally stratified program \( P \) is constructed
by transfinite induction as indicated in the following definition.

\begin{definition}
\noindent \label{def:proof_tree} \emph{{[}Proof Tree{]}} Let \( A \)
be a ground atom, \( P \) be a locally stratified program, and \( \sigma  \)
be any local stratification for \( P \). Let \( PT_{<A} \) be the
set of proof trees for ground atoms \( B \) and \( P \) with \( \sigma (B)<\sigma (A) \).
A \emph{proof tree} for \( A \) and \( P \) is a finite tree \( T \)
of goals such that: (i)~the root of \( T \) is \( A \), (ii)~a
node \( N \) of \( T \) has children \( L_{1},\ldots ,L_{r} \)
iff \( N \) is a ground atom \( B \) and there exists a clause \( \gamma \in P \)
and a valuation \( v \) such that \( v(\gamma ) \) is \( B\leftarrow c\wedge L_{1}\wedge \ldots \wedge L_{r} \)
and \( \mathcal{D}\models c \), and (iii)~every leaf of \( T \)
is either the empty conjunction \( \mathit{true} \) or a negated
ground atom \( \neg B \) such that there is no proof tree for \( B \)
and \( P \) in \( PT_{<A} \).
\end{definition}
The following theorem establishes that the operational semantics based
on proof trees is equivalent to the perfect model semantics.

\begin{theorem}
\noindent \emph{\label{th:proof_trees_perfect_model} {[}Proof Trees
and Perfect Models{]}} Let \( P \) be a locally stratified program.
For all ground atoms \( A \), there exists a proof tree for \( A \)
and \( P \) iff \( A\in M(P) \).
\end{theorem}
Our proofs of correctness use induction w.r.t.~suitable \emph{well-founded
measures} over proof trees, ground atoms, and ground goals (see, in
particular, the proofs of Propositions~\ref{prop:tc_of_pos_unf}
and \ref{prop:preserv-mu-consistency} in Appendices~B and C). We
now introduce these measures.

Let \( T \) be a proof tree for a ground atom \( A \) and a locally
stratified program \( P \). By \emph{\( \mathit{size}(T) \)} we
denote the number of atoms occurring at non-leaf nodes of \( T \).
For any ground atom \( A \), locally stratified program \( P \),
and local stratification \( \sigma  \) for \( P \), we define the
following measure:
\smallskip{}

\( \mu (A,P)=\mathit{min}_{\mathit{lex}}\{\langle \sigma (A),\mathit{size}(T)\rangle \, |\, T \)
is a proof tree for \( A \) and \( P\} \)
\smallskip{}

\noindent where \( \mathit{min}_{\mathit{lex}} \) denotes the minimum
w.r.t.~the lexicographic ordering \( <_{\mathit{lex}} \) over \( W\times N \),
where \( W \) is the set of countable ordinals and \( N \) is the
set of natural numbers. \( \mu (A,P) \) is undefined if there is
no proof tree for \( A \) and \( P \). The measure \( \mu  \) is
extended from ground atoms to ground literals as follows. Given a
ground literal \( L \), we define:
\smallskip{}

\( \mu (L,P)= \) \emph{if} \( L \) is an atom \( A \) \emph{then}
\( \mu (A,P) \) 

\emph{~~~~~~~~~~~~~else} \emph{if} \( L \) is a negated
atom \( \neg A \) \emph{then} \( \langle \sigma (A),0\rangle  \)
\smallskip{}

\noindent Now we extend \( \mu  \) to ground goals. First, we introduce
the binary, associative operation \( \oplus :\, (W\times N)^{2}\rightarrow (W\times N) \)
defined as follows: 
\smallskip{}

\( \langle \alpha _{1},m_{1}\rangle \oplus \langle \alpha _{2},m_{2}\rangle =\langle max(\alpha _{1},\alpha _{2}),\, m_{1}\! +\! m_{2}\rangle  \)
\smallskip{}

\noindent Then, given a ground goal \( L_{1}\wedge \ldots \wedge L_{n} \),
we define: 
\smallskip{}

\( \mu (L_{1}\wedge \ldots \wedge L_{n},\, P)=\mu (L_{1},P)\oplus \ldots \oplus \mu (L_{n},P) \)
\smallskip{}

\noindent The measure \( \mu  \) is well-founded in the sense that
there is no infinite sequence of ground goals \( G_{1},G_{2},\ldots  \)
such that \( \mu (G_{1},P)>\mu (G_{2},P)>\ldots  \)

In order to show that an ordered transformation sequence \( P_{0},\ldots ,P_{i},\ldots , \)
\( P_{j},\ldots ,P_{m},\ldots ,P_{n} \) (where the meaning of the
subscripts is the one of Definition~\ref{def:ordered}) preserves
the perfect model semantics, we will use Theorem~\ref{th:proof_trees_perfect_model}
and we will show that, for \( k=0,\ldots ,n \), given any ground
atom \( A \) whose predicate belongs to the set \( \mathit{Pred}_{int} \)
of predicates of interest, there exists a proof tree for \( A \)
and \( P_{k} \) iff there exists a proof tree for \( A \) and \( P_{0}\cup \mathit{Defs}_{n} \).
Since \( P_{i}=P_{0}\cup \mathit{Defs}_{n} \), it is sufficient to
show the following properties, for any ground atom \( A \): 
\smallskip{}

\noindent (P1)~there exists a proof tree for \( A \) and \( P_{i} \)
iff there exists a proof tree for \( A \) and \( P_{j} \),
\smallskip{}

\noindent (P2)~there exists a proof tree for \( A \) and \( P_{j} \)
iff there exists a proof tree for \( A \) and \( P_{m} \), and
\smallskip{}

\noindent (P3)~if the predicate of \( A \) is in \( \mathit{Pred}_{int} \),
then there exists a proof tree for \( A \) and \( P_{m} \) iff there
exists a proof tree for \( A \) and \( P_{n} \).
\smallskip{}

\noindent Property P1 is proved by the following proposition.

\begin{proposition}
\noindent \label{prop:tc_of_pos_unf} Let \( P_{0} \) be a locally
stratified program and let \( P_{0},\ldots ,P_{i},\ldots , \) \( P_{j},\ldots ,P_{m},\ldots ,P_{n} \)
be an ordered transformation sequence. Then there exists a proof tree
for a ground atom \( A \) and \( P_{i} \) iff there exists a proof
tree for \( A \) and \( P_{j} \).
\end{proposition}
The proof of Proposition~\ref{prop:tc_of_pos_unf} is given in Appendix~B.
It is a proof by induction on \( \sigma (A) \) and on the size of
the proof tree for \( A \).

In order to prove the only-if part of Property P2, we will show a
stronger invariant property based on the following consistency notion. 

\begin{definition}
\emph{{[}\( P_{j} \)-consistency{]}} Let \( P_{0},\ldots ,P_{i},\ldots , \)
\( P_{j},\ldots ,P_{m},\ldots ,P_{n} \) be an ordered transformation
sequence, \( P_{k} \) be a program in this sequence, and \( A \)
be a ground atom. We say that a proof tree \( T \) for \( A \) and
\( P_{k} \) is \emph{\( P_{j} \)-consistent} iff for every ground
atom \( B \) and ground literals \( L_{1},\ldots ,L_{r} \), if \( B \)
is the father of \( L_{1},\ldots ,L_{r} \) in \( T \), then \( \mu (B,P_{j})>\mu (L_{1}\wedge \ldots \wedge L_{r},P_{j}) \).
\end{definition}
The invariant property is as follows: for every program \( P_{k} \)
in the sequence \( P_{j},\ldots ,P_{m} \), if there exists a \mbox{\( P_{j} \)-consistent}
proof tree for \( A \) and \( P_{j} \), then there exists a \mbox{\( P_{j} \)-consistent}
proof tree for \( A \) and \( P_{k} \).

It is important that \( P_{j} \)-consistency refers to the program
\( P_{j} \) obtained by applying the positive unfolding rule to each
clause that belongs to \( \mathit{Defs}_{i} \) and is used in \( P_{j},\ldots ,P_{m} \)
for a folding step. Indeed, if the positive unfolding rule is not
applied to a clause in \( \mathit{Defs}_{i} \), and this clause is
then used (possibly, together with other clauses) in a folding step,
then the preservation of \( P_{j} \)-consistent proof trees may not
be ensured and the transformation sequence may not be correct. This
is shown by Example~\ref{ex:last_intro} of the Introduction where
we assume that the first clause \( p(X)\leftarrow \neg q(X) \) of
\( P_{0} \) has been added by the definition introduction rule in
a previous step.

We have the following.

\begin{proposition}
\noindent \label{prop:mu-consistency} If there exists a proof tree
for a ground atom \( A \) and program \( P_{j} \) then there exists
a \emph{\( P_{j} \)-}consistent proof tree for \( A \) and \( P_{j} \).
\end{proposition}
\begin{proof}
\noindent Let \( T \) be a proof tree for \( A \) and \( P_{j} \)
such that \( \langle \sigma (A),\mathit{size}(T)\rangle  \) is minimal
w.r.t.~\( <_{\mathit{lex}} \). Then \( T \) is \emph{\( P_{j} \)}-consistent.\hfill{}\( \Box  \)
\end{proof}
Notice that in the proof of Proposition~\ref{prop:mu-consistency}
we state the existence of a \emph{\( P_{j} \)-}consistent proof tree
for a ground atom \( A \) and program \( P_{j} \) without providing
an effective method for constructing this proof tree. In fact, it
should be noticed that no effective method can be given for constructing
the minimal proof tree for a given atom and program, because the existence
of such a proof tree is not decidable and not even semi-decidable. 

By Proposition~\ref{prop:mu-consistency}, in order to prove Property~P2
it is enough to show the following Proposition~\ref{prop:preserv-mu-consistency}.

\begin{proposition}
\noindent \label{prop:preserv-mu-consistency} Let \( P_{0} \) be
a locally stratified program and let \( P_{0},\ldots ,P_{i},\ldots , \)
\( P_{j},\ldots ,P_{m},\ldots ,P_{n} \) be an ordered transformation
sequence. Then, for every ground atom \( A \) we have that:
\smallskip{}

\noindent \textup{(Soundness)} if there exists a proof tree for \( A \)
and \( P_{m} \), then there exists a proof tree for \( A \) and
\( P_{j} \), and 
\smallskip{}

\noindent \textup{(Completeness)} if there exists a \( P_{j} \)-consistent
proof tree for \( A \) and \( P_{j} \), then there exists a \( P_{j} \)-consistent
proof tree for \( A \) and \( P_{m} \).
\end{proposition}
The proof of Proposition~\ref{prop:preserv-mu-consistency} is given
in Appendix~C. 

In order to prove Property~P3, it is enough to prove the following
Proposition~\ref{prop:tc_of_def_elim}, which is a straightforward
consequence of the fact that the existence of a proof tree for a ground
atom with predicate \( p \) is determined only by the existence of
proof trees for atoms with predicates on which \( p \) depends.

\begin{proposition}
\label{prop:tc_of_def_elim} Let \( P \) be a locally stratified
program and let \( \mathit{Pred}_{int} \) be a set of predicates
of interest. Suppose that program \( Q \) is derived from program
\( P \) by eliminating the definition of a predicate \( q \) such
that no predicate in \( \mathit{Pred}_{int} \) depends on \( q \).
Then, for every ground atom \( A \) whose predicate is in \( \mathit{Pred}_{int} \),
there exists a proof tree for \( A \) and \( P \) iff there exists
a proof tree for \( A \) and \( Q \).
\end{proposition}
Now, as a consequence of Propositions~\ref{prop:local_strat}--\ref{prop:tc_of_def_elim},
and Theorem~\ref{th:proof_trees_perfect_model}, we get the following
theorem which ensures that an admissible transformation sequence preserves
the perfect model semantics. 

\begin{theorem}
\noindent \emph{{[}Correctness of Admissible Transformation Sequences{]}}
\label{th:corr_of_rules}Let \( P_{0} \) be a locally stratified
program and let \( P_{0},\ldots ,P_{n} \) be an admissible transformation
sequence. Let \( \mathit{Pred}_{int} \) be the set of predicates
of interest. Then \( P_{0}\cup \mathit{Defs}_{n} \) and \( P_{n} \)
are locally stratified and for every ground atom \( A \) whose predicate
belongs to \( \mathit{Pred}_{int} \), \( A\in M(P_{0}\cup \mathit{Defs}_{n}) \)
iff \( A\in M(P_{n}) \).
\end{theorem}
This theorem does not hold if we add to the boolean laws listed in
rule R7 of Section~\ref{sec:rules} the inverse of law (4), as shown
by the following example.

\begin{example}
Let us consider the following transformation sequence:

\noindent \begin{tabular}{llllllll}
\( P_{0} \):~ &
\( p\leftarrow q\wedge q \)~~~~~~~~&
\( P_{1} \):~&
\( p\leftarrow q \)~~~~~~~~&
\( P_{2} \):~ &
\( p\leftarrow q\wedge q \)~~~~~~~~&
\( P_{3} \):~ &
\( p\leftarrow p \)\\
&
\( q\leftarrow  \)&
&
\( q\leftarrow  \)&
&
\( q\leftarrow  \)&
&
\( q\leftarrow  \)\\
\end{tabular} 

\noindent We assume that the clause for \( p \) in \( P_{0} \) is
added to \( P_{0} \) by the definition introduction rule, so that
it can be used for folding. Program \( P_{1} \) is derived from \( P_{0} \)
by unfolding, program \( P_{2} \) is derived from \( P_{1} \) by
replacement based on the reverse of law (4), and finally, program
\( P_{3} \) is derived by folding the first clause of \( P_{2} \)
using the first clause of \( P_{0} \). We have that \( p\in M(P_{0}) \),
while \( p\not \in M(P_{3}) \).
\end{example}
Analogously, the reader may verify that Theorem~\ref{th:corr_of_rules}
does not hold if we add to the boolean laws of rule R7 the inverse
of law (6).

\section{Examples of Use of the Transformation Rules\label{sec:examples}}

In this section we show some program derivations realized by applying
the transformation rules of Section~\ref{sec:rules}. These program
derivations are examples of the following three techniques: (1)~the
\emph{determinization} technique, which is used for deriving a deterministic
program from a nondeterministic one~\cite{Fi&02b,Pe&97a}, (2)~the
\emph{program synthesis} technique, which is used for deriving a program
from a first order logic specification (see, for instance, \cite{Hog81,SaT84}
and \cite{Ba&03} in this book for a recent survey), and (3)~the
\emph{program specialization} technique, which is used for deriving
a specialized program from a given program and a given portion of
its input data (see, for instance, \cite{Jo&93} and \cite{LeB02}
for a recent survey). 

Although we will \emph{not} provide in this paper any automatic transformation
strategy, the reader may realize that in the examples we will present,
there is a systematic way of performing the program derivations. In
particular, we perform all derivations according to the repeated application
of the following sequence of steps: (i)~first, we consider some predicate
definitions in the initial program or we introduce some new predicate
definitions, (ii)~then we unfold these definitions by applying the
positive and, possibly, the negative unfolding rules, (iii)~then
we manipulate the derived clauses by applying the rules of replacement,
constraint addition, and constraint deletion, and (iv)~finally, we
apply the folding rules. The final programs are derived by applying
the definition elimination rule, and keeping only those clauses that
are needed for computing the predicates of interest.

\subsection{Determinization: Comparing Even and Odd Occurrences of a List\label{ex:even-odd}}

Let us consider the problem of checking whether or not, for any given
list \( L \) of numbers, the following property \( r(L) \) holds:
every number occurring in \( L \) in an even position is greater
or equal than every number occurring in \( L \) in an odd position.
The locally stratified program \emph{EvenOdd} shown below, solves
the given problem by introducing a new predicate \( p(L) \) which
holds iff there is a pair \( \langle X,Y\rangle  \) of numbers such
that \( X \) occurs in the the list \( L \) in an even position,
\( Y \) occurs in \( L \) in an odd position, and \( X\! <\! Y \).
Thus, for any list \( L \), the property \( r(L) \) holds iff \( p(L) \)
does not hold.
\smallskip{}

\noindent \emph{EvenOdd}:
\smallskip{}

\begin{tabular}{ll}
1.~~\( r(L)\leftarrow  \)&
\( list(L)\wedge \neg p(L) \)\\
\end{tabular}

\begin{tabular}{ll}
2.~~\( p(L)\leftarrow  \)&
\( I\! \geq \! 1\wedge J\! \geq \! 1\wedge X\! <\! Y\wedge  \)\\
&
\( occurs(X,I,L)\wedge even(I)\wedge occurs(Y,J,L)\wedge \neg even(J) \)\\
\end{tabular}

\begin{tabular}{l}
3.~~\( even(X)\leftarrow X\! =\! 0 \)\\
4.~~\( even(X\! +\! 1)\leftarrow X\! \geq \! 0\wedge \neg even(X) \)\\
5.~~\( occurs(X,I,[H|T])\leftarrow I\! =\! 1\wedge X\! =\! H \)\\
6.~~\( occurs(X,I\! +\! 1,[H|T])\leftarrow I\! \geq \! 1\wedge occurs(X,I,T) \)\\
7.~~\( list([\, ])\leftarrow  \)\\
8.~~\( list([H|T])\leftarrow list(T) \)\\
\end{tabular}
\smallskip{}

\noindent In this program \( \mathit{occurs}(X,I,L) \) holds iff
\( X \) is the \( I \)-th element (with \( I\! \geq \! 1 \)) of
the list \( L \) starting from the left. When executed by using SLDNF
resolution, this \emph{EvenOdd} program may generate, in a nondeterministic
way, all possible pairs \( \langle X,Y\rangle  \), occurring in even
and odd positions, respectively. This program has an \( O(n^{2}) \)
time complexity in the worst case, where \( n \) is the length of
the input list.

We want to derive a more efficient \emph{definite} program that can
be executed in a \emph{deterministic} way, in the sense that for every
constrained goal \( c\wedge A\wedge G \) derived from a given ground
query by LD-resolution~\cite{Apt97} there exists at most one clause
\( H\leftarrow d\wedge K \) such that \( c\wedge A\! =\! H\wedge d \)
is satisfiable.

To give a sufficient condition for determinism we need the following
notion. We say that a variable \( X \) is a \emph{local variable}
of a clause \( \gamma  \) iff \( X\in FV(bd(\gamma ))-FV(hd(\gamma )) \).
The determinism of a program \( P \) can be ensured by the following
syntactic conditions: (i)~no clause in \( P \) has local variables
and (ii)~any two clauses \( H_{1}\leftarrow c_{1}\wedge G_{1} \)
and \( H_{2}\leftarrow c_{2}\wedge G_{2} \) in \( P \) are \emph{mutually
exclusive}, that is, the constraint \( H_{1}\! =\! H_{2}\wedge c_{1}\wedge c_{2} \)
is unsatisfiable.

Our derivation consists of two transformation sequences. The first
sequence starts from the program made out of clauses 2--8 and derives
a deterministic, definite program \( Q \) for predicate \( p \).
The second sequence starts from \( Q\cup \{1\} \) and derives a deterministic,
definite program \emph{\( \mathit{EvenOdd}_{\mathit{det}} \)} for
predicate \( r \).

Let us show the construction of the first transformation sequence.
Since clause~\( 2 \) has local variables, we want to transform it
into a set of clauses that have no local variables and are mutually
exclusive, and thus, they will constitute a deterministic, definite
program. We start off by applying the positive unfolding rule to clause~\( 2 \),
followed by applications of the replacement rule based on laws of
constraints and equality. We derive:
\smallskip{}

\begin{tabular}{c}
~9.~~\( p([A|L])\leftarrow J\! \geq \! 1\wedge Y\! <\! A\wedge \mathit{occurs}(Y,J,L)\wedge \mathit{even}(J\! +\! 1) \)\\
\end{tabular}

\begin{tabular}{cl}
10.~~\( p([A|L])\leftarrow  \)&
\( I\! \geq \! 1\wedge J\! \geq \! 1\wedge X\! <\! Y\wedge \mathit{occurs}(X,I,L)\wedge  \)\\
&
\( \mathit{even}(I\! +\! 1)\wedge \mathit{occurs}(Y,J,L)\wedge \neg \mathit{even}(J\! +\! 1) \)\\
\end{tabular}
\smallskip{}

\noindent Now, by applications of the positive unfolding rule, negative
unfolding, and replacement rules, we derive the following clauses
for \( p \):
\smallskip{}

\begin{tabular}{c}
11.~~\( p([A,B|L])\leftarrow B\! <\! A \)\\
\end{tabular}

\begin{tabular}{c}
12.~~\( p([A,B|L])\leftarrow B\! \geq \! A\wedge I\! \geq \! 1\wedge X\! <\! A\wedge \mathit{occurs}(X,I,L)\wedge \mathit{even}(I) \)\\
\end{tabular}

\begin{tabular}{c}
13.~~\( p([A,B|L])\leftarrow B\! \geq \! A\wedge I\! \geq \! 1\wedge B\! <\! X\wedge \mathit{occurs}(X,I,L)\wedge \neg \mathit{even}(I) \)\\
\end{tabular}

\begin{tabular}{cl}
14.~~\( p([A,B|L])\leftarrow  \)&
\( B\! \geq \! A\wedge I\! \geq \! 1\wedge J\! \geq \! 1\wedge X\! <\! Y\! \wedge \! \mathit{occurs}(X,I,L)\! \wedge \! \mathit{even}(I)\wedge  \)\\
&
~~~~~~~~~~~\( \mathit{occurs}(Y,J,L)\wedge \neg \mathit{even}(J) \)\\
\end{tabular}
\smallskip{}

\noindent Notice that the three clauses 12, 13, and 14, are not mutually
exclusive. In order to derive a deterministic program for \( p \),
we introduce the following new definition:
\smallskip{}

\begin{tabular}{c}
15.~~\( \mathit{new}1(A,B,L)\leftarrow I\! \geq \! 1\wedge X\! <\! A\wedge \mathit{occurs}(X,I,L)\wedge \mathit{even}(I) \)\\
\end{tabular}

\begin{tabular}{c}
16.~~\( \mathit{new}1(A,B,L)\leftarrow I\! \geq \! 1\wedge B\! <\! X\wedge \mathit{occurs}(X,I,L)\wedge \neg \mathit{even}(I) \)\\
\end{tabular}

\begin{tabular}{ll}
17.~~\( \mathit{new}1(A,B,L)\leftarrow  \)&
\( I\! \geq \! 1\wedge J\! \geq \! 1\wedge X\! <\! Y\wedge \mathit{occurs}(X,I,L)\wedge \mathit{even}(I)\wedge  \)\\
&
\( \mathit{occurs}(Y,J,L)\wedge \neg \mathit{even}(J) \)\\
\end{tabular}
\smallskip{}

\noindent and we fold clauses 12, 13, and 14 by using the definition
of \( \mathit{new}1 \), that is, clauses 15, 16, and 17. We derive:
\smallskip{}

\begin{tabular}{c}
18.~~\( p([A,B|L])\leftarrow B\! \geq \! A\wedge \mathit{new}1(A,B,L) \)\\
\end{tabular}
\smallskip{}

\noindent Clauses 11 and 18 have no local variables and are mutually
exclusive. We are left with the problem of deriving a deterministic
program for the newly introduced predicate \( \mathit{new}1 \). 

By applying the positive unfolding, negative unfolding, and replacement
rules, from clauses 15, 16, and 17, we get:
\smallskip{}

\begin{tabular}{c}
19.~~\( \mathit{new}1(A,B,[C|L])\leftarrow B\! <\! C \)\\
\end{tabular}

\begin{tabular}{c}
20.~~\( \mathit{new}1(A,B,[C|L])\leftarrow I\! \geq \! 1\wedge B\! <\! X\wedge \mathit{occurs}(X,I,L)\wedge \mathit{even}(I) \)\\
\end{tabular}

\begin{tabular}{c}
21.~~\( \mathit{new}1(A,B,[C|L])\leftarrow I\! \geq \! 1\wedge X\! <\! A\wedge \mathit{occurs}(X,I,L)\wedge \neg \mathit{even}(I) \)\\
\end{tabular}

\begin{tabular}{c}
22.~~\( \mathit{new}1(A,B,[C|L])\leftarrow I\! \geq \! 1\wedge X\! <\! C\wedge \mathit{occurs}(X,I,L)\wedge \neg \mathit{even}(I) \)\\
\end{tabular}

\begin{tabular}{ll}
23.~~\( \mathit{new}1(A,B,[C|L])\leftarrow  \)&
\( I\! \geq \! 1\wedge J\! \geq \! 1\wedge X\! <\! Y\wedge \mathit{occurs}(X,I,L)\wedge  \)\\
&
\( \neg \mathit{even}(I)\wedge \mathit{occurs}(Y,J,L)\wedge \mathit{even}(J) \)\smallskip{}
\\
\end{tabular}

\noindent In order to derive mutually exclusive clauses without local
variables we first apply the replacement rule and derive sets of clauses
corresponding to mutually exclusive cases, and then we fold each of
these sets of clauses. We use the replacement rule based on law (5)
and law (9) which is justified by the equivalence: \( \forall X\forall Y(true\leftrightarrow  \)
\( X\! \geq \! Y\vee X\! <\! Y) \). We get:
\smallskip{}

\begin{tabular}{c}
24.~~\( \mathit{new}1(A,B,[C|L])\leftarrow B\! <\! C \)\\
\end{tabular}

\begin{tabular}{ll}
25.~~\( \mathit{new}1(A,B,[C|L])\leftarrow  \)&
\( B\! \geq \! C\wedge A\! \geq \! C\wedge I\! \geq \! 1\wedge B\! <\! X\, \wedge  \)\\
&
\( \mathit{occurs}(X,I,L)\wedge \mathit{even}(I) \)\\
\end{tabular}

\begin{tabular}{cl}
26.~~\( \mathit{new}1(A,B,[C|L])\leftarrow  \)&
\( B\! \geq \! C\wedge A\! \geq \! C\wedge I\! \geq \! 1\wedge X\! <\! A\, \wedge  \)\\
&
\( \mathit{occurs}(X,I,L)\wedge \neg \mathit{even}(I) \)\\
\end{tabular}

\begin{tabular}{ll}
27.~~\( \mathit{new}1(A,B,[C|L])\leftarrow  \)&
\( B\! \geq \! C\wedge A\! \geq \! C\wedge I\! \geq \! 1\wedge J\! \geq \! 1\wedge X\! <\! Y\wedge  \)\\
&
\( \mathit{occurs}(X,I,L)\wedge \neg \mathit{even}(I)\wedge  \)\\
&
\( \mathit{occurs}(Y,J,L)\wedge \mathit{even}(J) \)\\
\end{tabular}

\begin{tabular}{cl}
28.~~\( \mathit{new}1(A,B,[C|L])\leftarrow  \)&
\( B\! \geq \! C\wedge A\! <\! C\wedge I\! \geq \! 1\wedge B\! <\! X\wedge  \)\\
&
\( \mathit{occurs}(X,I,L)\wedge \mathit{even}(I) \)\\
\end{tabular}

\begin{tabular}{cl}
29.~~\( \mathit{new}1(A,B,[C|L])\leftarrow  \)&
\( B\! \geq \! C\wedge A\! <\! C\wedge I\! \geq \! 1\wedge X\! <\! C\, \wedge  \)\\
&
\( \mathit{occurs}(X,I,L)\wedge \neg \mathit{even}(I) \)\\
\end{tabular}

\begin{tabular}{ll}
30.~~\( \mathit{new}1(A,B,[C|L])\leftarrow  \)&
\( B\! \geq \! C\wedge A\! <\! C\wedge I\! \geq \! 1\wedge J\! \geq \! 1\wedge X\! <\! Y\wedge  \)\\
&
\( \mathit{occurs}(X,I,L)\wedge \neg \mathit{even}(I)\wedge  \)\\
&
\( \mathit{occurs}(Y,J,L)\wedge \mathit{even}(J) \)\\
\end{tabular}
\smallskip{}

\noindent The three sets of clauses: \{24\}, \{25, 26, 27\}, and \{28,
29, 30\} correspond to the mutually exclusive cases: \( (B\! <\! C) \),
\( (B\! \geq \! C\wedge A\! \geq \! C) \), and \( (B\! \geq \! C\wedge A\! <\! C) \),
respectively. Now, in order to fold each set \{25, 26, 27\} and \{28,
29, 30\} and derive mutually exclusive clauses without local variables,
we introduce the following new definition: 
\smallskip{}

\begin{tabular}{c}
31.~~\( \mathit{new}2(A,B,L)\leftarrow I\! \geq \! 1\wedge B\! <\! X\wedge \mathit{occurs}(X,I,L)\wedge \mathit{even}(I) \)\\
\end{tabular}

\begin{tabular}{c}
32.~~\( \mathit{new}2(A,B,L)\leftarrow I\! \geq \! 1\wedge X\! <\! A\wedge \mathit{occurs}(X,I,L)\wedge \neg \mathit{even}(I) \)\\
\end{tabular}

\begin{tabular}{ll}
33.~~\( \mathit{new}2(A,B,L)\leftarrow  \)&
\( I\! \geq \! 1\wedge J\! \geq \! 1\wedge X\! <\! Y\wedge \mathit{occurs}(X,I,L)\wedge \neg \mathit{even}(I)\wedge  \)\\
&
\( \mathit{occurs}(Y,J,L)\wedge \mathit{even}(J) \)\\
\end{tabular}
\smallskip{}

\noindent By folding clauses 25, 26, 27 and 28, 29, 30 using clauses
31, 32, and 33, for predicate \( \mathit{new}1 \) we get the following
mutually exclusive clauses without local variables:
\smallskip{}

\begin{tabular}{c}
34.~~\( \mathit{new}1(A,B,[C|L])\leftarrow B\! <\! C \)\\
\end{tabular}

\begin{tabular}{c}
35.~~\( \mathit{new}1(A,B,[C|L])\leftarrow B\! \geq \! C\wedge A\! \geq \! C\wedge \mathit{new}2(A,B,L) \)\\
\end{tabular}

\begin{tabular}{c}
36.~~\( \mathit{new}1(A,B,[C|L])\leftarrow B\! \geq \! C\wedge A\! <\! C\wedge \mathit{new}2(C,B,L) \)\smallskip{}
\\
\end{tabular}

\noindent Unfortunately, the clauses for the new predicate \( \mathit{new}2 \)
have local variables and are not mutually exclusive. Thus, we continue
our derivation and, by applying the positive unfolding, negative unfolding,
replacement, and folding rules, from clauses 31, 32, and 33 we derive
the following clauses (this derivation is similar to the derivation
that lead from \{15, 16, 17\} to \{34, 35, 36\} and we omit it):
\smallskip{}

\begin{tabular}{c}
37.~~\( \mathit{new}2(A,B,[C|L])\leftarrow C\! <\! A \)\\
\end{tabular}

\begin{tabular}{c}
38.~~\( \mathit{new}2(A,B,[C|L])\leftarrow C\! \geq \! A\wedge B\! \geq \! C\wedge \mathit{new}1(A,C,L) \)\\
\end{tabular}

\begin{tabular}{c}
39.~~\( \mathit{new}2(A,B,[C|L])\leftarrow C\! \geq \! A\wedge B\! <\! C\wedge \mathit{new}1(A,B,L) \)\smallskip{}
\\
\end{tabular}

\noindent The set of clauses derived so far starting from the initial
clause~2, that is, \{11, 18, 34, 35, 36, 37, 38, 39\} constitutes
a deterministic program for \( p \), call it \( Q \). 

Now we construct the second transformation sequence starting from
\( Q\cup \{1\} \) for deriving a deterministic, \emph{definite} program
for \( r \). We start off by considering clause 1 which defines \( r \)
and, by positive unfolding, negative unfolding, and replacement we
derive:
\smallskip{}

\begin{tabular}{c}
40.~~\( r([\, ])\leftarrow  \)\\
\end{tabular}

\begin{tabular}{c}
41.~~\( r([A])\leftarrow  \)\\
\end{tabular}

\begin{tabular}{c}
42.~~\( r([A,B|L])\leftarrow \mathit{list}(L)\wedge B\! \geq \! A\wedge \neg \mathit{new}1(A,B,L) \)\\
\end{tabular}
\smallskip{}

\noindent By introducing the following definition:
\smallskip{}

\begin{tabular}{c}
43.~~\( \mathit{new}3(A,B,L)\leftarrow \mathit{list}(L)\wedge B\! \geq \! A\wedge \neg \mathit{new}1(A,B,L) \)\\
\end{tabular}
\smallskip{}

\noindent and then folding clause 42 using clause 43, we derive the
following definite clauses:
\smallskip{}

\begin{tabular}{c}
44.~~\( r([\, ])\leftarrow  \)\\
\end{tabular}

\begin{tabular}{c}
45.~~\( r([A])\leftarrow  \)\\
\end{tabular}

\begin{tabular}{c}
46.~~\( r([A,B|L])\leftarrow \mathit{B}\! \geq \! \mathit{A}\wedge \mathit{new}3(A,B,L) \)\\
\end{tabular}
\smallskip{}

\noindent Now, we want to transform clause 43 into a set of definite
clauses. By positive unfolding, negative unfolding, and replacement,
from clause 43 we derive:
\smallskip{}

\begin{tabular}{c}
47.~~\( \mathit{new}3(A,B,[\, ])\leftarrow B\! \geq \! A \)\\
\end{tabular}

\begin{tabular}{c}
48.~~\( \mathit{new}3(A,B,[C|L])\leftarrow B\! \geq \! C\wedge A\! <\! C\wedge \mathit{list}(L)\wedge B\! \geq \! C\wedge \neg \mathit{new}2(C,B,L) \)\\
\end{tabular}

\begin{tabular}{c}
49.~~\( \mathit{new}3(A,B,[C|L])\leftarrow B\! \geq \! C\wedge A\! \geq \! C\wedge \mathit{list}(L)\wedge B\! \geq \! A\wedge \neg \mathit{new}2(A,B,L) \)\\
\end{tabular}~
\smallskip{}

\noindent In order to transform clauses 48 and 49 into definite clauses,
we introduce the following definition:
\smallskip{}

\begin{tabular}{c}
50.~~\( \mathit{new}4(A,B,L)\leftarrow \mathit{list}(L)\wedge B\! \geq \! A\wedge \neg \mathit{new}2(A,B,L) \)\\
\end{tabular}
\smallskip{}

\noindent and we fold clauses 48 and 49 using clause 50. We get:
\smallskip{}

\begin{tabular}{c}
51.~~\( \mathit{new}3(A,B,[\, ])\leftarrow B\! \geq \! A \)\\
\end{tabular}

\begin{tabular}{c}
52.~~\( \mathit{new}3(A,B,[C|L])\leftarrow B\! \geq \! C\wedge A\! <\! C\wedge \mathit{new}4(C,B,L) \)\\
\end{tabular}

\begin{tabular}{c}
53.~~\( \mathit{new}3(A,B,[C|L])\leftarrow B\! \geq \! C\wedge A\! \geq \! C\wedge \mathit{new}4(A,B,L) \)\\
\end{tabular}
\smallskip{}

\noindent Now we are left with the task of transforming clause 50
into a set of definite clauses. By applying the positive unfolding,
negative unfolding, replacement, and folding rules, we derive:
\smallskip{}

\begin{tabular}{c}
54.~~\( \mathit{new}4(A,B,[\, ])\leftarrow B\! \geq \! A \)\\
\end{tabular}

\begin{tabular}{c}
55.~~\( \mathit{new}4(A,B,[C|L])\leftarrow B\! <\! C\wedge C\! \geq \! A\wedge \mathit{new}3(A,B,L) \)\\
\end{tabular}

\begin{tabular}{c}
56.~~\( \mathit{new}4(A,B,[C|L])\leftarrow B\! \geq \! C\wedge C\! \geq \! A\wedge \mathit{new}3(A,C,L) \)\\
\end{tabular}
\smallskip{}

\noindent Finally, by eliminating the definitions of the predicates
on which \( r \) does not depend, we get, as desired, the following
final program which is a deterministic, definite program.
\medskip{}

\noindent \emph{\( \mathit{EvenOdd}_{\mathit{det}} \):}
\smallskip{}

\begin{tabular}{c}
44.~~\( r([\, ])\leftarrow  \)\\
\end{tabular}

\begin{tabular}{c}
45.~~\( r([A])\leftarrow  \)\\
\end{tabular}

\begin{tabular}{c}
46.~~\( r([A,B|L])\leftarrow \mathit{B}\! \geq \! \mathit{A}\wedge \mathit{new}3(A,B,L) \)\\
\end{tabular}

\begin{tabular}{c}
51.~~\( \mathit{new}3(A,B,[\, ])\leftarrow B\! \geq \! A \)\\
\end{tabular}

\begin{tabular}{c}
52.~~\( \mathit{new}3(A,B,[C|L])\leftarrow B\! \geq \! C\wedge A\! <\! C\wedge \mathit{new}4(C,B,L) \)\\
\end{tabular}

\begin{tabular}{c}
53.~~\( \mathit{new}3(A,B,[C|L])\leftarrow B\! \geq \! C\wedge A\! \geq \! C\wedge \mathit{new}4(A,B,L) \)\\
\end{tabular}

\begin{tabular}{c}
54.~~\( \mathit{new}4(A,B,[\, ])\leftarrow B\! \geq \! A \)\\
\end{tabular}

\begin{tabular}{c}
55.~~\( \mathit{new}4(A,B,[C|L])\leftarrow B\! <\! C\wedge C\! \geq \! A\wedge \mathit{new}3(A,B,L) \)\\
\end{tabular}

\begin{tabular}{c}
56.~~\( \mathit{new}4(A,B,[C|L])\leftarrow B\! \geq \! C\wedge C\! \geq \! A\wedge \mathit{new}3(A,C,L) \)\\
\end{tabular}
\smallskip{}

\noindent Given a list of numbers \( L \) of length \( n \), the
\emph{\( \mathit{EvenOdd}_{\mathit{det}} \)} program checks that
\( r(L) \) holds by performing at most \( 2n \) comparisons between
numbers occurring in \( L \). Program \emph{\( \mathit{EvenOdd}_{\mathit{det}} \)}
works by traversing the input list \( L \) only once (without backtracking)
and storing, for every initial portion \( L_{1} \) of the input list
\( L \), the maximum number \( A \) occurring in an odd position
of \( L_{1} \) and the minimum number \( B \) occurring in an even
position of \( L_{1} \) (see the first two arguments of the predicates
\( new3 \) and \( new4 \)). When looking at the first element \( C \)
of the portion of the input list still to be visited (i.e., the third
argument of \( new3 \) or \( new4 \)), the following two cases are
possible: either (Case~1) the element \( C \) occurs in an odd position
of the input list \( L \), i.e., a call of the form \( new3(A,B,[C|L_{2}]) \)
is executed, or (Case~2) the element \( C \) occurs in an even position
of the input list \( L \), i.e., a call of the form \( new4(A,B,[C|L_{2}]) \)
is executed. In Case~(1) program \emph{\( \mathit{EvenOdd}_{\mathit{det}} \)}
checks that \( B\! \geq \! C \) holds and then updates the value
of the maximum number occurring in an odd position with the maximum
between \( A \) and \( C \). In Case~(2) program \emph{\( \mathit{EvenOdd}_{\mathit{det}} \)}
checks that \( C\! \geq \! A \) holds and then updates the value
of the minimum number occurring in an even position with the minimum
between \( B \) and \( C \).

\subsection{Program Synthesis: The N-queens Problem\label{ex:nqueens}}

The \( N \)-queens problem has been often considered in the literature
for presenting various programming techniques, such as recursion and
backtracking. We consider it here as an example of the program synthesis
technique, as it has been done in~\cite{SaT84}. Our derivation is
different from the one presented in~\cite{SaT84}, because the derivation
in~\cite{SaT84} makes use of the unfold/fold transformation rules
for definite programs together with an \emph{ad hoc} transformation
rule (called \emph{negation technique}) for transforming general programs
(with negation) into definite programs. In contrast, we use unfold/fold
transformation rules for general programs, and in particular, our
negative unfolding rule of Section~\ref{sec:rules}.

The \( N \)-queens problem can be informally specified as follows.
We are required to place \( N(\geq \! 0) \) queens on an \( N\times N \)
chess board, so that no two queens attack each other, that is, they
do not lie on the same row, column, or diagonal. A board configuration
with this property is said to be \emph{safe}. By using the fact that
no two queens should lie on the same row, we represent an \( N\times N \)
chess board as a list \( L \) of \( N \) positive integers: the
\( k \)-th element on \( L \) represents the column of the queen
on row \( k \).

In order to give a formal specification of the \( N \)-queens problem
we follow the approach presented in \cite{PeP02a}, which is based
on first order logic. We introduce the following constraint logic
program:
\smallskip{}

\noindent \begin{tabular}{cl}
\( P:\, \, \,  \) &
\( \mathit{nat}(0)\leftarrow  \)\\
&
\( \mathit{nat}(N)\leftarrow N\! =\! M\! +\! 1\wedge M\! \geq \! 0\wedge \mathit{nat}(M) \) \\
&
\( \mathit{nat}_{-}\! \mathit{list}([\, ])\leftarrow  \)\\
&
\( \mathit{nat}_{-}\! \mathit{list}([H|T])\leftarrow \mathit{nat}(H)\wedge \mathit{nat}_{-}\! \mathit{list}(T) \) \\
&
\( \mathit{length}([\, ],0)\leftarrow  \)\\
&
\( \mathit{length}([H|T],N)\leftarrow N\! =\! M\! +\! 1\wedge M\! \geq \! 0\wedge \mathit{length}(T,M) \) \\
&
\( \mathit{member}(X,[H|T])\leftarrow X\! =\! H \)\\
&
\( \mathit{member}(X,[H|T])\leftarrow \mathit{member}(X,T) \) \\
&
\( \mathit{in}_{-}\! \mathit{range}(X,M,N)\leftarrow X\! =\! N\wedge M\! \leq \! N \)\\
&
\( \mathit{in}_{-}\! \mathit{range}(X,M,N)\leftarrow N\! =\! K\! +\! 1\wedge M\! \leq \! K\wedge \mathit{in}_{-}\! \mathit{range}(X,M,K) \) \\
&
\( \mathit{occurs}(X,I,[H|T])\leftarrow I\! =\! 1\wedge X\! =\! H \)\\
&
\( \mathit{occurs}(X,I\! +\! 1,[H|T])\leftarrow I\! \geq \! 1\wedge \mathit{occurs}(X,I,T) \) \\
\end{tabular}
\smallskip{}

\noindent and the following first order formula:
\smallskip{}

\noindent \begin{tabular}{lll}
\( \varphi (N,L): \)&
\( \mathit{nat}(N)\wedge \mathit{nat}_{-}\! \mathit{list}(L)\wedge  \)&
(1)\\
&
\( \mathit{length}(L,N)\wedge \forall X\, (\mathit{member}(X,L)\rightarrow \mathit{in}_{-}\! \mathit{range}(X,1,N))\wedge  \) &
(2)\\
&
\( \forall A,\! B,\! M,\! N\, ((1\! \leq \! M\wedge M\! \leq \! N\wedge \! \mathit{occurs}(A,M,L)\wedge \! \mathit{occurs}(B,\! N,\! L)) \)&
(3)\\
&
 ~~~~~~~~~~~~~~~~~~\( \rightarrow (A\! \neq \! B\wedge A\! -\! B\! \neq \! N\! -\! M\wedge B\! -\! A\! \neq \! N\! -\! M)) \)~~&
(4)\\
\end{tabular}
\smallskip{}

\noindent In the above program and formula \( \mathit{in}_{-}\! \mathit{range}(X,M,N) \)
holds iff \( X\in \{M,M\! +\! 1, \) \( \ldots ,N\} \) and \( N\! \geq \! 0 \).
The other predicates have been defined in previous programs or do
not require explanation. Now we define the relation \( \mathit{queens}(N,L) \)
where \( N \) is a nonnegative integer and \( L \) is a list of
positive integers, as follows: 
\smallskip{}

\( \mathit{queens}(N,L) \) ~iff~ \( M(P)\models \varphi (N,L) \)
\smallskip{}

\noindent Line (2) of the formula \( \varphi (N,L) \) above specifies
a chess board as a list of \( N \) integers each of which is in the
range \( [1,\ldots ,N] \). If \( N\! =\! 0 \) the list is empty.
Lines~(3) and (4) of \( \varphi (N,L) \) specify the safety property
of board configurations. Now, we would like to derive a constraint
logic program \( R \) which computes the relation \( \mathit{queens}(N,L) \),
that is, \( R \) should define a predicate \( \mathit{queens}(N,L) \)
such that:
\smallskip{}

\( (\pi ) \)~~~~~\( M(R)\models \mathit{queens}(N,L) \) ~iff~
\( M(P)\models \varphi (N,L) \)
\smallskip{}

\noindent Following the approach presented in \cite{PeP02a}, we start
from the formula (called a \emph{statement}) \( \mathit{queens}(N,L)\leftarrow \varphi (N,L) \)
and, by applying a variant of the \emph{Lloyd-Topor transformation}~\cite{Llo87},
we derive the following stratified logic program:
\smallskip{}

\noindent \begin{tabular}{ll}
\begin{tabular}{c}
\( F: \)\\
\\
\end{tabular}&
\begin{tabular}{ccl}
1.~~&
\( \mathit{queens}(N,L)\leftarrow  \)&
\( \mathit{nat}(N)\wedge \mathit{nat}_{-}\! \mathit{list}(L)\wedge \mathit{length}(L,N)\wedge  \)\\
&
&
\( \neg \mathit{aux}1(L,N)\wedge \neg \mathit{aux}2(L) \)\\
\end{tabular}\\
&
\begin{tabular}{ccc}
2.~~&
\( \mathit{aux}1(L,N)\leftarrow  \)&
\( \mathit{member}(X,L)\wedge \neg \mathit{in}_{-}\! \mathit{range}(X,1,N) \)\\
\end{tabular} \\
&
\begin{tabular}{lll}
3.~~&
\( \mathit{aux}2(L)\leftarrow  \)&
\( 1\! \leq \! K\wedge K\! \leq \! M\wedge  \)\\
&
&
\( \neg (A\! \neq \! B\wedge A\! -\! B\! \neq \! M\! -\! K\wedge B\! -\! A\! \neq \! M\! -\! K)\wedge  \)\\
&
&
\( \mathit{occurs}(A,K,L)\wedge \mathit{occurs}(B,M,L) \)\\
\end{tabular} \\
\end{tabular}
\smallskip{}

\noindent This variant of the Lloyd-Topor transformation is a fully
automatic transformation, but it cannot be performed by using our
transformation rules, because it operates on first order formulas.
It can be shown that this variant of the Lloyd-Topor transformation
preserves the perfect model semantics and, thus, we have that: \( M(P\cup F)\models \mathit{queens}(N,L) \)
~iff~ \( M(P)\models \varphi (N,L) \). 

The derived program \( P\cup F \) is not very satisfactory from a
computational point of view because, when using SLDNF resolution with
the left-to-right selection rule, it may not terminate for calls of
the form \( \mathit{queens}(n,L) \) where \( n \) is a nonnegative
integer and \( L \) is a variable. Thus, the process of program synthesis
proceeds by applying the transformation rules listed in Section~\ref{sec:rules},
thereby transforming program \( P\cup F \) into a program \( R \)
such that: (i)~Property \( (\pi ) \) holds, (ii)~\( R \) is a
definite program, and (iii)~\( R \) terminates for all calls of
the form \( \mathit{queens}(n,L) \), where \( n \) is any nonnegative
integer and \( L \) is a variable. Actually, the derivation of the
final program \( R \) is performed by constructing two transformation
sequences: (i)~a first one, which starts from the initial program
\( P \), introduces clauses 2 and 3 by definition introduction, and
ends with a program \( Q \), and (ii)~a second one, which starts
from program \( Q \), introduces clause 1 by definition introduction,
and ends with program \( R \). 

We will illustrate the application of the transformation rules for
deriving program \( R \) without discussing in detail how this derivation
can be performed in an automatic way using a particular strategy.
As already mentioned, the design of suitable transformation strategies
for the automation of program derivations for constraint logic programs,
is beyond the scope of the present paper. 

The program transformation process starts off from program \( P\cup \{2,3\} \)
by transforming clauses 2 and 3 into a set of clauses without local
variables, so that they can be subsequently used for unfolding clause
1 w.r.t. \( \neg \mathit{aux}1(L,N) \) and \( \neg \mathit{aux}2(L) \)
(see the negative unfolding rule R4).

By positive unfolding, replacement, and positive folding, from clause
2 we derive:
\smallskip{}

\noindent \begin{tabular}{rll}
\makebox[8mm][r]{4. }&
\( \mathit{aux}1([H|T],N)\leftarrow  \)&
\( \neg \mathit{in}_{-}\! \mathit{range}(X,1,N) \)\\
\makebox[8mm][r]{5. }&
\( \mathit{aux}1([H|T],N)\leftarrow  \)&
\( \mathit{aux}1(T,N) \)\\
\end{tabular} 
\smallskip{}

\noindent Similarly, by positive unfolding, replacement, and positive
folding, from clause 3 we derive:
\smallskip{}

\noindent \begin{tabular}{rll}
\makebox[8mm][r]{6. }&
\( \mathit{aux}2([A|T])\leftarrow  \)&
\( M\! \geq \! 1\wedge \neg (A\! \neq \! B\wedge A\! -\! B\! \neq \! M\wedge B\! -\! A\! \neq \! M)\wedge  \)\\
&
&
\( \mathit{occurs}(B,M,T) \)\\
\makebox[8mm][r]{7. }&
\( \mathit{aux}2([A|T])\leftarrow  \)&
\( \mathit{aux}2(T) \)\\
\end{tabular} 
\smallskip{}

\noindent In order to eliminate the local variables \( B \) and \( M \)
occurring in clause 6, by the definition introduction rule we introduce
the following new clause, whose body is a generalization of the body
of clause 6:
\smallskip{}

\noindent \begin{tabular}{rll}
\makebox[8mm][r]{8. }&
\( \mathit{new}1(A,T,K)\leftarrow  \)&
\( M\! \geq \! 1\wedge \neg (A\! \neq \! B\wedge A\! -\! B\! \neq \! M\! +\! K\wedge B\! -\! A\! \neq \! M\! +\! K)\wedge  \)\\
&
&
\( \mathit{occurs}(B,M,T) \)\\
\end{tabular}
\smallskip{}

\noindent By replacement and positive folding, from clause 6 we derive:
\smallskip{}

\noindent \begin{tabular}{rll}
\makebox[8mm][r]{6f. }&
\( \mathit{aux}2([A|T])\leftarrow  \)&
\( \mathit{new}1(A,T,0) \)\\
\end{tabular} 
\smallskip{}

\noindent Now, by positive unfolding, replacement, and positive folding,
from clause 8 we derive:
\smallskip{}

\noindent \begin{tabular}{cll}
\makebox[8mm][r]{9. }&
\( \mathit{new}1(A,[B|T],K)\leftarrow  \)&
\( \neg (A\! \neq \! B\wedge A\! -\! B\! \neq \! K\! +\! 1\wedge B\! -\! A\! \neq \! K\! +\! 1) \)\\
\makebox[8mm][r]{10. }&
\( \mathit{new}1(A,[B|T],K)\leftarrow  \)&
\( \mathit{new}1(A,T,K\! +\! 1) \)\\
\end{tabular}
\smallskip{}

\noindent The program, call it \( Q \), derived so far is \( P\cup \{4,5,6{\rm f},7,9,10\} \),
and clauses 4, 5, 6f, 7, 9, and 10 have no local variables.

Now we construct a new transformation sequence which takes \( Q \)
as initial program. We start off by applying the definition introduction
rule and adding clause \( 1 \) to program \( Q \). Our objective
is to transform clause 1 into a set of definite clauses. We first
apply the definition rule and we introduce the following clause, whose
body is a generalization of the body of clause 1:
\smallskip{}

\noindent \begin{tabular}{ccl}
\makebox[8mm][r]{11. }&
\( \mathit{new}2(N,L,K)\leftarrow  \)&
\( \mathit{nat}(M)\wedge \mathit{nat}_{-}\! \mathit{list}(L)\wedge \mathit{length}(L,M)\wedge  \)\\
&
&
\( \neg \mathit{aux}1(L,N)\wedge \neg \mathit{aux}2(L)\wedge N\! =\! M\! +\! K \)\\
\end{tabular}
\smallskip{}

\noindent By replacement and positive folding, from clause 11 we derive:
\smallskip{}

\noindent \begin{tabular}{ccl}
\makebox[8mm][r]{1f. }&
\( \mathit{queens}(N,L)\leftarrow  \)&
\( \mathit{new}2(N,L,0) \)\\
\end{tabular}
\smallskip{}

\noindent By positive and negative unfolding, replacement, constraint
addition, and positive folding, from clause 11 we derive:

\noindent \begin{tabular}{rll}
\makebox[8mm][r]{12. }&
\( \mathit{new}2(N,[\: ],K)\leftarrow  \)&
\( N\! =\! K \)\\
\end{tabular}

\noindent \begin{tabular}{rll}
\makebox[8mm][r]{13. }&
\( \mathit{new}2(N,[H|T],K)\leftarrow  \)&
\( \mathit{N}\! \geq \! \mathit{K}\! +\! 1\wedge \mathit{new}2(N,T,K\! +\! 1)\wedge  \)\\
&
&
\( \mathit{nat}(H)\wedge \mathit{nat}_{-}\! \mathit{list}(T)\wedge \mathit{in}_{-}\! \mathit{range}(H,1,N)\wedge  \)\\
&
&
\( \neg \mathit{new}1(H,T,0) \)\\
\end{tabular}
\smallskip{}

\noindent In order to derive a definite program we introduce a new
predicate \( \mathit{new}3 \) defined by the following clause:
\smallskip{}

\noindent \begin{tabular}{rll}
\makebox[8mm][r]{14. }&
\( \mathit{new}3(A,T,N,M)\leftarrow  \)&
\( \mathit{nat}(A)\wedge \mathit{nat}_{-}\! \mathit{list}(T)\wedge \mathit{in}_{-}\! \mathit{range}(A,1,N)\wedge  \)\\
&
&
\( \neg \mathit{new}1(A,T,M) \)\\
\end{tabular}
\smallskip{}

\noindent We fold clause 13 using clause 14 and we derive the following
definite clause:
\smallskip{}

\noindent \begin{tabular}{rll}
\makebox[8mm][r]{13f. }&
\( \mathit{new}2(N,[H|T],K)\leftarrow  \)&
\( \mathit{N}\! \geq \! \mathit{K}\! +\! 1\wedge \mathit{new}2(N,T,K\! +\! 1)\wedge \mathit{new}3(H,T,N,0) \)\\
\end{tabular}
\smallskip{}

\noindent By positive and negative unfolding, replacement, and positive
folding, from clause 14 we derive the following definite clauses:
\smallskip{}

\noindent \begin{tabular}{rll}
\makebox[8mm][r]{15. }&
\( \mathit{new}3(A,[\, ],N,M)\leftarrow  \)&
\( \mathit{in}_{-}\! \mathit{range}(A,1,N)\wedge \mathit{nat}(A) \)\\
\end{tabular}

\noindent \begin{tabular}{rll}
\makebox[8mm][r]{16. }&
\( \mathit{new}3(A,[B|T],N,M)\leftarrow  \)&
\( A\! \neq \! B\wedge A\! -\! B\! \neq \! M\! +\! 1\wedge B\! -\! A\! \neq \! M\! +\! 1\, \wedge  \)\\
&
&
\( \mathit{nat}(B)\wedge \mathit{new}3(A,T,N,M\! +\! 1) \)\\
\end{tabular}
\smallskip{}

\noindent Finally, by assuming that the set of predicates of interest
is the singleton \{\emph{queens}\}, by definition elimination we derive
the following program:
\smallskip{}

\noindent \begin{tabular}{cl}
\( R \):&
\begin{tabular}{ccl}
\makebox[5mm][r]{1f.}&
\( \mathit{queens}(N,L)\leftarrow  \)&
\( \mathit{new}2(N,L,0) \)\\
\end{tabular}\\
&
\begin{tabular}{lll}
\makebox[5mm][r]{12.}&
\( \mathit{new}2(N,[\: ],K)\leftarrow  \)&
\( N\! =\! K \)\\
\end{tabular}\\
&
\begin{tabular}{rll}
\makebox[5mm][r]{13f.}&
\( \mathit{new}2(N,[H|T],K)\leftarrow  \)&
\( \mathit{N}\! \geq \! \mathit{K}\! +\! 1\wedge \mathit{new}2(N,T,K\! +\! 1)\wedge \mathit{new}3(H,T,N,0) \)\\
\end{tabular}\\
&
\begin{tabular}{rll}
\makebox[5mm][r]{15.}&
\( \mathit{new}3(A,[\, ],N,M)\leftarrow  \)&
\( \mathit{in}_{-}\! \mathit{range}(A,1,N)\wedge \mathit{nat}(A) \)\\
\end{tabular}\\
&
\begin{tabular}{rll}
\makebox[5mm][r]{16.}&
\( \mathit{new}3(A,[B|T],N,M)\leftarrow  \)&
\( A\! \neq \! B\wedge A\! -\! B\! \neq \! M\! +\! 1\wedge B\! -\! A\! \neq \! M\! +\! 1\, \wedge  \)\\
&
&
\( \mathit{nat}(B)\wedge \mathit{new}3(A,T,N,M\! +\! 1) \)\\
\end{tabular}\\
\end{tabular}
\smallskip{}

\noindent together with the clauses for the predicates \( \mathit{in}_{-}\! \mathit{range} \)
and \( \mathit{nat} \).

Program \( R \) is a definite program and, by Theorem~\ref{th:corr_of_rules},
we have that \( M(R)\models \mathit{queens}(N,L) \) iff \( M(P\cup F\cup \mathit{Defs})\models \mathit{queens}(N,L) \),
where \( \mathit{F}\cup \mathit{Defs} \) is the set of all clauses
introduced by the definition introduction rule during the transformation
sequences from \( P \) to \( R \). Since \emph{queens} does not
depend on \( \mathit{Defs} \) in \( P\cup F\cup \mathit{Defs} \),
we have that \( M(R)\models \mathit{queens}(N,L) \) iff \( M(P\cup F)\models \mathit{queens}(N,L) \)
and, thus, Property~\( (\pi ) \) holds. Moreover, it can be shown
that \( R \) terminates for all calls of the form \( \mathit{queens}(n,L) \),
where \( n \) is any nonnegative integer and \( L \) is a variable.

Notice that program \( R \) computes a solution of the \emph{N}-queens
problem in a clever way: each time a queen is placed on the board,
program \( R \) checks that it does not attack any other queen already
placed on the board.

\subsection{Program Specialization: Derivation of Counter Machines from Constrained
Regular Expressions \label{ex:constrregexp}}

Given a set \( \mathcal{N} \) of variables ranging over natural numbers,
a set \( \mathcal{C} \) of constraints over natural numbers, and
a set \( K \) of identifiers, we define a \emph{constrained regular
expression} \( e \) over the alphabet \( \{a,b\} \) as follows:
\smallskip{}

\begin{tabular}{ll}
\( e \)&
\( ::=\, \, \, a\, \, |\, \, b\, \, |\, \, e_{1}\cdot e_{2}\, \, |\, \, e_{1}+e_{2}\, \, |\, \, e^{\wedge }\! N\, \, |\, \, not(e)\, \, |\, \, k \)\\
\end{tabular}
\smallskip{}

\noindent where \( N\in \mathcal{N} \) and \( k\in K \). An identifier
\( k\in K \) is defined by a \emph{definition} of the form \( k\, \equiv \, (c\! :\! e) \),
where \( c\in \mathcal{C} \) and \( e \) is a constrained regular
expression. For instance, the set \( \{a^{m}b^{n}\, |\, m\! =\! n\! \geq \! 0\} \)
of strings in \( \{a,b\}^{*} \) is denoted by the identifier \( k \)
which is defined by the following definition: 
\smallskip{}

\( k\, \equiv \, (M\! =\! N:(a^{\wedge }\! M\! \cdot b^{\wedge }\! N)) \).
\smallskip{}

\noindent Obviously, constrained regular expressions may denote languages
which are \emph{not} regular.

Given a string \( S \) and a constrained regular expression \( e \),
the following locally stratified program \( P \) checks whether or
not \( S \) belongs to the language denoted by \( e \). We assume
that constraints are definable as conjunctions of equalities and disequalities
over natural numbers.
\smallskip{}

\smallskip{}
\noindent \begin{tabular}{cl}
\( P \) :&
\( \mathit{string}([\, ])\leftarrow  \)\\
~~~\( \, \, \,  \)&
\( \mathit{string}([a|S])\leftarrow \mathit{string}(S) \)\\
&
\( \mathit{string}([b|S])\leftarrow \mathit{string}(S) \)\\
&
\( \mathit{symbol}(a)\leftarrow  \) \\
&
\( \mathit{symbol}(b)\leftarrow  \) \\
&
\( \mathit{app}([\, ],L,L)\leftarrow  \)\\
&
\( \mathit{app}([A|X],Y,[A|Z])\leftarrow \mathit{app}(X,Y,Z) \)\\
\end{tabular}

\noindent \begin{tabular}{cl}
~~~\( \, \, \,  \)&
\( \mathit{in}_{-}\! \mathit{language}([A],A)\leftarrow \mathit{symbol}(A) \) \\
\end{tabular}

\noindent \begin{tabular}{ccl}
~~~\( \, \, \,  \)&
\( \mathit{in}_{-}\! \mathit{language}(S,(E1\! \cdot \! E2))\leftarrow  \)&
\( \mathit{app}(S1,S2,S)\wedge  \)\\
&
&
\( \mathit{in}_{-}\! \mathit{language}(S1,E1)\wedge \mathit{in}_{-}\! \mathit{language}(S2,E2) \)\\
\end{tabular}

\noindent \begin{tabular}{cl}
~~~\( \, \, \,  \)&
\( \mathit{in}_{-}\! \mathit{language}(S,E1\! +\! E2)\leftarrow \mathit{in}_{-}\! \mathit{language}(S,E1) \)\\
&
\( \mathit{in}_{-}\! \mathit{language}(S,E1\! +\! E2)\leftarrow \mathit{in}_{-}\! \mathit{language}(S,E2) \)\\
&
\( \mathit{in}_{-}\! \mathit{language}(S,\mathit{not}(E))\leftarrow \neg \, \mathit{in}_{-}\! \mathit{language}(S,E) \)\\
&
\( \mathit{in}_{-}\! \mathit{language}([\, ],E^{\wedge }\! I)\leftarrow I\! =\! 0 \)\\
\end{tabular}

\noindent \begin{tabular}{cll}
~~~\( \, \, \,  \)&
\( \mathit{in}_{-}\! \mathit{language}(S,E^{\wedge }\! I)\leftarrow  \)&
\( I\! =\! J\! +\! 1\wedge J\! \geq \! 0\wedge \mathit{app}(S1,S2,S)\wedge  \)\\
&
&
\( \mathit{in}_{-}\! \mathit{language}(S1,E)\wedge \mathit{in}_{-}\! \mathit{language}(S2,E^{\wedge }\! J) \)\\
\end{tabular}

\noindent \begin{tabular}{cl}
~~~\( \, \, \,  \)&
\( \mathit{in}_{-}\! \mathit{language}(S,K)\leftarrow (K\equiv (C\! :\! E))\wedge \mathit{solve}(C)\wedge \mathit{in}_{-}\! \mathit{language}(S,E) \) \\
\end{tabular}

\noindent \begin{tabular}{cl}
~~~\( \, \, \,  \)&
\( \mathit{solve}(X\! =\! Y)\leftarrow X\! =\! Y \)\\
&
\( \mathit{solve}(X\! \geq \! Y)\leftarrow X\! \geq \! Y \)\\
&
\( \mathit{solve}(C_{1}\wedge C_{2})\leftarrow \mathit{solve}(C_{1})\wedge \mathit{solve}(C_{2}) \)\\
\end{tabular}
\smallskip{}

\noindent For example, in order to check whether a string \( S \)
does \emph{not} belong to the language denoted by \( k \), where
\( k \) is defined by the following definition: \( k\, \equiv \, (M\! =\! N:(a^{\wedge }\! M\cdot b^{\wedge }\! N)) \),
we add to program \( P \) the clause:

\smallskip{}
\makebox[5mm][r]~\( (k\, \equiv \, (M\! =\! N:(a^{\wedge }\! M\! \cdot b^{\wedge }\! N)))\, \leftarrow  \)
\smallskip{}

\noindent and we evaluate a query of the form: 

\smallskip{}
\makebox[5mm][r]{}~\( \mathit{string}(S)\wedge \mathit{in}_{-}\! \mathit{language}(S,\mathit{not}(k)) \)
\smallskip{}

\noindent Now, if we want to specialize program \( P \) w.r.t. this
query, we introduce the new definition:

\smallskip{}
\makebox[5mm][r]{1.}~\( \mathit{new}1(S)\leftarrow \mathit{string}(S)\wedge \mathit{in}_{-}\! \mathit{language}(S,\mathit{not}(k)) \)
\smallskip{}

\noindent By unfolding clause 1 we get:

\smallskip{}
\makebox[5mm][r]{2.}~\( \mathit{new}1(S)\leftarrow \mathit{string}(S)\wedge \neg \, \mathit{in}_{-}\! \mathit{language}(S,k) \)
\smallskip{}

\noindent We cannot perform the negative unfolding of clause 2 w.r.t.~\( \neg \, \mathit{in}_{-}\! \mathit{language}(S,k) \)
because of the local variables in the clauses for \( \mathit{in}_{-}\! \mathit{language}(S,k) \).
In order to derive a predicate which is equivalent to \( \mathit{in}_{-}\! \mathit{language}(S,k) \)
and is defined by clauses without local variables, we introduce the
following clause:

\smallskip{}
\makebox[5mm][r]{3.}~\( \mathit{new}2(S)\leftarrow \mathit{in}_{-}\! \mathit{language}(S,k) \)
\smallskip{}

\noindent By unfolding clause 3 we get: 

\smallskip{}
\makebox[5mm][r]{4.}~\( \mathit{new}2(S)\leftarrow \mathit{M}\! =\! N\wedge \mathit{app}(S1,S2,S)\wedge  \)

\makebox[24mm][r]{}~\( \mathit{in}_{-}\! \mathit{language}(S1,a^{\wedge }\! M)\wedge \mathit{in}_{-}\! \mathit{language}(S2,b^{\wedge }\! N) \)
\smallskip{}

\noindent We generalize clause 4 and we introduce the following clause
5:

\smallskip{}
\makebox[5mm][r]{5.}~\( \mathit{new}3(S,I)\leftarrow \mathit{M}\! =\! N\! +\! I\wedge \mathit{app}(S1,S2,S)\wedge  \)

\makebox[24mm][r]{}~\( \mathit{in}_{-}\! \mathit{language}(S1,a^{\wedge }\! M)\wedge \mathit{in}_{-}\! \mathit{language}(S2,b^{\wedge }\! N) \)

\noindent By unfolding clause 5, performing replacements based on
laws of constraints, and folding, we get:

\smallskip{}
\makebox[5mm][r]{6.}~\( \mathit{new}3(S,N)\leftarrow \mathit{in}_{-}\! \mathit{language}(S,b^{\wedge }\! N) \)

\makebox[5mm][r]{7.}~\( \mathit{new}3([a|S],N)\leftarrow \mathit{new}3(S,N\! +\! 1) \)

\smallskip{}
\noindent In order to fold clause 6 we introduce the following definition:

\smallskip{}
\makebox[5mm][r]{8.}~\( \mathit{new}4(S,N)\leftarrow \mathit{in}_{-}\! \mathit{language}(S,b^{\wedge }\! N) \)
\smallskip{}

\noindent By unfolding clause 8, performing some replacements based
on laws of constraints, and folding, we get:

\smallskip{}
\makebox[5mm][r]{9.}~\( \mathit{new}4([\, ],0)\leftarrow  \)

\makebox[5mm][r]{10.}~\( \mathit{new}4([b|S],N)\leftarrow \mathit{N}\! \geq \! 1\wedge \mathit{new}4(S,N\! -\! 1) \)

\smallskip{}
\noindent By negative folding of clause 2 and positive folding of
clauses 4 and 6 we get the following program:

\smallskip{}
\makebox[5mm][r]{2f.}~\( \mathit{new}1(S)\leftarrow \mathit{string}(S)\wedge \neg \, \mathit{new}2(S) \)

\makebox[5mm][r]{4f.}~\( \mathit{new}2(S)\leftarrow \mathit{new}3(S,0) \)

\makebox[5mm][r]{6f.}~\( \mathit{new}3(S,N)\leftarrow \mathit{new}4(S,N) \)

\makebox[5mm][r]{7.}~\( \mathit{new}3([a|S],N)\leftarrow \mathit{new}3(S,N\! +\! 1) \)

\makebox[5mm][r]{9.}~\( \mathit{new}4([\, ],0)\leftarrow  \)

\makebox[5mm][r]{10.}~\( \mathit{new}4([b|S],N)\leftarrow \mathit{N}\! \geq \! 1\wedge \mathit{new}4(S,N\! -\! 1) \)

\smallskip{}
\noindent Now from clause 2f, by positive and negative unfoldings,
replacements based on laws of constraints, and folding, we get:

\smallskip{}
\makebox[5mm][r]{11.}~\( \mathit{new}1([a|S])\leftarrow \mathit{string}(S)\wedge \neg \, \mathit{new}3(S,1) \)

\makebox[5mm][r]{12.}~\( \mathit{new}1([b|S])\leftarrow \mathit{string}(S) \)

\smallskip{}
\noindent In order to fold clause 11 we introduce the following definition:

\smallskip{}
\makebox[5mm][r]{13.}~\( \mathit{new}5(S,N)\leftarrow \mathit{string}(S)\wedge \neg \, \mathit{new}3(S,N) \)
\smallskip{}

\noindent By positive and negative unfolding and folding we get:

\smallskip{}
\makebox[5mm][r]{14.}~\( \mathit{new}5([\, ],N)\leftarrow  \)

\makebox[5mm][r]{15.}~\( \mathit{new}5([a|S],N)\leftarrow \mathit{new}5(S,N\! +\! 1) \)

\makebox[5mm][r]{16.}~\( \mathit{new}5([a|S],N)\leftarrow \mathit{string}(S)\wedge \neg \, N\! \geq \! 1 \)

\makebox[5mm][r]{17.}~\( \mathit{new}5([b|S],N)\leftarrow \mathit{string}(S)\wedge \neg \, \mathit{new}4(S,N\! -\! 1) \)

\smallskip{}
\noindent In order to fold clause 17 we introduce the following definition:

\smallskip{}
\makebox[5mm][r]{18.}~\( \mathit{new}6(S,N)\leftarrow \mathit{string}(S)\wedge \neg \, \mathit{new}4(S,N) \)
\smallskip{}

\noindent Now, starting from clause 18, by positive and negative unfolding,
replacements based on laws of constraints, folding, and elimination
of the predicates on which \( new1 \) does not depend, we get the
following final, specialized program:

\medskip{}
\noindent \begin{tabular}{crl}
\( P_{\mathit{spec}} \) :~&
11f.~~&
\( new1([a|S])\leftarrow \mathit{new}5(S,1) \)\\
&
12.~~&
\( new1([b|S])\leftarrow \mathit{string}(S) \)\\
&
14.~~&
\( new5([\, ],N)\leftarrow  \)\\
&
15.~~&
\( new5([a|S],N)\leftarrow \mathit{new}5(S,N\! +\! 1) \)\\
&
16.~~&
\( new5([b|S],0)\leftarrow \mathit{string}(S) \)\\
&
17f.~~&
\( new5([b|S],N)\leftarrow \mathit{new}6(S,N\! -\! 1) \)\\
&
19.~~&
\( new6([\, ],N)\leftarrow N\not =0 \)\\
&
20.~~&
\( new6([a|S],N)\leftarrow \mathit{string}(S) \)\\
&
21.~~&
\( new6([b|S],0)\leftarrow \mathit{string}(S) \)\\
&
22.~~&
\( new6([b|S],N)\leftarrow \mathit{new}6(S,N\! -\! 1) \)\\
\end{tabular}

\medskip{}
\noindent This specialized program corresponds to a one-counter machine
(that is, a pushdown automaton where the stack alphabet contains one
letter only~\cite{Au&97}) and it takes \( O(n) \) time to test
that a string of length \( n \) does \emph{not} belong to the language
\( \{a^{m}\! \cdot \! b^{n}\, |\, m=n\geq 0\} \).

\section{Related Work and Conclusions \label{sec:related_work}}

During the last two decades various sets of unfold/fold transformation
rules have been proposed for different classes of logic programs.
The authors who first introduced the unfold/fold rules for logic programs
were Tamaki and Sato in their seminal paper~\cite{TaS84}. That paper
presents a set of rules for transforming definite logic programs and
it also presents the proof that those rules are correct w.r.t.~the
least Herbrand model semantics. Most of the subsequent papers in the
field have followed Tamaki and Sato's approach in that: (i)~the various
sets of rules which have been published can be seen as extensions
or variants of Tamaki and Sato's rules, and (ii)~the techniques used
for proving the correctness of the rules are similar to those used
by Tamaki and Sato (the reader may look at the references given later
in this section, and also at~\cite{PeP94} for a survey). In the
present paper we ourselves have followed Tamaki and Sato's approach,
but we have considered the more complex framework of locally stratified
constraint logic programs with the perfect model semantics. 

Among the rules we have presented, the following ones were initially
introduced in~\cite{TaS84} (in the case of definite logic programs):
(R1) definition introduction, restricted to one clause only (that
is, with \( m\! =\! 1 \)), (R3) positive unfolding, (R5) positive
folding, restricted to one clause only (that is, with \( m\! =\! 1 \)).
Our rules of replacement, deletion of useless predicates, constraint
addition, and constraint deletion (that is, rules R7, R8, R9, and
R10, respectively) are extensions to the case of constraint logic
programs with negation of the \emph{goal replacement} and \emph{clause
addition/deletion} rules presented in~\cite{TaS84}. In comparing
the rules in~\cite{TaS84} and the corresponding rules we have proposed,
let us highlight also the following important difference. The goal
replacement and clause addition/deletion of~\cite{TaS84} are very
general, but their applicability conditions are based on properties
of the least Herbrand model and properties of the proof trees (such
as goal equivalence or clause implication) which, in general, are
very difficult to prove. On the contrary, (i)~the applicability conditions
of our replacement rule require the verification of (usually decidable)
properties of the constraints, (ii)~the property of being a useless
predicate is decidable, because it refers to predicate symbols only
(and not to the value of their arguments), and (iii)~the applicability
conditions for constraint addition and constraint deletion can be
verified in most cases by program analysis techniques based on abstract
interpretation~\cite{Ga&96}. 

For the correctness theorem (see Theorem~\ref{th:corr_of_rules})
relative to admissible transformation sequences we have followed Tamaki
and Sato's approach, and as in~\cite{TaS84}, the correctness is
ensured by assuming the validity of some suitable conditions on the
construction of the transformation sequences.

Let us now relate our work here to that of other authors who have
extended in several ways the work by Tamaki and Sato and, in particular,
those who have extended it to the cases of: (i)~general logic programs,
and (ii)~constraint logic programs.

Tamaki and Sato's unfolding and folding rules have been extended to
general logic programs (without constraints) by Seki. He proved his
extended rules correct w.r.t.~various semantics, including the perfect
model semantics \cite{Sek91,Sek93}. Building upon previous work for
definite logic programs reported in \cite{GeK94,KaF86,Ro&99a}, paper
\cite{Ro&02} extended Seki's folding rule by allowing: (i)~\emph{multiple}
folding, that is, one can fold \( m \) (\( \geq 1 \)) clauses at
a time using a definition consisting of \( m \) clauses, and (ii)~\emph{recursive}
folding, that is, the definition used for folding can contain recursive
clauses. 

Multiple folding can be performed by applying our rule R5, but recursive
folding \emph{cannot}. Indeed, by rule R5 we can fold using a definition
introduced by rule R1, and this rule does \emph{not} allow the introduction
of recursive clauses. Thus, in this respect the folding rule presented
in this paper is less powerful than the folding rule considered in
\cite{Ro&02}. On the other hand, the set of rules presented here
is more powerful than the one in~\cite{Ro&02} because it includes
negative unfolding (R4) and negative folding (R6). These two rules
are very useful in practice, and both are needed for the program derivation
examples we have given in Section~\ref{sec:examples}. They are also
needed in the many examples of program verification presented in~\cite{Fi&01a}.
For reasons of simplicity, we have presented our non-recursive version
of the positive folding rule because it has much simpler applicability
conditions. In particular, the notion of admissible transformation
sequence is much simpler for non-recursive folding. We leave for future
research the problem of studying the correctness of a set of transformation
rules which includes positive and negative unfolding, as well as recursive
positive folding and recursive negative folding.

Negative unfolding and negative folding were also considered in our
previous work \cite{PeP02a}. The present paper extends the transformation
rules presented in \cite{PeP02a} by adapting them to a logic language
with constraints. Moreover, in \cite{PeP02a} we did not present the
proof of correctness of the transformation rules and we only showed
some applications of our transformation rules to theorem proving and
program synthesis.

In \cite{Sat92} Sato proposed a set of transformation rules for \emph{first
order programs}, that is, for a logic language that extends general
logic programs by allowing arbitrary first order formulas in the bodies
of the clauses. However, the semantics considered in \cite{Sat92}
is based on a three valued logic with the three truth values \emph{true},
\emph{false}, and \emph{undefined} (corresponding to non terminating
computations). Thus, the results presented in \cite{Sat92} cannot
be directly compared with ours. In particular, for instance, the rule
for eliminating useless predicates (R8) does not preserve the three
valued semantics proposed in \cite{Sat92}, because this rule may
transform a program that does not terminate for a given query, into
a program that terminates for that query. Moreover, the conditions
for the applicability of the folding rule given in \cite{Sat92} are
based on the chosen three valued logic and cannot be compared with
those presented in this paper.

Various other sets of transformation rules for general logic programs
(including several variants of the goal replacement rule) have been
proved correct w.r.t.~other semantics, such as, the operational semantics
based on SLDNF resolution~\cite{GaS91,Sek91}, Clark's completion~\cite{GaS91},
and Kunen's and Fitting's three valued extensions of Clark's completion~\cite{Bo&92a}.
We will not enter into a detailed comparison with these works here.
It will suffice to say that these works are not directly comparable
with ours because of the different set of rules (in particular, none
of these works considers the negative unfolding rule) and the different
semantics considered.

The unfold/fold transformation rules have also been extended to constraint
logic programs in~\cite{BeG98,EtG96,Fio02,Mah93}. Papers~\cite{BeG98,EtG96}
deal with definite programs, while \cite{Mah93} considers locally
stratified programs and proves that, with suitable restrictions, the
unfolding and folding rules preserve the perfect model semantics.
Our correctness result presented here extends that in~\cite{Mah93}
because: (i)~the rules of~\cite{Mah93} include neither negative
unfolding nor negative folding, and (ii)~the folding rule of~\cite{Mah93}
is \emph{reversible}, that is, it can only be applied for folding
a set of clauses in a program \( P \) by using a set of clauses that
occur in \( P \). As already mentioned in Section~\ref{sec:rules},
our folding rule is not reversible, because we may fold clauses in
program \( P_{k} \) of a transformation sequence by using definitions
occurring in \( \mathit{Defs}_{k} \), but possibly not in \( P_{k} \).
Reversibility is a very strong limitation, because it does not allow
the derivation of recursive clauses from non-recursive clauses. In
particular, the derivations presented in our examples of Section~\ref{sec:examples}
could not be performed by using the reversible folding rule of \cite{Mah93}.

Finally, \cite{Fio02} proposes a set of transformation rules for
locally stratified constraint logic programs tailored to a specific
task, namely, program specialization and its application to the verification
of infinite state reactive systems. Due to their specific application,
the transformation rules of \cite{Fio02} are much more restricted
than the ones presented here. In particular, by using the rules of~\cite{Fio02}:
(i)~we can only introduce \emph{constrained atomic definitions},
that is, definitions that consist of single clauses whose body is
a constrained atom, (ii)~we can unfold clauses w.r.t.~a negated
atom only if that atom succeeds or fails in one step, and (iii)~we
can apply the positive and negative folding rules by using constrained
atomic definitions only.

We envisage several lines for further development of the work presented
in this paper. As a first step forward, one could design strategies
for automating the application of the transformation rules proposed
here. In our examples of Section~\ref{sec:examples} we have demonstrated
that some strategies already considered in the literature for the
case of definite programs, can be extended to general constraint logic
programs. This extension can be done, in particular, for the following
strategies: (i)~the elimination of local variables~\cite{PrP95a},
(ii)~the derivation of deterministic programs~\cite{Pe&97a}, and
(iii)~the rule-based program specialization~\cite{LeB02}. 

It has been pointed out by recent studies that there is a strict relationship
between program transformation and various other methodologies for
program development and software verification (see, for instance,
\cite{Fi&01a,FrO97b,LeM99,PeP99a,PeP00a,Ro&00}). Thus, strategies
for the automatic application of transformation rules can be exploited
in the design of automatic techniques in these related fields and,
in particular, in program synthesis and theorem proving. We believe
that transformation methodologies for logic and constraint languages
can form the basis of a very powerful framework for machine assisted
software development.

\section*{Acknowledgements}

We would like to thank Maurice Bruynooghe and Kung-Kiu Lau for inviting
us to contribute to this volume. We would like also to acknowledge
the very stimulating conversations we have had over the years with
the members of the LOPSTR community since the beginning of the series
of the LOPSTR workshops. Finally, we express our thanks to the anonymous
referees for their helpful comments and suggestions.

\section{Appendices}

\subsection{Appendix A\label{app:localstratification}}

In this Appendix~A we will use the fact that, given any two atoms
\( A \) and \( B \), and any valuation \( v \), if \( \sigma (v(A))\geq \sigma (v(B)) \)
then for every substitution \( \vartheta  \), \( \sigma (v(A\vartheta ))\geq \sigma (v(B\vartheta )) \).
The same holds with \( > \), instead of~\( \geq  \).

\medskip{}
\noindent \emph{Proof of Proposition}~\emph{\ref{prop:local_strat}}.
{[}Preservation of Local \emph{}Stratification{]}\emph{.} We will
prove that, for \( k=0,\ldots ,n \), \( P_{k} \) is locally stratified
w.r.t.~\( \sigma  \) by induction on \( k \). 

\noindent \emph{Base case} (\( k=0 \)). By hypothesis \( P_{0} \)
is locally stratified w.r.t.~\( \sigma  \). 

\noindent \emph{Induction step}. We assume that \( P_{k} \) is locally
stratified w.r.t.~\( \sigma  \) and we show that \( P_{k+1} \)
is locally stratified w.r.t.~\( \sigma  \). We proceed by cases
depending on the transformation rule which is applied to derive \( P_{k+1} \)
from \( P_{k} \).

\noindent \emph{Case} 1. Program \( P_{k+1} \) is derived by definition
introduction (rule R1). We have that \( P_{k+1}=P_{k}\cup \{\delta _{1},\ldots ,\delta _{m}\} \),
where \( P_{k} \) is locally stratified w.r.t.~\( \sigma  \) by
the inductive hypothesis and \( \{\delta _{1},\ldots ,\delta _{m}\} \)
is locally stratified w.r.t.~\( \sigma  \) by Condition~(iv) of
R1. Thus, \( P_{k+1} \) is locally stratified w.r.t.~\( \sigma  \). 
\smallskip{}

\noindent \emph{Case} 2. Program \( P_{k+1} \) is derived by definition
elimination (rule R2). Then \( P_{k+1} \) is locally stratified w.r.t.~\( \sigma  \)
because \( P_{k+1}\subseteq P_{k} \). 
\smallskip{}

\noindent \emph{Case} 3. Program \( P_{k+1} \) is derived by positive
unfolding (rule R3). We have that \( P_{k+1}=(P_{k}-\{\gamma \})\cup \{\eta _{1},\ldots ,\eta _{m}\} \),
where \( \gamma  \) is a clause in \( P_{k} \) of the form \( H\leftarrow c\wedge G_{L}\wedge A\wedge G_{R} \)
and clauses \( \eta _{1},\ldots ,\eta _{m} \) are derived by unfolding
\( \gamma  \) w.r.t.~\( A \). Since, by the induction hypothesis,
\( (P_{k}-\{\gamma \}) \) is locally stratified w.r.t.~\( \sigma  \),
it remains to show that, for every valuation \( v \), for \( i=1,\ldots ,m \),
clause \( v(\eta _{i}) \) is locally stratified w.r.t.~\( \sigma  \).
Take any valuation \( v \). For \( i=1,\ldots ,m \), there exists
a clause \( \gamma _{i} \) in a variant of \( P_{k} \) of the form
\( K_{i}\leftarrow c_{i}\wedge B_{i} \) such that \( \eta _{i} \)
is of the form \( H\leftarrow c\wedge A\! =\! K_{i}\wedge c_{i}\wedge G_{L}\wedge B_{i}\wedge G_{R} \).
By the inductive hypothesis, \( v(H\leftarrow c\wedge G_{L}\wedge A\wedge G_{R}) \)
and \( v(K_{i}\leftarrow c_{i}\wedge B_{i}) \) are locally stratified
w.r.t.~\( \sigma  \). We consider two cases: (a)~\( \mathcal{D}\models \neg v(c\wedge A\! =\! K_{i}\wedge c_{i}) \)
and (b)~\( \mathcal{D}\models v(c\wedge A\! =\! K_{i}\wedge c_{i}) \).
In Case~(a), \( v(\eta _{i}) \) is locally stratified w.r.t.~\( \sigma  \)
by definition. In Case~(b), we have that: (i)~\( \mathcal{D}\models v(c) \),
(ii)~\( \mathcal{D}\models v(A)\! =\! v(K_{i}) \), and (iii)~\( \mathcal{D}\models v(c_{i}) \).
Let us consider a literal \( v(L) \) occurring in the body of \( v(\eta _{i}) \).
If \( v(L) \) is an atom occurring positively in \( v(G_{L}\wedge G_{R}) \)
then \( \sigma (v(H))\! \geq \! \sigma (v(L)) \) because \( v(H\leftarrow c\wedge G_{L}\wedge A\wedge G_{R}) \)
is locally stratified w.r.t.~\( \sigma  \) and \( \mathcal{D}\models v(c) \).
Similarly, if \( v(L) \) is a negated atom occurring in \( v(G_{L}\wedge G_{R}) \)
then \( \sigma (v(H))\! >\! \sigma (\overline{v({L})}) \). If \( v(L) \)
is an atom occurring positively in \( v(B_{i}) \) then \( \sigma (v(H))\! \geq \! \sigma (v(L)) \).
Indeed:

\smallskip{}
\begin{tabular}{cll}
\( \sigma (v(H)) \)&
\( \! \geq \! \sigma (v(A)) \) &
(because \( v(H\leftarrow c\wedge G_{L}\wedge A\wedge G_{R}) \) is
locally stratified\\
&
&
~w.r.t.~\( \sigma  \) and \( \mathcal{D}\models v(c) \))\\
&
\( \! =\! \sigma (v(K_{i})) \)&
(because \( v(A)\! =\! v(K_{i}) \))\\
&
\( \! \geq \! \sigma (v(L)) \)&
(because \( v(K_{i}\leftarrow c_{i}\wedge B_{i}) \) is locally stratified
w.r.t.~\( \sigma  \) \\
&
&
~and \( \mathcal{D}\models v(c_{i}) \))\\
\end{tabular}

\smallskip{}
\noindent Similarly, if \( v(L) \) is a negated atom occurring in
\( v(B) \) then \( \sigma (v(H))\! >\! \sigma (\overline{v({L})}) \).
Thus, the clause \( v(\eta _{i}) \) is locally stratified w.r.t.~\( \sigma  \).

\smallskip{}
\noindent \emph{Case} 4. Program \( P_{k+1} \) is derived by negative
unfolding (rule R4). As in Case~3, we have that \( P_{k+1}=(P_{k}-\{\gamma \})\cup \{\eta _{1},\ldots ,\eta _{s}\} \),
where \( \gamma  \) is a clause in \( P_{k} \) of the form \( H\leftarrow c\wedge G_{L}\wedge \neg A\wedge G_{R} \)
and clauses \( \eta _{1},\ldots ,\eta _{s} \) are derived by negative
unfolding \( \gamma  \) w.r.t.~\( \neg A \). Since, by the induction
hypothesis, \( (P_{k}-\{\gamma \}) \) is locally stratified w.r.t.~\( \sigma  \),
it remains to show that, for every valuation \( v \), for \( j=1,\ldots ,s \),
clause \( v(\eta _{j}) \) is locally stratified w.r.t.~\( \sigma  \).
Take any valuation \( v \). Let \( K_{1}\leftarrow c_{1}\wedge B_{1},\ldots ,\, K_{m}\leftarrow c_{m}\wedge B_{m} \)
be the clauses in a variant of \( P_{k} \) such that, for \( i=1,\ldots ,m \),
\( \mathcal{D}\models \exists (c\wedge A\! =\! K_{i}\wedge c_{i}) \).
Then, we have that, for \( j=1,\ldots ,s \), the clause \( v(\eta _{j}) \)
is of the form \( v(H\leftarrow c\wedge e_{j}\wedge G_{L}\wedge Q_{j}\wedge G_{R}) \),
where \( v(Q_{j}) \) is a conjunction of literals. By the applicability
conditions of the negative unfolding rule and by construction (see
Steps 1--4 of R4), we have that there exist \( m \) substitutions
\( \vartheta _{1},\ldots ,\vartheta _{m} \) such that the following
two properties hold:

\smallskip{}
\noindent (P.1) for every literal \( v(L) \) occurring in \( v(Q_{j}) \)
there exists a (positive or negative) literal \( v(M) \) occurring
in \( v(B_{i}\vartheta _{i}) \) for some \( i\in \{1,\ldots ,m\} \),
such that \( v(L) \) is \( \overline{v(M)} \), and 

\smallskip{}
\noindent (P.2) if \( v(L) \) occurs in \( v(Q_{j}) \) and \( v(L) \)
is \( \overline{v(M)} \) with \( v(M) \) occurring in \( v(B_{i}\vartheta _{i}) \)
for some \( i\in \{1,\ldots ,m\} \), then \( \mathcal{D}\models v((c\wedge e_{j})\rightarrow (A\! =\! K_{i}\vartheta _{i}\wedge c_{i}\vartheta _{i})) \). 

\smallskip{}
\noindent We will show that \( v(\eta _{j}) \) is locally stratified
w.r.t.~\( \sigma  \). By the inductive hypothesis, we have that
\( v(H\leftarrow c\wedge G_{L}\wedge \neg A\wedge G_{R}) \) and \( v(K_{i}\vartheta _{i}\leftarrow c_{i}\vartheta _{i}\wedge B_{i}\vartheta _{i}) \)
are locally stratified w.r.t.~\( \sigma  \). 

\noindent We consider two cases: (a)~\( \mathcal{D}\models \neg v(c\wedge e_{j}) \)
and (b)~\( \mathcal{D}\models v(c\wedge e_{j}) \). In Case~(a),
\( v(\eta _{j}) \) is locally stratified w.r.t.~\( \sigma  \) by
definition. In Case~(b), take any literal \( v(L) \) occurring in
\( v(Q_{j}) \). By Properties~(P.1) and (P.2), \( v(L) \) is \( \overline{v(M)} \)
for some \( v(M) \) occurring in \( v(B_{i}) \). We also have that:
(i)~\( \mathcal{D}\models v(A)\! =\! v(K_{i}\vartheta _{i}) \) and
(ii)~\( \mathcal{D}\models v(c_{i}\vartheta _{i}) \). Moreover \( \mathcal{D}\models v(c) \),
because we are in Case~(b). Now, if \( v(M) \) is a positive literal
occurring in \( v(B_{i}) \) we have:

\smallskip{}
\noindent \begin{tabular}{cll}
\( \sigma (v(H)) \)&
\( \! >\! \sigma (v(A)) \) &
(because \( v(H\leftarrow c\wedge G_{L}\wedge \neg A\wedge G_{R}) \)
is locally stratified\\
&
&
~w.r.t.~\( \sigma  \) and \( \mathcal{D}\models v(c) \))\\
&
\( \! =\! \sigma (v(K_{i}\vartheta _{i})) \)&
(because \( v(A)\! =\! v(K_{i}\vartheta _{i}) \))\\
\( (\dagger ) \)&
\( \! \geq \! \sigma (v(M)) \)&
(because \( v(K_{i}\vartheta _{i}\leftarrow c_{i}\vartheta _{i}\wedge B_{i}\vartheta _{i}) \)
is locally stratified\\
&
&
~w.r.t.~\( \sigma  \) ~and \( \mathcal{D}\models v(c_{i}\vartheta _{i}) \)).\\
\end{tabular}

\smallskip{}
\noindent Thus, we get: \( \sigma (v(H))>\sigma (v(M)) \), and we
conclude that \( v(\eta _{j}) \) is locally stratified w.r.t.~\( \sigma  \).
Similarly, if \( v(M) \) is a negative literal occurring in \( v(B_{i}\vartheta _{i}) \),
we also get: \( \sigma (v(H))>\sigma (\overline{{v(M)}}) \). (In
particular, if \( v(M) \) is a negative literal, at Point \( (\dagger ) \)
above, we have \( \sigma (v(K_{i}\vartheta _{i}))>\sigma (\overline{v(M)}) \).)
Thus, we also conclude that \( v(\eta _{j}) \) is locally stratified
w.r.t.~\( \sigma  \).

\smallskip{}
\noindent \emph{Case} 5. Program \( P_{k+1} \) is derived by positive
folding (rule R5). For reasons of simplicity, we assume that we fold
one clause only, that is, \( m=1 \) in rule R5. The general case
where \( m\geq 1 \) is analogous. We have that \( P_{k+1}=(P_{k}-\{\gamma \})\cup \{\eta \} \),
where \( \eta  \) is a clause of the form \( H\leftarrow c\wedge G_{L}\wedge K\vartheta \wedge G_{R} \)
derived by positive folding of clause \( \gamma  \) of the form \( H\leftarrow c\wedge d\vartheta \wedge G_{L}\wedge B\vartheta \wedge G_{R} \)
using a clause \( \delta  \) of the form \( K\leftarrow d\wedge B \)
introduced by rule R1. We have to show that, for every valuation \( v \),
\( v(H\leftarrow c\wedge G_{L}\wedge K\vartheta \wedge G_{R}) \)
is locally stratified w.r.t.~\( \sigma  \). By the inductive hypothesis,
we have that: (i)~for every valuation \( v \), \( v(\gamma ) \)
is locally stratified w.r.t.~\( \sigma  \), and (ii)~for every
valuation \( v \), \( v(\delta ) \) is locally stratified w.r.t.~\( \sigma  \).
Take any valuation \( v \). There are two cases: (a)~\( \mathcal{D}\models \neg v(c) \)
and (b)~\( \mathcal{D}\models v(c) \). In Case~(a), \( v(\eta ) \)
is locally stratified w.r.t.~\( \sigma  \) by definition. In Case~(b),
take any literal \( v(L) \) occurring in \( v(B\vartheta ) \). Now,
\emph{either} (b1)~\( v(L) \) is a positive literal, \emph{or} (b2)~\( v(L) \)
is a negative literal. In Case~(b1) there are two subcases: (b1.1)~\( \mathcal{D}\models \neg v(d\vartheta ) \),
and (b1.2)~\( \mathcal{D}\models v(d\vartheta ) \). In Case~(b1.1)
by Condition (iv) of rule R1, \( \sigma (v(K\vartheta ))=0 \) and
thus, \( \sigma (v(H))\geq \sigma (v(K\vartheta )) \). Hence, \( v(\eta ) \)
is locally stratified w.r.t.~\( \sigma  \). In Case~(b1.2), we
have that \( \mathcal{D}\models v(c\wedge d\vartheta ) \) and, by
the inductive hypothesis, \( \sigma (v(H))\geq \sigma (v(L\vartheta )) \).
Thus, \( \sigma (v(H))\geq \sigma (v(K\vartheta )) \), because by
Condition~(iv) of rule~R1, \( \sigma (v(K\vartheta )) \) is the
smallest ordinal \( \alpha  \) such that \( \alpha \geq \sigma (v(L\vartheta )) \).
Thus, \( v(\eta ) \) is locally stratified w.r.t.~\( \sigma  \).

\noindent Case~(b2), when \( v(L) \) is a negative literal occurring
in \( v(B\vartheta ) \), has a proof similar to the one of Case~(b1),
except that \( \sigma (v(H))>\sigma (\overline{v(L\vartheta )}) \),
instead of \( \sigma (v(H))\geq \sigma (v(L\vartheta )) \). 

\smallskip{}
\noindent \emph{Case} 6. Program \( P_{k+1} \) is derived by negative
folding (rule R6). We have that \( P_{k+1}=(P_{k}-\{\gamma \})\cup \{\eta \} \),
where \( \eta  \) is a clause of the form \( H\leftarrow c\wedge d\vartheta \wedge G_{L}\wedge \neg K\vartheta \wedge G_{R} \)
derived by negative folding of clause \( \gamma  \) of the form \( H\leftarrow c\wedge d\vartheta \wedge G_{L}\wedge \neg A\vartheta \wedge G_{R} \)
using a clause \( \delta  \) of the form \( K\leftarrow d\wedge A \)
introduced by rule R1. We have to show that, for every valuation \( v \),
\( v(\eta ) \) is locally stratified w.r.t.~\( \sigma  \). By the
inductive hypothesis, we have that: (i)~for every valuation \( v \),
\( v(H\leftarrow c\wedge d\vartheta \wedge G_{L}\wedge \neg A\vartheta \wedge G_{R}) \)
is locally stratified w.r.t.~\( \sigma  \), and (ii)~for every
valuation \( v \), \( v(K\leftarrow d\wedge A) \) is locally stratified
w.r.t.~\( \sigma  \). Take any valuation \( v \). There are two
cases: (a)~\( \mathcal{D}\models \neg v(c\wedge d\vartheta ) \),
and (b)~\( \mathcal{D}\models v(c\wedge d\vartheta ) \). In Case~(a),
\( v(\eta ) \) is locally stratified w.r.t.~\( \sigma  \) by definition.
In Case~(b), by the inductive hypothesis, we have only to show that
\( \sigma (v(H))>\sigma (v(K\vartheta )) \). Since \( \mathcal{D}\models v(c\wedge d\vartheta ) \),
by the inductive hypothesis we have that \( \sigma (v(H))>\sigma (v(A\vartheta )) \).
By Condition~(iv) of the rule R1, we have that \( \sigma (v(H))>\sigma (v(K\vartheta )) \).
Hence, \( v(\eta ) \) is locally stratified w.r.t.~\( \sigma  \).~
\smallskip{}

\noindent \emph{Case} 7. Program \( P_{k+1} \) is derived by replacement
(rule R7). We have that \( P_{k+1}=(P_{k}-\Gamma _{1})\cup \Gamma _{2} \),
where \( (P_{k}-\Gamma _{1}) \) is locally stratified w.r.t.~\( \sigma  \)
by the inductive hypothesis and \( \Gamma _{2} \) is locally stratified
w.r.t.~\( \sigma  \) by the applicability conditions of rule R7.
Thus, \( P_{k+1} \) is locally stratified w.r.t.~\( \sigma  \).

\smallskip{}
\noindent \emph{Case} 8. Program \( P_{k+1} \) is derived by deletion
of useless clauses (rule R8). \( P_{k+1} \) is locally stratified
w.r.t.~\( \sigma  \) by the inductive hypothesis because \( P_{k+1}\subseteq P_{k} \). 

\smallskip{}
\noindent \emph{Case} 9. Program \( P_{k+1} \) is derived by constraint
addition (rule R9). We have that \( P_{k+1}=(P_{k}-\{\gamma _{1}\})\cup \{\gamma _{2}\} \),
where \( \gamma _{2}:\, H\leftarrow c\wedge d\wedge G \) is the clause
in \( P_{k+1} \) derived by constraint addition from the clause \( \gamma _{1}:\, H\leftarrow c\wedge G \)
in \( P_{k} \). For every valuation \( v \), \( v(H\leftarrow c\wedge d\wedge G) \)
is locally stratified w.r.t.~\( \sigma  \) because: (i)~by the
induction hypothesis \( v(H\leftarrow c\wedge G) \) is locally stratified
w.r.t.~\( \sigma  \) and (ii)~if \( \mathcal{D}\models v(c\wedge d) \)
then \( \mathcal{D}\models v(c) \). Since, by the inductive hypothesis,
\( (P_{k}-\{\gamma _{1}\}) \) is locally stratified w.r.t.~\( \sigma  \),
also \( P_{k+1} \) is locally stratified w.r.t.~\( \sigma  \).

\smallskip{}
\noindent \emph{Case} 10. Program \( P_{k+1} \) is derived by constraint
deletion (rule R10). We have that \( P_{k+1}=(P_{k}-\{\gamma _{1}\})\cup \{\gamma _{2}\} \),
where \( \gamma _{2} \): \( H\leftarrow c\wedge G \) is the clause
in \( P_{k+1} \) derived by constraint deletion from clause \( \gamma _{1} \):
\( H\leftarrow c\wedge d\wedge G \) in \( P_{k} \). By the applicability
conditions of R10, \( \gamma  \) is locally stratified w.r.t.~\( \sigma  \).
Since, by the inductive hypothesis, \( (P_{k}-\{\gamma _{1}\}) \)
is locally stratified w.r.t.~\( \sigma  \), also \( P_{k+1} \)
is locally stratified w.r.t.~\( \sigma  \).

\noindent Finally, \( P_{0}\cup \mathit{Defs}_{n} \) is locally stratified
w.r.t.~\( \sigma  \) by the hypothesis that \( P_{0} \) is locally
stratified w.r.t.~\( \sigma  \) and by Condition~(iv) of rule R1.\hfill{}\( \Box  \)

\subsection{Appendix B}

In the proofs of Appendices B and C we use the following notions.
Given a clause \( \gamma  \):~\( H\leftarrow c\wedge L_{1}\wedge \ldots \wedge L_{m} \)
and a valuation \( v \) such that \( \mathcal{D}\models v(c) \),
we denote by \( \gamma _{v} \) the clause \( v(H\leftarrow L_{1}\wedge \ldots \wedge L_{m}) \).
We define \( \mathit{ground}(\gamma )=\{\gamma _{v}\, |\, v \) is
a valuation and \( \mathcal{D}\models v(c)\} \). Given a set \( \Gamma  \)
of clauses, we define \( \mathit{ground}(\Gamma )=\bigcup _{\gamma \in \Gamma }\mathit{ground}(\gamma ) \).\\

\noindent \emph{Proof of Proposition~}\ref{prop:tc_of_pos_unf}.
Recall that \( P_{0},\ldots ,P_{i} \) is constructed by \( i\, (\geq 0) \)
applications of the definition rule, that is, \( P_{i}=P_{0}\cup \mathit{Defs}_{i} \),
and \( P_{i},\ldots ,P_{j} \) is constructed by applying once the
positive unfolding rule to each clause in \( \mathit{Defs}_{i} \).
Let \emph{\( \sigma  \)} be the fixed stratification function considered
at the beginning of the construction of the transformation sequence.
By Proposition~\ref{prop:local_strat}, each program in the sequence
\( P_{i},\ldots ,P_{j} \) is locally stratified w.r.t.~\( \sigma  \). 
\smallskip{}

Let us consider a ground atom \( A \). By complete induction on the
ordinal \( \sigma (A) \) we prove that, for \( k=i,\ldots ,j\! -\! 1 \),
there exists a proof tree for \( A \) and \( P_{k} \) iff there
exists a proof tree for \( A \) and \( P_{k+1} \). The inductive
hypothesis is: 
\smallskip{}

\noindent (I1) for every ground atom \( A' \), if \( \sigma (A')\! <\! \sigma (A) \)
then there exists a proof tree for \( A' \) and \( P_{k} \) iff
there exists a proof tree for \( A' \) and \( P_{k+1} \).
\medskip{}

\noindent (\emph{If} \emph{Part}) We consider a proof tree \( U \)
for \( A \) and \( P_{k+1} \), and we show that we can construct
a proof tree \( T \) for \( A \) and \( P_{k} \). We proceed by
complete induction on \( \mathit{size}(U) \). The inductive hypothesis
is:
\smallskip{}

\noindent (I2) given any proof tree \( U_{1} \) for a ground atom
\( A_{1} \) and \( P_{k+1} \), if \( \mathit{size}(U_{1})\! <\! \mathit{size}(U) \)
then there exists a proof tree \( T_{1} \) for \( A_{1} \) and \( P_{k} \). 
\smallskip{}

Let \( \gamma  \) be a clause of \( P_{k+1} \) and let \( \gamma _{v} \):
\( A\leftarrow L_{1}\wedge \ldots \wedge L_{r} \) be the clause in
\( \mathit{ground}(\gamma ) \) used at the root of \( U \). Thus,
\( L_{1},\ldots ,L_{r} \) are the children of \( A \) in \( U \).
For \( h=1,\ldots ,r \), if \( L_{h} \) is an atom then the subtree
\( U_{h} \) of \( U \) rooted at \( L_{h} \) is a proof tree for
\( L_{h} \) and \( P_{k+1} \). Since \( \mathit{size}(U_{h})\! <\! \mathit{size}(U) \),
by the inductive hypothesis (I2) there exists a proof tree \( T_{h} \)
for \( L_{h} \) and \( P_{k} \). For \( h=1,\ldots ,r \), if \( L_{h} \)
is a negated atom \( \neg A_{h} \) then, by the definition of proof
tree, there exists no proof tree for \( A_{h} \) and \( P_{k+1} \).
Since \( \sigma  \) is a local stratification for \( P_{k+1} \),
we have that \( \sigma (A_{h})\! <\! \sigma (A) \) and, by the inductive
hypothesis (I1) there exists no proof tree for \( A_{h} \) and \( P_{k} \).

Now, we proceed by cases.

\noindent \emph{Case} 1. \( \gamma \in P_{k} \). We construct \( T \)
as follows. The root of \( T \) is \( A \). We use \( \gamma _{v} \):
\( A\leftarrow L_{1}\wedge \ldots \wedge L_{r} \) to construct the
children of \( A \). If \( r=0 \) then \emph{true} is the only child
of \( A \) in \( T \), and \( T \) is a proof tree for \( A \)
and \( P_{k} \). Otherwise \( r\! \geq \! 1 \) and, for \( h=1,\ldots ,r \),
if \( L_{h} \) is an atom \( A_{h} \) then \( T_{h} \) is the subtree
of \( T \) at \( A_{h} \), and if \( L_{h} \) is a negated atom
then \( L_{h} \) is a leaf of \( T \). By construction we have that
\( T \) is a proof tree for \( A \) and \( P_{k} \).

\noindent \emph{Case} 2. \( \gamma \not \in P_{k} \) and \( \gamma \in P_{k+1} \)
because \( \gamma  \) is derived by positive unfolding. Thus, there
exist: a clause \( \alpha  \) in \( P_{k} \) of the form \( H\leftarrow c\wedge G_{L}\wedge A_{S}\wedge G_{R} \)
and a variant \( \beta  \) of a clause in \( P_{k} \) of the form
\( K\leftarrow d\wedge B \) such that clause \( \gamma  \) is of
the form \( H\leftarrow c\wedge A_{S}\! =\! K\wedge d\wedge G_{L}\wedge B\wedge G_{R} \).
Thus, (i)~\( v(H)=A \), (ii) \( \mathcal{D}\models v(c\wedge A_{S}\! =\! K\wedge d) \),
and (iii)~\( v(G_{L}\wedge B\wedge G_{R})=L_{1},\ldots ,L_{r} \).
By (ii) we have that \( \alpha _{v}\in \mathit{ground}(P_{k}) \)
and \( \beta _{v}\in \mathit{ground}(P_{k}) \). (Notice that, since
\( \beta  \) is a variant of a clause in \( P_{k} \), then \( \beta _{v}\in \mathit{ground}(P_{k}) \).) 

\noindent We construct \( T \) as follows. The root of \( T \) is
\( A \). We use \( \alpha _{v} \) to construct the children of \( A \)
and then we use \( \beta _{v} \) to construct the children of \( A_{S} \).
The leaves of the tree constructed in this way are \( L_{1},\ldots ,L_{r} \).
If \( r=0 \) then \emph{true} is the only leaf of \( T \), and \( T \)
is a proof tree for \( A \) and \( P_{k} \). Otherwise \( r\! \geq \! 1 \)
and, for \( h=1,\ldots ,r \), if \( L_{h} \) is an atom then \( T_{h} \)
is the subtree of \( T \) rooted at \( L_{h} \), and if \( L_{h} \)
is a negated atom then \( L_{h} \) is a leaf of \( T \). By construction
we have that \( T \) is a proof tree for \( A \) and \( P_{k} \).
\smallskip{}

\noindent (\emph{Only-if} \emph{Part}) We consider a proof tree \( T \)
for a ground atom \( A \) and program \( P_{k} \), for \( k=i,\ldots \, j\! -\! 1 \),
and we show that we can construct a proof tree \( U \) for \( A \)
and \( P_{k+1} \). We proceed by complete induction on \( \mathit{size}(T) \).
The inductive hypothesis is:
\smallskip{}

\noindent (I3) given any proof tree \( T_{1} \) for a ground atom
\( A_{1} \) and \( P_{k} \), if \( \mathit{size}(T_{1})\! <\! \mathit{size}(T) \)
then there exists a proof tree \( U_{1} \) for \( A_{1} \) and \( P_{k+1} \). 
\smallskip{}

Let \( \gamma  \) be a clause of \( P_{k} \) and let \( \gamma _{v} \):
\( A\leftarrow L_{1}\wedge \ldots \wedge L_{r} \) be the clause in
\( \mathit{ground}(\gamma ) \) used at the root of \( T \). Now
we proceed by cases.

\noindent \emph{Case} 1. \( \gamma \in P_{k+1} \). We construct the
proof tree \( U \) for \( A \) and \( P_{k+1} \) as follows. We
use \( \gamma _{v} \) to construct the children \( L_{1},\ldots ,L_{r} \)
of the root \( A \). If \( r=0 \) then \emph{true} is the only child
of \( A \) in \( U \), and \( U \) is a proof tree for \( A \)
and \( P_{k+1} \). Otherwise, \( r\! \geq \! 1 \) and, for \( h=1,\ldots ,r \),
if \( L_{h} \) is an atom, we consider the subtree \( T_{h} \) of
\( T \) rooted at \( L_{h} \). We have that \( T_{h} \) is a proof
tree for \( L_{h} \) and \( P_{k} \) with \( \mathit{size}(T_{h})\! <\! \mathit{size}(T) \)
and, therefore, by the inductive hypothesis (I3), there exists a proof
tree \( U_{h} \) for \( L_{h} \) and \( P_{k+1} \). For \( h=1,\ldots ,r \),
if \( L_{h} \) is a negated atom \( \neg A_{h} \), then \( \sigma (A)\! >\! \sigma (A_{h}) \)
because \( \sigma  \) is a stratification function for \( P_{k} \).
Thus, by the inductive hypothesis (I1) we have that there is no proof
tree for \( A_{h} \) and \( P_{k+1} \). The construction of \( U \)
continues as follows. For \( h=1,\ldots ,r \), if \( L_{h} \) is
an atom then we use \( U_{h} \) as a subtree of \( U \) rooted at
\( L_{h} \) and, if \( L_{h} \) is a negated atom, then \( L_{h} \)
is a leaf of \( U \). Thus, by construction we have that \( U \)
is a proof tree for \( A \) and \( P_{k+1} \).

\noindent \emph{Case} 2. \( \gamma \in P_{k} \) and \( \gamma \not \in P_{k+1} \)
because \( \gamma  \) has been unfolded w.r.t.~an atom in its body.
Let us assume that \( \gamma  \) is of the form \( H\leftarrow c\wedge G_{L}\wedge A_{S}\wedge G_{R} \)
and \( \gamma  \) has been unfolded w.r.t.~\( A_{S} \). We have
that: (i)~\( v(H)=A \), (ii) \( \mathcal{D}\models v(c) \), and
(iii)~the ground literals \( L_{1},\ldots ,L_{r} \) such that \( L_{1}\wedge \ldots \wedge L_{r}=v(G_{L}\wedge A_{S}\wedge G_{R}) \)
are the children of \( A \) in \( T \). Let \( \beta  \): \( K\leftarrow d\wedge B \)
be the clause in \( P_{k} \) which has been used for constructing
the children of \( v(A_{S}) \) in \( T \). Thus, there exists a
valuation \( v' \) such that: (iv)~\( v(A_{S})=v'(K) \), (v)~\( \mathcal{D}\models v'(d) \),
and (vi)~the literals in \( v'(B) \) are the children of \( v(A_{S}) \)
in \( T \). Without loss of generality we may assume that \( \gamma  \)
and \( \beta  \) have no variables in common and \( v=v' \). Thus,
the ground literals \( M_{1},\ldots ,M_{s} \) such that \( M_{1}\wedge \ldots \wedge M_{s}=v(G_{L}\wedge B\wedge G_{R}) \)
are descendants of \( A \) in \( T \). For \( h=1,\ldots ,s \),
if \( M_{h} \) is an atom, let us consider the subtree \( T_{h} \)
of \( T \) rooted at \( M_{h} \). We have that \( T_{h} \) is a
proof tree for \( M_{h} \) and \( P_{k} \) with \( \mathit{size}(T_{h})\! <\! \mathit{size}(T) \)
and, therefore, by the inductive hypothesis (I3), there exists a proof
tree \( U_{h} \) for \( M_{h} \) and \( P_{k+1} \). For \( h=1,\ldots ,s \),
if \( M_{h} \) is a negated atom \( \neg A_{h} \) then \( M_{h} \)
is a leaf of \( T \) and there exists no proof tree for \( A_{h} \)
and \( P_{k} \). Since \( \sigma  \) is a stratification function
for \( P_{k} \), we have that \( \sigma (A)\! >\! \sigma (A_{h}) \)
and thus, by the inductive hypothesis (I1), there exists no proof
tree for \( A_{h} \) and \( P_{k+1} \).

Now let us consider the clause \( \eta : \) \( H\leftarrow c\wedge A_{S}\! =\! K\wedge d\wedge G_{L}\wedge B\wedge G_{R} \).
\( \eta  \) is one of the clauses derived by unfolding \( \gamma  \)
because \( \beta \in P_{k} \) and, by (ii), (iv), (v) and the assumption
that \( v=v' \), we have that \( \mathcal{D}\models v(c\wedge A_{S}\! =\! K\wedge d) \)
and hence \( \mathcal{D}\models \exists (c\wedge A_{S}\! =\! K\wedge d) \).
Thus, we construct a proof tree \( U \) for \( A \) and \( P_{k+1} \)
as follows. Since \( A=v(H) \) and \( M_{1}\wedge \ldots \wedge M_{s}=v(G_{L}\wedge B\wedge G_{R}) \),
we can use \( \eta _{v} \): \( v(H\leftarrow G_{L}\wedge B\wedge G_{R}) \)
to construct the children \( M_{1},\ldots ,M_{s} \) of \( A \) in
\( U \). If \( s=0 \) then \emph{true} is the only child of \( A \)
in \( U \), and \( U \) is a proof tree for \( A \) and \( P_{k+1} \).
Otherwise, \( s\! \geq \! 1 \) and, for \( h=1,\ldots ,s \), if
\( M_{h} \) is an atom then \( U_{h} \) is the proof tree rooted
at \( M_{h} \) in \( U \). If \( M_{h} \) is a negated atom then
\( M_{h} \) is a leaf of \( U \). The proof tree \( U \) is the
proof tree for \( A \) and \( P_{k+1} \) to be constructed.\hfill{}\( \Box  \)

\subsection{Appendix C}

\noindent \emph{Proof of Proposition~}\ref{prop:preserv-mu-consistency}.
Recall that the transformation sequence \( P_{0},\ldots ,P_{i}, \)
\( \ldots ,P_{j}, \) \( \ldots ,P_{m} \) is constructed as follows
(see Definition~\ref{def:ordered}):

\noindent (1) the sequence \( P_{0},\ldots ,P_{i} \), with \( i\! \geq \! 0 \),
is constructed by applying \( i \) times the definition introduction
rule, that is, \( P_{i}=P_{0}\cup \mathit{Defs}_{i} \);

\noindent (2) the sequence \( P_{i},\ldots ,P_{j} \) is constructed
by applying once the positive unfolding rule to each clause in \( \mathit{Defs}_{i} \)
which is used for applications of the folding rule in \( P_{j},\ldots ,P_{m} \);

\noindent (3) the sequence \( P_{j},\ldots ,P_{m} \), with \( j\! \leq \! m \),
is constructed by applying any rule, except the definition introduction
and definition elimination rules.

\noindent Let \emph{\( \sigma  \)} be the fixed stratification function
considered at the beginning of the construction of the transformation
sequence. By Proposition~\ref{prop:local_strat}, each program in
the sequence \( P_{0}\cup \mathit{Defs}_{i},\ldots ,P_{j},\ldots ,P_{m} \)
is locally stratified w.r.t.~\( \sigma  \). 

We will prove by induction on \( k \) that, for \( k=j,\ldots ,m \), 

\noindent (\emph{Soundness}) if there exists a proof tree for a ground
atom \( A \) and \( P_{k} \) then there exists a proof tree for
\( A \) and \( P_{j} \), and

\noindent (\emph{Completeness}) if there exists a \( P_{j} \)-consistent
proof tree for a ground atom \( A \) and \( P_{j} \) then there
exists a \( P_{j} \)-consistent proof tree for \( A \) and \( P_{k} \).

\noindent The base \textbf{\emph{}}case \textbf{}(\textbf{\( k=j \)})
is trivial.

\noindent For proving the induction \textbf{\emph{}}step, consider
any \( k \) in \( \{j,\ldots ,m\! -\! 1\} \). We assume that the
soundness and completeness properties hold for that \( k \), and
we prove that they hold for \( k\! +\! 1 \). For the soundness property
it is enough to prove that:
\smallskip{}

\noindent - if there exists a proof tree for a ground atom \( A \)
and \( P_{k+1} \) then there exists a proof tree for \( A \) and
\( P_{k} \), 
\smallskip{}

\noindent and for the completeness property it is enough to prove
that:
\smallskip{}

\noindent - if there exists a \( P_{j} \)-consistent proof tree for
a ground atom \( A \) and \( P_{k} \) then there exists a \( P_{j} \)-consistent
proof tree for \( A \) and \( P_{k+1} \).
\smallskip{}

\noindent We proceed by complete induction on the ordinal \( \sigma (A) \)
associated with the ground atom \( A \). The inductive hypotheses
are: 
\smallskip{}

\sloppy

\noindent (IS) for every ground atom \( A' \) such that \( \sigma (A')\! <\! \sigma (A) \),
if there exists a proof tree for \( A' \) and \( P_{k+1} \) then
there exists a proof tree for \( A' \) and \( P_{k} \), and
\medskip{}

\noindent (IC) for every ground atom \( A' \) such that \( \sigma (A')\! <\! \sigma (A) \),
if there exists a \mbox{\( P_{j} \)-consistent} proof tree for \( A' \)
and \( P_{k} \) then there exists a \( P_{j} \)-consistent proof
tree for \( A' \) and \( P_{k+1} \).
\smallskip{}

By the inductive hypotheses on soundness and completeness for \( k \),
(IS), (IC), and Proposition~\ref{prop:mu-consistency}, we have that:
\smallskip{}

\noindent (ISC) for every ground atom \( A' \) such that \( \sigma (A')\! <\! \sigma (A) \),
there exists a proof tree for \( A' \) and \( P_{k} \) iff there
exists a proof tree for \( A' \) and \( P_{k+1} \).
\smallskip{}

Now we give the proofs for the soundness and the completeness properties.
\medskip{}

\noindent \emph{Proof of Soundness.} Given a proof tree \( U \) for
\( A \) and \( P_{k+1} \) we have to prove that there exists a proof
tree \( T \) for \( A \) and \( P_{k} \). The proof is by complete
induction on \emph{\( \mathit{size}(T) \)}. The inductive hypothesis
is:
\smallskip{}

\noindent (Isize) Given any proof tree \( U' \) for a ground atom
\( A' \) and \( P_{k+1} \), if \( \mathit{size}(U')<\mathit{size}(U) \)
then there exists a proof tree \( T' \) for \( A' \) and \( P_{k} \).
\smallskip{}

Let \( \gamma  \) be a clause in \( P_{k+1} \) and \( v \) be a
valuation. Let \( \gamma _{v}\in \mathit{ground}(\gamma ) \) be the
ground clause of the form \( A\leftarrow L_{1}\wedge \ldots \wedge L_{r} \)
used at the root of \( U \). We proceed by considering the following
cases: \emph{either} (Case 1)~\( \gamma  \) belongs to \( P_{k} \)
\emph{or} (Case 2)~\( \gamma  \) does not belong to \( P_{k} \)
and it has been derived from some clauses in \( P_{k} \) by applying
a transformation rule among R3, R4, R5, R6, R7, R9, R10. (Recall that
R1 and R2 are not applied in \( P_{j},\ldots ,P_{m} \), and by R8
we delete clauses.)
\smallskip{}

The proof of Case 1 and the proofs of Case 2 for rules R3, R4, R9,
and R10 are left to the reader. Now we present the proofs of Case~2
for rules R5, R6, and R7.
\smallskip{}

\noindent \emph{Case} 2, rule \emph{}R5. Clause \( \gamma  \) is
derived by positive folding. Let \( \gamma  \) be derived by folding
clauses \( \gamma _{1},\ldots ,\gamma _{m} \) in \( P_{k} \) using
clauses \( \delta _{1},\ldots ,\delta _{m} \) where, for \( i=1,\ldots ,m \),
clause \( \delta _{i} \) is of the form \( K\leftarrow d_{i}\wedge B_{i} \)
and clause \( \gamma _{i} \) is of the form \( H\leftarrow c\wedge d_{i}\vartheta \wedge G_{L}\wedge B_{i}\vartheta \wedge G_{R} \),
for a substitution \( \vartheta  \) satisfying Conditions (i) and
(ii) given in (R5). Thus, \( \gamma  \) is of the form: \( H\leftarrow c\wedge G_{L}\wedge K\vartheta \wedge G_{R} \)
and we have that: (a)~\( v(H)=A \), (b)~\( \mathcal{D}\models v(c) \),
and (c)~\( v(G_{L}\wedge K\vartheta \wedge G_{R})=L_{1}\wedge \ldots \wedge L_{r} \).
Since program \( P_{k+1} \) is locally stratified w.r.t.~\( \sigma  \),
by the inductive hypotheses (ISC) and (Isize) we have that: for \( h=1,\ldots ,r \),
if \( L_{h} \) is an atom then there exists a proof tree \( T_{h} \)
for \( L_{h} \) and \( P_{k} \), and if \( L_{h} \) is a negated
atom \( \neg A_{h} \) then there is no proof tree for \( A_{h} \)
and \( P_{k} \). The atom \( v(K\vartheta ) \) is one of the literals
\( L_{1},\ldots ,L_{r} \), say \( L_{f} \), and thus, there exists
a proof tree for \( v(K\vartheta ) \) and \( P_{k} \). By the inductive
hypothesis (Soundness) for \( P_{k} \) and Proposition~\ref{prop:tc_of_pos_unf},
there exists a proof tree for \( v(K\vartheta ) \) and \( P_{i} \).
Since \( P_{i}=P_{0}\cup \mathit{Defs}_{n} \) and \( \delta _{1},\ldots ,\delta _{m} \)
are all clauses in (a variant of) \( P_{0}\cup \mathit{Defs}_{n} \)
which have the same predicate symbol as \( K \), there exists \( \delta _{p}\in \delta _{1},\ldots ,\delta _{m} \)
such that \( \delta _{p} \) is of the form \( K\leftarrow d_{p}\wedge B_{p} \)
and \( \delta _{p} \) is used to construct the children of \( v(K\vartheta ) \)
in the proof tree for \( v(K\vartheta ) \) and \( P_{i} \). By Conditions
(i) and (ii) on \( \vartheta  \) given in (R5), we have that: (d)~\( \mathcal{D}\models v(d_{p}\vartheta ) \)
and (e)~\( v(B_{p}\vartheta )=M_{1}\wedge \ldots \wedge M_{s} \).
By the definition of proof tree, for \( h=1,\ldots ,s \), if \( M_{h} \)
is an atom then there exists a proof tree for \( M_{h} \) and \( P_{i} \),
else if \( M_{h} \) is a negated atom \( \neg E_{h} \) then there
is no proof tree for \( E_{h} \) and \( P_{i} \). By Propositions~\ref{prop:tc_of_pos_unf}
and \ref{prop:mu-consistency} and the inductive hypotheses (Soundness
and Completeness) we have that, for \( h=1,\ldots ,s \), if \( M_{h} \)
is an atom then there exists a proof tree \( \widehat{T_{h}} \) for
\( M_{h} \) and \( P_{k} \), else if \( M_{h} \) is a negated atom
\( \neg E_{h} \) then there is no proof tree for \( E_{h} \) and
\( P_{k} \). 

Now we construct the proof tree \( T \) for \( A \) and \( P_{k} \)
as follows. By (a), (b), and (d), we have that \( v(H)=A \) and \( \mathcal{D}\models v(c\wedge d_{p}\vartheta ) \).
Thus, we construct the children of \( A \) in \( T \) by using the
clause \( \gamma _{p} \): \( H\leftarrow c\wedge d_{p}\vartheta \wedge G_{L}\wedge B_{p}\vartheta \wedge G_{R} \).
Since \( v(G_{L}\wedge B_{p}\vartheta \wedge G_{R})=L_{1}\wedge \ldots \wedge L_{f-1}\wedge M_{1}\wedge \ldots \wedge M_{s}\wedge L_{f+1}\wedge \ldots \wedge L_{r} \),
the children of \( A \) in \( T \) are: \( L_{1},\ldots ,L_{f-1},M_{1},\ldots ,M_{s},L_{f+1},\ldots ,L_{r} \).
By the applicability conditions of the positive folding rule, we have
that \( s>0 \) and \( A \) has a child different from the empty
conjunction \emph{true}. The children of \( A \) are constructed
as follows. For \( h=1,\ldots ,r \), if \( L_{h} \) is an atom then
\( T_{h} \) is the subtree of \( T \) rooted in \( L_{h} \), else
if \( L_{h} \) is a negated atom then \( L_{h} \) is a leaf of \( T \).
For \( h=1,\ldots ,s \), if \( M_{h} \) is an atom then \( \widehat{T_{h}} \)
is the subtree of \( T \) rooted in \( M_{h} \), else if \( M_{h} \)
is a negated atom then \( M_{h} \) is a leaf of \( T \). 
\smallskip{}

\noindent \emph{Case} 2, rule R6. Clause \( \gamma  \) is derived
by negative folding. Let \( \gamma  \) be derived by folding a clause
\( \alpha  \) in \( P_{k} \) of the form \( H\leftarrow c\wedge G_{L}\wedge \neg A_{F}\vartheta \wedge G_{R} \)
by using a clause \( \delta \in \mathit{Defs}_{i} \) of the form
\( K\leftarrow d\wedge A_{F} \). Thus, \( \gamma  \) is of the form
\( H\leftarrow c\wedge G_{L}\wedge \neg K\vartheta \wedge G_{R} \). 

Let \( \gamma _{v} \) be of the form \( A\leftarrow L_{1}\wedge \ldots \wedge L_{f-1}\wedge \neg v(K\vartheta )\wedge L_{f+1}\wedge \ldots \wedge L_{r} \),
that is, \( v(H)=A \) and \( \mathcal{D}\models v(c) \). By the
conditions on the applicability of rule R6, we also have that \( \mathcal{D}\models v(d\vartheta ) \).
Since program \( P_{k+1} \) is locally stratified w.r.t.~\( \sigma  \),
we have that \( \sigma (v(K\vartheta ))<\sigma (A) \). By the definition
of proof tree, there is no proof tree for \( v(K\vartheta ) \) and
\( P_{k+1} \). Thus, by hypothesis (ISC) there exists no proof tree
for \( v(K\vartheta ) \) and \( P_{k} \). By the inductive hypothesis
(Completeness) and Propositions~\ref{prop:tc_of_pos_unf} and \ref{prop:mu-consistency},
there exists no proof tree for \( v(K\vartheta ) \) and \( P_{0}\cup \mathit{Defs}_{i} \)
and thus, since \( K\leftarrow d\wedge A_{F} \) is the only clause
defining the head predicate of \( K \) and \( \mathcal{D}\models v(d\vartheta ) \),
there is no proof tree for \( v(A_{F}\vartheta ) \) and \( P_{0}\cup \mathit{Defs}_{i} \).
By Proposition~\ref{prop:tc_of_pos_unf} and the inductive hypothesis
(Soundness), there exists no proof tree for \( v(A_{F}\vartheta ) \)
and \( P_{k} \). Since \( \mathcal{D}\models v(c) \) there exists
a clause \( \alpha _{v} \) in \( \mathit{ground}(\alpha ) \) of
the form \( A\leftarrow L_{1}\wedge \ldots \wedge L_{f-1}\wedge \neg v(A_{F}\vartheta )\wedge L_{f+1}\wedge \ldots \wedge L_{r} \).
We begin the construction of \( T \) by using \( \alpha _{v} \)
at the root. For all \( h=1,\ldots ,f-1,f+1,\ldots ,r \) such that
\( L_{h} \) is an atom and \( U_{h} \) is the subtree of \( U \)
rooted in \( L_{h} \), we have that \( \mathit{size}(U_{h})<\mathit{size}(U) \).
By hypothesis (Isize) there exists a proof tree \( T_{h} \) for \( L_{h} \)
and \( P_{k} \) which we use as a subtree of \( T \) rooted in \( L_{h} \).
For all \( h=1,\ldots ,f-1,f+1,\ldots ,r \) such that \( L_{h} \)
is a negated atom \( \neg A_{h} \) we have that \( \sigma (A_{h})<\sigma (A) \),
because program \( P_{k+1} \) is locally stratified w.r.t.~\( \sigma  \).
Moreover, there is no proof tree for \( A_{h} \) in \( P_{k+1} \),
because \( U \) is a proof tree. By hypothesis (ISC) we have that
there is no proof tree for \( A_{h} \) in \( P_{k} \). Thus, for
all \( h=1,\ldots ,f-1,f+1,\ldots ,r \) such that \( L_{h} \) is
a negated atom we take \( L_{h} \) to be a leaf of \( T \).
\smallskip{}

\noindent \emph{Case}2, rule R7. Clause \( \gamma  \) is derived
by replacement. We only consider the case where \( P_{k+1} \) is
derived from program \( P_{k} \) by applying the replacement rule
based on law (8). The other cases are left to the reader. Suppose
that a clause \( \eta  \): \( H\leftarrow c_{1}\wedge G \) in \( P_{k} \)
is replaced by clause \( \gamma  \): \( H\leftarrow c_{2}\wedge G \)
and \( \mathcal{D}\models \forall \, (\exists Y\, c_{1}\leftrightarrow \exists Z\, c_{2}) \),
where: (i) \( Y=\mathit{FV}(c_{1})\! -\! FV(\{H,G\}) \) and (ii)
\( Z=\mathit{FV}(c_{2})\! -\! FV(\{H,G\}) \). Thus, \( \mathit{ground}(\gamma )=\mathit{ground}(\eta ) \)
and we can construct a proof tree for the ground atom \( A \) and
\( P_{k} \) by using a clause in \( \mathit{ground}(\eta ) \), instead
of a clause in \( \mathit{ground}(\gamma ) \). 
\medskip{}

\noindent \emph{Proof of Completeness.} Given a \( P_{j} \)-consistent
proof tree for \( A \) and \( P_{k} \), we prove that there exists
a \( P_{j} \)-consistent proof tree for \( A \) and \( P_{k+1} \).
The proof is by well-founded induction on \emph{\( \mu (A,P_{j}) \)}.
The inductive hypothesis is:
\smallskip{}

\noindent (I\( \mu  \)) for every ground atom \( A' \) such that
\( \mu (A',P_{j})<\mu (A,P_{j}) \), if there exists a \( P_{j} \)-consistent
proof tree \( T' \) for \( A' \) and \( P_{k} \) then there exists
a \( P_{j} \)-consistent proof tree \( U' \) for \( A' \) and \( P_{k+1} \).
\smallskip{}

Let \( \gamma  \) be a clause in \( P_{k} \) and \( v \) be a valuation
such that \( \gamma _{v}\in \mathit{ground}(\gamma ) \) is the ground
clause of the form \( H\leftarrow L_{1}\wedge \ldots \wedge L_{r} \)
used at the root of \( T \). 

The proof proceeds by considering the following cases: \emph{either}
\( \gamma  \) belongs to \( P_{k+1} \) \emph{or} \( \gamma  \)
does not belong to \( P_{k+1} \) because it has been replaced (together
with other clauses in \( P_{k} \)) with new clauses derived by an
application of a transformation rule among R3, R4, R5, R6, R7, R8,
R9, R10 (recall that R1 and R2 are not applied in \( P_{j},\ldots ,P_{m} \)).
We present only the case where \( P_{k+1} \) is derived from \( P_{k} \)
by positive folding (rule R5). The other cases are similar and are
left to the reader.

Suppose that \( P_{k+1} \) is derived from \( P_{k} \) by folding
clauses \( \gamma _{1},\ldots ,\gamma _{m} \) in \( P_{k} \) using
clauses \( \delta _{1},\ldots ,\delta _{m} \) in (a variant of) \( \mathit{Defs}_{k} \),
and let \( \gamma  \) be \( \gamma _{p} \), with \( 1\leq p\leq m \).
Suppose also that, for \( i=1,\ldots ,m \), clause \( \delta _{i} \)
is of the form \( K\leftarrow d_{i}\wedge B_{i} \) and clause \( \gamma _{i} \)
is of the form \( H\leftarrow c\wedge d_{i}\vartheta \wedge G_{L}\wedge B_{i}\vartheta \wedge G_{R} \),
for a substitution \( \vartheta  \) satisfying Conditions (i) and
(ii) given in (R5). The clause \( \eta  \) derived by folding \( \gamma _{1},\ldots ,\gamma _{m} \)
using \( \delta _{1},\ldots ,\delta _{m} \) is of the form: \( H\leftarrow c\wedge G_{L}\wedge K\vartheta \wedge G_{R} \).
Since we use \( \gamma _{v} \) at the root of \( T \), we have that:
(a)~\( v(H)=A \), (b)~\( \mathcal{D}\models v(c\wedge d_{p}\vartheta ) \),
and (c)~\( v(G_{L}\wedge B_{p}\vartheta \wedge G_{R})=L_{1}\wedge \ldots \wedge L_{r} \),
that is, for some \( f1 \), \( f2 \), \( v(G_{L})=L_{1}\wedge \ldots \wedge L_{f1} \),
\( v(B_{p}\vartheta )=L_{f1+1}\wedge \ldots \wedge L_{f2} \), and
\( v(G_{R})=L_{f2+1}\wedge \ldots \wedge L_{r} \). By Proposition~\ref{prop:mu-consistency}
and the inductive hypotheses (Soundness and Completeness), for \( h=f1+1,\ldots ,f2 \),
if \( L_{h} \) is an atom then there exists a proof tree for \( L_{h} \)
and \( P_{j} \), and if \( L_{h} \) is a negated atom \( \neg A_{h} \)
then there is no a proof tree for \( A_{h} \) and \( P_{j} \). By
Proposition~\ref{prop:tc_of_pos_unf}, by the fact that (by ii) \( \mathcal{D}\models v(d_{p}\vartheta ) \),
and by the fact that \( \delta _{p}\in P_{i} \) (recall that \( \mathit{Defs}_{k}\subseteq P_{i} \)),
we have that there exists a proof tree for \( v(K\vartheta ) \) and
\( P_{j} \). Moreover, since \( K\leftarrow d_{p}\wedge B_{p} \)
has been unfolded w.r.t.~a positive literal, we have that:

\smallskip{}
(\( \dagger  \))~~~~\( \mu (v(B_{p}\vartheta ),P_{j})\geq \mu (v(K\vartheta ),P_{j}) \)
\smallskip{}

\noindent By Proposition~\ref{prop:mu-consistency} and the inductive
hypothesis (Completeness), there exists a proof tree for \( v(K\vartheta ) \)
and \( P_{k} \). Since \( T \) is \( P_{j} \)-consistent we have
that, for \( h=1,\ldots ,r \), \( \mu (A,P_{j})>\mu (L_{h},P_{j}) \).
Moreover, we have that:
\smallskip{}

\noindent \begin{tabular}{cl}
\( \mu (A,P_{j}) \)&
\( >\mu (v(G_{L}\wedge B_{p}\vartheta \wedge G_{R}),P_{j}) \) ~~~~~~~~~~~~~~~~~~(because
\( T \) is \( P_{j} \)-consistent)\\
&
\( =\mu (v(G_{L}),P_{j})\oplus \mu (v(B_{p}\vartheta ),P_{j})\oplus \mu (v(G_{R}),P_{j}) \)~~~~~(by
definition of \( \mu  \))\\
&
\( \geq \mu (v(G_{L}),P_{j})\oplus \mu (v(K\vartheta ),P_{j})\oplus \mu (v(G_{R}),P_{j}) \)~~~~~~(by
(\( \dagger  \)))\\
&
\( \geq \mu (v(K\vartheta ),P_{j}) \) ~~~~~~~~~~~~~~~~~~~~~~~~~~~~~~~~~~~~~~~~~~~~~~(by
definition of \( \mu  \))\\
\end{tabular}
\smallskip{}

\noindent By the inductive hypotheses (I\( \mu  \)) and (IS), for
\( h=1,\ldots ,f1,f2\! +\! 1,\ldots ,r \), if \( L_{h} \) is an
atom then there exists a \( P_{j} \)-consistent proof tree \( U_{h} \)
for \( L_{h} \) and \( P_{k+1} \), and if \( L_{h} \) is a negated
atom \( \neg A_{h} \) then there is no a proof tree for \( A_{h} \)
and \( P_{k+1} \). Moreover, by the inductive hypothesis (I\( \mu  \)),
there exists a \( P_{j} \)-consistent proof tree \( \widehat{U} \)
for \( v(K\vartheta ) \) and \( P_{k+1} \). 

Now we construct a \( P_{j} \)-consistent proof tree \( U \) for
\( A \) and \( P_{k+1} \) as follows. By (a) and (b) we have that
\( v(H)=A \) and \( \mathcal{D}\models v(c) \). Thus, we construct
the children of \( A \) in \( U \) by using the clause \( \eta  \):
\( H\leftarrow c\wedge G_{L}\wedge K\vartheta \wedge G_{R} \). Since
\( v(G_{L}\wedge K\vartheta \wedge G_{R})=L_{1}\wedge \ldots \wedge L_{f1}\wedge v(K\vartheta )\wedge L_{f2+1}\wedge \ldots \wedge L_{r} \),
the children of \( A \) in \( U \) are: \( L_{1},\ldots ,L_{f1},v(K\vartheta ),L_{f2+1},\ldots ,L_{r} \).
The construction of \( U \) continues as follows. For \( h=1,\ldots ,f1,f2\! +\! 1,\ldots ,r \),
if \( L_{h} \) is an atom then \( U_{h} \) is the \mbox{\( P_{j} \)-consistent}
subtree of \( U \) rooted in \( L_{h} \), else if \( L_{h} \) is
a negated atom then \( L_{h} \) is a leaf of \( U \). Finally, the
subtree of \( U \) rooted in \( v(K\vartheta ) \) is the \mbox{\( P_{j} \)-consistent}
proof tree \( \widehat{U} \). 

The proof tree \( U \) is indeed \( P_{j} \)-consistent because:
(i)~for \( h=1,\ldots ,f1, \) \( {f2\! +\! 1},\ldots ,r \),~ \( \mu (A,P_{j})>\mu (L_{h},P_{j}) \),
(ii) \( \mu (A,P_{j}) \)\( \geq \mu (v(K\vartheta ),P_{j}) \), and
(iii)~every subtree rooted in one of the literals \( L_{1},\ldots ,L_{f1},v(K\vartheta ),L_{f2+1},\ldots ,L_{r} \)
is \( P_{j} \)-consistent. \hfill{}\( \Box  \)


\begin{thebibliography}{10}

\bibitem{Al&99}
M.~Alpuente, M.~Falaschi, G.~Moreno, and G.~Vidal.
\newblock A transformation system for lazy functional logic programs.
\newblock In A.~Middeldorp and T.~Sato, editors, {\em Proceedings of the 4th
  Fuji International Symposium on Functional and Logic Programming, FLOPS'99},
  Lecture Notes in Computer Science 631, pages 147--162. Springer-Verlag, 1999.

\bibitem{Apt90}
K.~R. Apt.
\newblock Introduction to logic pro\-gramming.
\newblock In J.~van Leeuwen, editor, {\em Handbook of Theoretical Computer
  Science}, pages 493--576. Elsevier, 1990.

\bibitem{Apt97}
K.~R. Apt.
\newblock {\em From Logic Programming to {P}rolog}.
\newblock Prentice Hall, London, UK, 1997.

\bibitem{ApB94}
K.~R. Apt and R.~N. Bol.
\newblock Logic programming and negation: A survey.
\newblock {\em Journal of Logic Programming}, 19, 20:9--71, 1994.

\bibitem{Au&97}
J.-M. Autebert, J.~Berstel, and L.~Boasson.
\newblock Context-free languages and pushdown automata.
\newblock In G.~Rozenberg and A.~Salomaa, editors, {\em Handbook of Formal
  Languages}, volume~1, pages 111--174. Springer, Berlin, 1997.

\bibitem{Ba&03}
D.~Basin, Y.~Deville, P.~Flener, A.~Hamfelt, and J.F. Nilsson.
\newblock Synthesis of programs in computational logic.
\newblock In M.~Bruynooghe and K.-K. Lau, editors, {\em Program Development in
  Computational Logic}. Springer, 2004.
\newblock This volume.

\bibitem{BeG98}
N.~Bensaou and I.~Guessarian.
\newblock Transforming constraint logic pro\-grams.
\newblock {\em Theoretical Computer Science}, 206:81--125, 1998.

\bibitem{Bo&92a}
A.~Bossi, N.~Cocco, and S.~Etalle.
\newblock Trans\-forming normal pro\-grams by replacement.
\newblock In A.~Pettorossi, editor, {\em Proceedings 3rd International Workshop
  on Meta-Pro\-gramming in Logic, Meta '92, Uppsala, Sweden}, Lecture Notes in
  Computer Science 649, pages 265--279, Berlin, 1992. Springer-Verlag.

\bibitem{BuD77}
R.~M. Burstall and J.~Darlington.
\newblock A trans\-form\-ation system for developing recursive pro\-grams.
\newblock {\em Journal of the {ACM}}, 24(1):44--67, January 1977.

\bibitem{Ga&96}
M.~Garcia de~la Banda, M.~Hermenegildo, M.~Bruynooghe, V.~Dumortier,
  G.~Janssens, and W.~Simoens.
\newblock Global analysis of constraint logic programs.
\newblock {\em ACM Transactions on Programming Languages and Systems},
  18(5):564--614, 1996.

\bibitem{EtG96}
S.~Etalle and M.~Gabbrielli.
\newblock Trans\-form\-ations of {CLP} modules.
\newblock {\em Theoretical Computer Science}, 166:101--146, 1996.

\bibitem{Fio02}
F.~Fioravanti.
\newblock {\em Transformation of Constraint Logic Programs for Software
  Specialization and Verification}.
\newblock PhD thesis, Universit\`a di Roma ``La Sapienza", Italy, 2002.

\bibitem{Fi&01a}
F.~Fioravanti, A.~Pettorossi, and M.~Proietti.
\newblock Verifying {CTL} properties of infinite state systems by specializing
  constraint logic programs.
\newblock In {\em Proceedings of the ACM Sigplan Workshop on Verification and
  Computational Logic VCL'01, Florence (Italy)}, Technical Report
  DSSE-TR-2001-3, pages 85--96. University of Southampton, UK, 2001.

\bibitem{Fi&02b}
F.~Fioravanti, A.~Pettorossi, and M.~Proietti.
\newblock Specialization with clause splitting for deriving deterministic
  constraint logic programs.
\newblock In {\em Proceedings of the IEEE International Conference on Systems,
  Man and Cybernetics, Hammamet (Tunisia)}. IEEE Computer Society Press, 2002.

\bibitem{FrO97b}
L.~Fribourg and H.~Ols\'{e}n.
\newblock Proving safety properties of infinite state systems by compilation
  into {P}resburger arithmetic.
\newblock In {\em CONCUR '97}, Lecture Notes in Computer Science 1243, pages
  96--107. Springer-Verlag, 1997.

\bibitem{GaS91}
P.~A. Gardner and J.~C. Shepherdson.
\newblock Unfold/\-fold trans\-form\-ations of logic pro\-grams.
\newblock In J.-L. Lassez and G.~Plotkin, editors, {\em Computational Logic,
  Essays in Honor of Alan Robin\-son}, pages 565--583. MIT, 1991.

\bibitem{GeK94}
M.~Gergatsoulis and M.~Katzouraki.
\newblock Unfold/fold trans\-form\-ations for definite clause pro\-grams.
\newblock In M.~Hermenegildo and J.~Penjam, editors, {\em Proceedings Sixth
  International Symposium on Pro\-gramming Language Implementation and Logic
  Pro\-gramming (PLILP '94)}, Lecture Notes in Computer Science 844, pages
  340--354. Springer-Verlag, 1994.

\bibitem{Hog81}
C.~J. Hogger.
\newblock Derivation of logic pro\-grams.
\newblock {\em Journal of the {ACM}}, 28(2):372--392, 1981.

\bibitem{JaM94}
J.~Jaffar and M.~Maher.
\newblock Constraint logic programming: A survey.
\newblock {\em Journal of Logic Programming}, 19/20:503--581, 1994.

\bibitem{Ja&98}
J.~Jaffar, M.~Maher, K.~Marriott, and P.~Stuckey.
\newblock The semantics of constraint logic programming.
\newblock {\em Journal of Logic Programming}, 37:1--46, 1998.

\bibitem{Jo&93}
N.~D. Jones, C.~K. Gomard, and P.~Sestoft.
\newblock {\em Partial Evaluation and Automatic Pro\-gram Generation}.
\newblock Prentice Hall, 1993.

\bibitem{KaF86}
T.~Kanamori and H.~Fujita.
\newblock Unfold/fold trans\-form\-ation of logic pro\-grams with counters.
\newblock Technical Report 179, ICOT, Tokyo, Japan, 1986.

\bibitem{KaH87}
T.~Kanamori and K.~Horiuchi.
\newblock Construction of logic pro\-grams based on generalized unfold/fold
  rules.
\newblock In {\em Proceedings of the Fourth International Conference on Logic
  Pro\-gramming}, pages 744--768. The MIT Press, 1987.

\bibitem{LeB02}
M.~Leuschel and M.~Bruynooghe.
\newblock Logic program specialisation through partial deduction: Control
  issues.
\newblock {\em Theory and Practice of Logic Programming}, 2(4\&5):461--515,
  2002.

\bibitem{LeM99}
M.~Leuschel and T.~Massart.
\newblock Infinite state model checking by abstract interpretation and program
  specialization.
\newblock In A.~Bossi, editor, {\em Proceedings of LOPSTR '99, Venice, Italy},
  Lecture Notes in Computer Science 1817, pages 63--82. Springer, 1999.

\bibitem{Llo87}
J.~W. Lloyd.
\newblock {\em Foundations of Logic Pro\-gramming}.
\newblock Springer-Verlag, Berlin, 1987.
\newblock Second Edition.

\bibitem{Mah93}
M.~J. Maher.
\newblock A trans\-form\-ation system for deductive data\-base modules with
  perfect model semantics.
\newblock {\em Theoretical Computer Science}, 110:377--403, 1993.

\bibitem{MaS98}
K.~Marriott and P.~Stuckey.
\newblock {\em Programming with Constraints: {A}n Introduction}.
\newblock The MIT Press, 1998.

\bibitem{PeP94}
A.~Pettorossi and M.~Proietti.
\newblock Trans\-form\-ation of logic pro\-grams: Foundations and techniques.
\newblock {\em Journal of Logic Programming}, 19,20:261--320, 1994.

\bibitem{PeP99a}
A.~Pettorossi and M.~Proietti.
\newblock Synthesis and transformation of logic programs using unfold/fold
  proofs.
\newblock {\em Journal of Logic Programming}, 41(2\&3):197--230, 1999.

\bibitem{PeP00a}
A.~Pettorossi and M.~Proietti.
\newblock Perfect model checking via unfold/fold transformations.
\newblock In J.~W. Lloyd, editor, {\em First International Conference on
  Computational Logic, CL'2000, London, UK, 24-28 July, 2000}, Lecture Notes in
  Artificial Intelligence 1861, pages 613--628. Springer, 2000.

\bibitem{PeP02a}
A.~Pettorossi and M.~Proietti.
\newblock Program {D}erivation = {R}ules + {S}trategies.
\newblock In A.~Kakas and F.~Sadri, editors, {\em Computational Logic: Logic
  Programming and Beyond {\rm (}Essays in honour of Bob Kowalski, Part I{\rm
  )}}, Lecture Notes in Computer Science 2407, pages 273--309. Springer, 2002.

\bibitem{Pe&97a}
A.~Pettorossi, M.~Proietti, and S.~Renault.
\newblock Reducing nondeterminism while specializing logic programs.
\newblock In {\em Proc. 24-th {ACM} Symposium on Principles of Pro\-gramming
  Languages, Paris, France}, pages 414--427. ACM Press, 1997.

\bibitem{PrP95a}
M.~Proietti and A.~Pettorossi.
\newblock Unfolding-definition-folding, in this order, for avoiding unnecessary
  variables in logic pro\-grams.
\newblock {\em Theoretical Computer Science}, 142(1):89--124, 1995.

\bibitem{Prz87}
T.~C. Przymusinski.
\newblock On the declarative semantics of stratified deductive data\-bases and
  logic pro\-grams.
\newblock In J.~Minker, editor, {\em Foundations of Deductive Data\-bases and
  Logic Pro\-gramming}, pages 193--216. Morgan Kaufmann, 1987.

\bibitem{Ro&00}
A.~Roychoudhury, K.~Narayan Kumar, C.~R. Ramakrishnan, I.~V. Ramakrishnan, and
  S.~A. Smolka.
\newblock Verification of parameterized systems using logic program
  transformations.
\newblock In {\em Proceedings of the Sixth International Conference on Tools
  and Algorithms for the Construction and Analysis of Systems, TACAS 2000,
  Berlin, Germany}, Lecture Notes in Computer Science 1785, pages 172--187.
  Springer, 2000.

\bibitem{Ro&02}
A.~Roychoudhury, K.~Narayan Kumar, C.~R. Ramakrishnan, and I.V. Ramakrishnan.
\newblock Beyond {T}amaki-{S}ato style unfold/fold transformations for normal
  logic programs.
\newblock {\em International Journal on Foundations of Computer Science},
  13(3):387--403, 2002.

\bibitem{Ro&99a}
A.~Roychoudhury, K.~Narayan Kumar, C.R. Ramakrishnan, and I.V. Ramakrishnan.
\newblock A parameterized unfold/fold transformation framework for definite
  logic programs.
\newblock In {\em Proceedings of {P}rinciples and {P}ractice of {D}eclarative
  {P}rogramming ({PPDP})}, Lecture Notes in Computer Science 1702, pages
  396--413. Springer-Verlag, 1999.

\bibitem{San96}
D.~Sands.
\newblock Total correctness by local improvement in the transformation of
  functional programs.
\newblock {\em {ACM} Toplas}, 18(2):175--234, 1996.

\bibitem{Sat92}
T.~Sato.
\newblock An equivalence preserving first order unfold/fold trans\-form\-ation
  system.
\newblock {\em Theoretical Computer Science}, 105:57--84, 1992.

\bibitem{SaT84}
T.~Sato and H.~Tamaki.
\newblock Transformational logic program synthesis.
\newblock In {\em Proceedings of the International Conference on Fifth
  Generation Computer Systems}, pages 195--201. ICOT, 1984.

\bibitem{Sek91}
H.~Seki.
\newblock Unfold/fold trans\-form\-ation of stratified pro\-grams.
\newblock {\em Theoretical Computer Science}, 86:107--139, 1991.

\bibitem{Sek93}
H.~Seki.
\newblock Unfold/fold trans\-form\-ation of general logic pro\-grams for
  well-founded semantics.
\newblock {\em Journal of Logic Pro\-gramming}, 16(1{\&}2):5--23, 1993.

\bibitem{TaS84}
H.~Tamaki and T.~Sato.
\newblock Unfold/fold trans\-form\-ation of logic pro\-grams.
\newblock In S.-{\AA}. T{{\"a}}rnlund, editor, {\em Proceedings of the Second
  International Conference on Logic Pro\-gramming}, pages 127--138, Uppsala,
  Sweden, 1984. Uppsala University.

\end{thebibliography}
\end{document}